\documentclass[journal,twoside]{IEEEtran}
\usepackage{amsmath,amsfonts}
\usepackage{algorithmic}
\usepackage{algorithm}
\usepackage{array}
\usepackage{textcomp}
\usepackage{stfloats}
\usepackage{url}
\usepackage{verbatim}
\usepackage{graphicx}
\usepackage{cite}
\usepackage{subcaption}
\usepackage{soul}
\usepackage{cleveref}
\usepackage{mathrsfs}
\usepackage{xcolor}
\usepackage{multirow}
\usepackage{booktabs}
\usepackage{amssymb}
\usepackage{bbding}
\newcommand{\eg}{e.\,g., }
\newcommand{\ie}{i.\,e., }

\newcommand{\shuo}[1] {{\color{black} #1}}

\begin{document}

\title{Audio Self-supervised Learning: A Survey}

\author{Shuo Liu, Adria Mallol-Ragolta, Emilia Parada-Cabaleiro, Kun Qian,  \IEEEmembership{Senior Member, IEEE,} \\ Xin Jing, Alexander Kathan, Bin Hu, \IEEEmembership{Senior Member, IEEE,} and Bj\"orn W.\ Schuller, \IEEEmembership{Fellow, IEEE}
        % <-this % stops a space
%\thanks{This paper was produced by the IEEE Publication Technology Group. They are in Piscataway, NJ.}% <-this % stops a space
\thanks{Manuscript received Jan, 2022.}
\thanks{Shuo Liu, Adria Mallol-Ragolta, Xin Jing, Alexander Kathan are with the Chair of Embedded Intelligence for Health Care \& Wellbeing, University of Augsburg, Augsburg 86159, Germany (e-mail:~\{shuo.liu,~adria.mallol-ragolta,~xin.jing,~alexander.kathan\}@uni-a.de). }
\thanks{Emilia Parada-Cabaleiro is with the Institute of Computational Perception, Johannes Kepler University Linz, Linz  4040, Austria  (e-mail:~emilia.parada-cabaleiro@jku.at).}
\thanks{Kun Qian and Bin Hu are with the School of Medical Technology, Beijing Institute of Technology, China (e-mail:~\{qian,~bh\}@bit.edu.cn)} 
\thanks{Bj\"orn W.\ Schuller is leading the Chair of Embedded Intelligence for Health Care \& Wellbeing, University of Augsburg, Augsburg 86159, Germany. He is also the head of GLAM -- the Group on Language, Audio, \& Music, Imperial College London, London SW7 2AZ, UK. (e-mail:~bjoern.schuller@imperial.ac.uk).}
}

%IEEE Communications Survey \& Tutorials
\markboth{}{Liu \MakeLowercase{\textit{et al.}}: Audio SSL}
% ~Vol.~xx, No.~x, xxxx~2022

\maketitle

\begin{abstract}
Inspired by the humans' cognitive ability to generalise knowledge and skills,   
Self-Supervised Learning (SSL) targets at discovering general representations from large-scale data without requiring human annotations, which is an expensive and time consuming task. 
Its success in the fields of computer vision and natural language processing have prompted its recent adoption into the field of audio and speech processing.  
Comprehensive reviews summarising the knowledge in audio SSL are currently missing.
To fill this gap, in the present work, we provide an overview of the SSL methods used for audio and speech processing applications.  
Herein, we also summarise the empirical works that exploit the audio modality in multi-modal SSL frameworks, and the existing suitable benchmarks to evaluate the power of SSL in the computer audition domain. 
Finally, we discuss some open problems and point out the future directions on the development of audio SSL.
\end{abstract}


\section{Introduction}
\label{sec:intro}
\noindent
\IEEEPARstart{A}{ccording} to Piaget's theory of cognitive development \cite{piaget1964part,huitt2003piaget}, since their birth  up to approximately 18 months, children acquire knowledge  from sensory and motor  experiences.    During this stage,  \ie the `sensorimotor' stage, through basic actions such as sucking, grasping, looking, and listening, the   early representational thought emerges \cite{baillargeon1991object}.  Along with the acquisition  of  knowledge, over the different developmental stages until the last one, \ie  the `formal operational'  (adolescence  and adulthood) stage,   children's reasoning progressively moves towards the acquisition of  abstract ideas and  the  use of  deductive logic, \ie   subtracting specific information from a general principle  \cite{oesterdiekhoff2016child}.  During this process, in order to   understand the world,   the so-called `schemas', \ie \textit{higher-order cognitive structures that have been hypothesised to underlie many aspects of human knowledge and skill} \cite{brewer1984nature}, emerge.  According to Piaget, children development is interpreted trough  an equilibration mechanism which explains how  new information is balanced according to old knowledge. Equilibration involves `assimilation'   (the process of taking in new information to fit in with the pre-existing schemas) and  `accommodation' (the process of  modifying the pre-existing schemas  as a result of new information)  \cite{piaget1964part,wadsworth1996piaget}.   In this view, learning is possible if  complex structures are based on simpler ones, \ie when  a natural  development between structures exists instead of a simple  external reinforcement  \cite{piaget1964part}. Indeed, the interesting
aspect 
of learning (and one of the main goals in education) is to create dynamic structures which can lead to generalisation,  \ie the ability   to apply  learnt knowledge and skills for understanding a different context. This is known as  `transfer of learning'    \cite{perkins1992transfer}.  Inspired by  the cognitive process of developing   dynamic structures with the capability of generalisaiton,  the  Self-Supervised Learning (SSL) paradigm has been presented \cite{jing2020self,bengio2013representation,raina2007self,liu2021self} ---  a machine and deep  learning
technique  rapidly evolving in the last years.  SSL  targets at learning a model that is able to produce universal representations. This is  approached   by first solving some pretext tasks (also known as upstream tasks in literature), \ie a procedure which, similarly to the sensorimotor stage, enables to artificially  learn representations directly from the data attributes  without the need of human-annotations %labelled  data 
\cite{bansal2021}. %\emi{How does the pretext task relate to the cognitive developmental theory? is the corresponding to the pre-existing schemas?} \sure{The pretext tasks do not need } 
Then, with a %versatile 
pre-trained model generated on the pretext task, feature representations are extracted to understand new data, %,is used to extract features (representations) \textcolor{olive}{and use them on a different context. 
\ie similarly to cognitive development, a pre-trained model (previous knowledge) can be used through generalisation to understand a new context, a process  known as  %properly perform for diverse 
downstream task \cite{teng2021}. %\emi{you need to define downstream task and say how this relates to the cognitive development. Is the downstream task the different context?}. 

SSL mitigates two difficulties that currently %impede the progress of \shuo{applying} 
limit the application of deep learning: the need of human annotations and the difficulty in designing effective network architectures for specific tasks. 
First, the current success of deep learning reckons on big data which typically consumes uninhibited human efforts in annotations. This  faces  the issue of annotation bias as well as the fact that annotation procedures often cannot optimally preserve  data privacy. As SSL learns representations from the data itself without the need of labels \cite{lee2021} (sometimes creates pseudo-labels for self-supervision), it overtakes  the challenges derived by the use of human annotations.  
%\textcolor{red}{Second, designing efficient network architectures bla bla bla} \emi{can you repeat the limitation? rephrased or maybe (if possible)  more precisely?} 
Second, as long as the pretext model can generate proper representations of the data, these can be used for multiple downstream tasks, reducing, at the same time, the difficulty in designing reliable downstream models. For instance, 
a Multi-layer Perceptron (MLP) is commonly used for this step,  reaching  state-of-the-art results for different research areas in computer science. 
%Furthermore, a well-trained upstream model 
%\am{would it clarify the reading describing what you mean by upstream and downstream models?} 
As the main effort of SSL concentrates on the development of well-trained upstream models, it  guarantees to extract data representations with a sufficient level of generalisation and  distinctiveness. %%%EMI%%%\hl{, by this enabling to discriminate  between different instances as well as between  instances from different classes.} \emi{can we drop this? it does not add much and makes the sentence so long and complicated}
%\hl{Hence, the technique requires much less (a small subset of) annotated data in downstream tasks for achieving good performance.}   \emi{I would also drop this one... it seems contradictory, as we said that SSL does not need human annotations...} 
% S5: Hence, the main effort of SSL focus on pretext training.
%Therefore, the main effort of SSL concentrates on the upstream training.  
% S5+: SSL gets use of negative samples in order to learn the difference boundary between objects.
Furthermore, as a way to increase %achieving sufficient 
distinctiveness of the learnt representations, when solving pretext tasks, negative examples can be additionally provided in order to contrast the target sample with negative examples \cite{wang2021understanding}.  %like distinguish cups from pencil jars and cans, which 
This process formalises the SSL into a contrastive learning framework  \cite{le2020contrastive,saunshi2019theoretical,jaiswal2021survey,tosh2021contrastive}.  
Finally,  it has been shown in many works, such as \cite{simclr} that the amount of labelled data needed to fine-tune a model during the downstream task considerably decreases when taking upstream models into account, which makes using SSL effective in the design of efficient architectures.
 %as the technique requires much less (a small subset of)  annotated data in downstream tasks for achieving good performance.  

% \emi{We might integrate the following paragraph here. Although, without more one-to-one connection with SSL, I do not think it is needed to include it.}
% \textcolor{green}{During the learning, children understand the associations between seen objects, heard sounds, etc and able to discriminate between them. To distinguish between similar objects but of different classes, such as cups and pencil jars, we are often given a pencil jar as an example for comparison. The difference learnt is not constrained on shapes, colors, materials, but higher-level observations of whether the container is proper for drinks or pencils.  
% The quantitative improvement of cognitive development results in qualitative changes, which explains the qualitative and quantitative differences between the thinking of young children versus older children.}

% S5++: The knowledge mismatch between prefix and downstream tasks mainly affect the performance for classification. However, the prior knowledge is beneficial to identify new objects, the benefit also originates from the generalisability of the representation learning, as can be seen for research of transfer learning, domain adaptation, ...

The fitness of upstream and downstream tasks, \ie  how much the knowledge learnt from pretext tasks is applicable to the downstream tasks, is partially determined by the data relevance  used in both steps.  From a cognitive point of view, this is comparable %is similar 
to the aforementioned `transfer of learning', 
%for which near transfer of knowledge, such as the one shown when 
as a speaker of a given  language would find it easier to learn a  related language (near transfer) than an unrelated one (far transfer). 
Thus, near transfer of knowledge is expected to ensure the downstream tasks to particularly benefit %as much as possible 
from the upstream training. 
However, far transfer may also occur. Therefore,  %benefit from prior knowledge. Similarly, 
downstream tasks that use data from a different domain can still benefit from learning representations of sufficient generalisability.  %%%
% S6: Methods have been used in different domains.
%The technique \emi{Which technique?} 
The versatility of SSL has yielded to a  superior performance in several research fields, such as,  Natural Language Processing (NLP) \cite{qiu2020pre}  and %later been found effective for 
Computer Vision (CV)  %tasks 
\cite{jing2020self,liu2021self}, as well as in  a variety of deep learning methods, \eg graphical neural networks \cite{wu2021self} and reinforcement learning \cite{shelhamer2017}, to name a few.
% S7: Audio SSL and their relations to others, mention the difficulty of audio to other modalities and this difficulty for audio SSL methods.
Nevertheless, processing audio sources increases further the difficulty of applying deep learning methods, as in real-world, this modality is typically characterised by many uncertainties. 
%However, for deep learning, audio is an relative more difficult modality to analyse, as its recordings usually contain many uncertainties in real-world conditions. 
Speech, for instance,  due to within- and cross-speaker variations, such as those produced by disfluencies, as well as differences in language, acoustic environments, or recording setups, presents usually a considerable variability. 
%
%This makes it difficult to infer relevant latent structures without any supervision guidance. 
%EMI%
This  makes it difficult  deducing relevant latent structures without taking into account any supervision guidance.
%EMI%
In addition, unlike for 
%And different from 
images,   overlapping  noises %with the audio content of interest 
are typical of  recordings. Through its masking properties, surrounding noises   
%The overlapping can 
limit (and even  impede)  understanding, in some cases distorting the  spectrogram of the the audio content of interest. %, as strong noises can submerge the target audio content. From the perspective of the frequency domain, altering the spectrogram of the audio can also be distorted by the uncontrolled noise components. 
Indeed, as each pixel (time-frequency bin) of the spectrogram can be deteriorated,  noise reduction and removal  is still an open challenge in the field  \cite{liu2021n}.  %difficult problem in the field. Although efforts have been paid for tackling this issue in many works \cite{liu2021n}, however, they can still introduce distortions to the audio of interest.
Similarly,  in comparison to NLP tasks, which (despite their inherent difficulties) process texts that are comprised of limited possible words and characters, the infinite possibilities of audio that represent the same meaning creates more uncertainties in audio understanding.  %\emi{Are we sure nobody from NLP will be offended?... not sure, maybe it is ok}       
These might be indicators that %consideration 
explain why SSL has achieved lower performance for audio signal processing than for CV and NLP. %it has more difficulties in achieving comparable performance to CV and NLP tasks. 

% As SSL requires a model to have generalisation and discrimination abilities in parallel, audio processing faces a more challenging situation. A proper SSL model should be able to extract representations that are distributed - more expressive as the dimensionality increases, abstract - aggregate more abstract features and invariant to local changes in input audio, disentangled - each factor of the representation vector should be interpretable \cite{bengio2013representation}.  

% As SSL requires from a model both,  generalisation and discrimination (in parallel), using SSL for audio processing becomes  particularly challenging. 
Specifically, a proper SSL model should be able to extract representations that are: (i) distributed, \ie more expressive as the dimensionality increases; (ii) abstract, \ie aggregate more abstract features which are invariant to local changes in the input; % audio; 
(iii) disentangled, \ie each factor of the representation vector should be interpretable  \cite{bengio2013representation}.  %\emi{This previous sentence is generic, right? Is not just about audio processing} \sure{Yes, very generic.} \emi{perfect}
As SSL requires from a model both,  generalisation and discrimination (in parallel), using SSL for audio processing becomes  particularly challenging.  
%We notice that several overview papers 
Although several survey articles aimed to give an overview of the existing literature  on SSL have been presented to the research community, due to the prominent use of SSL in CV and NLP,  these works show a clear bias towards these two fields \cite{jing2020self,liu2021self,qiu2020pre}. %to help researchers better grab the SSL knowledge and existing works in the direction of 
%CV \cite{jing2020self,liu2021self} and NLP \cite{qiu2020pre}.   
However, despite the challenges, recent research  has shown an always increasing interest in applying SSL to audio sources. As this rapidly developing area has not been  systematically investigated yet, to fill this gap, we present a survey on SSL  
%
% S8: Recent evolvement but so far no systematic overview paper
%The approaches have recently boosted the research of SSL for audio processing, 
%we survey this rapidly developing area 
with a special emphasis on the recent progress, by this %and discuss their theoretical soundness by 
including for the first time an overview of SSL in  audio  within unified  frameworks. % of audio  SSL, combing the existing works, and presenting our prospect for future research of this topic.   
% S9: Our hope for readers (contributions)
By providing  an overview of the existing techniques  as well as a disambiguation  between approaches, this work is especially  thought  to support both  beginners and more experienced researchers interested in the use of  SSL for audio signal processing. 
% \hl{We hope, after reading the paper, beginners of audio SSL can grab an overview of the technique, and eliminate the conceptual confusion in approaches and with other similar techniques.  
% The experienced researchers can also benefit from the combing of the existing methods and inspired by our outlook.} \emi{Do we need this?} 
% S10: therefore, paper organisation 
%Hence, t
%

The rest of the manuscript is organised as follows. 
% - section1: unify the format of SSL 
We first introduce SSL in \Cref{sec:ssl}, aiming to unify its frameworks and cover the basic useful blocks and operations that lead to its success. 
%\hl{This  provides a good basis for %sorting out 
%introducing audio SSL  later on.} \emi{I guess I you could remove this sentence.}
% - section2: how audio tasks fit into the format and the design of pretext tasks, a) different training strategies, b) loss functions, c) network architecture
Then, we describe how this unified framework can fit for audio processing in \Cref{sec:audio} with discussion focus on training objectives, network architectures, and the training framework. 
% - section3: Multi-modality and Cross-modality
Next, SSL approaches exploiting audio as one of the modalities will be discussed in \Cref{sec:mm}.  
% - section4: Downstream tasks and data used for audio only
In \Cref{sec:ds}, we summarise the downstream tasks considered in the literature and list the databases  and benchmarks that are used for evaluating the performance of pretext tasks.
% - section5: Discussion
At last, we discuss several aspects of SSL, including its relations and difference to other similar deep learning techniques, 
% - section6: Conclusion by pointing out future reseach directions
before drawing a conclusion and pointing out the potential research directions. 

\begin{figure*}[ht!]
  \centering
    \hspace*{-1.3cm}
    \begin{subfigure}{0.25\textwidth}
        \centering
        \vspace{0.3cm}
        \hspace*{1cm}
        \includegraphics[height=2.4in]{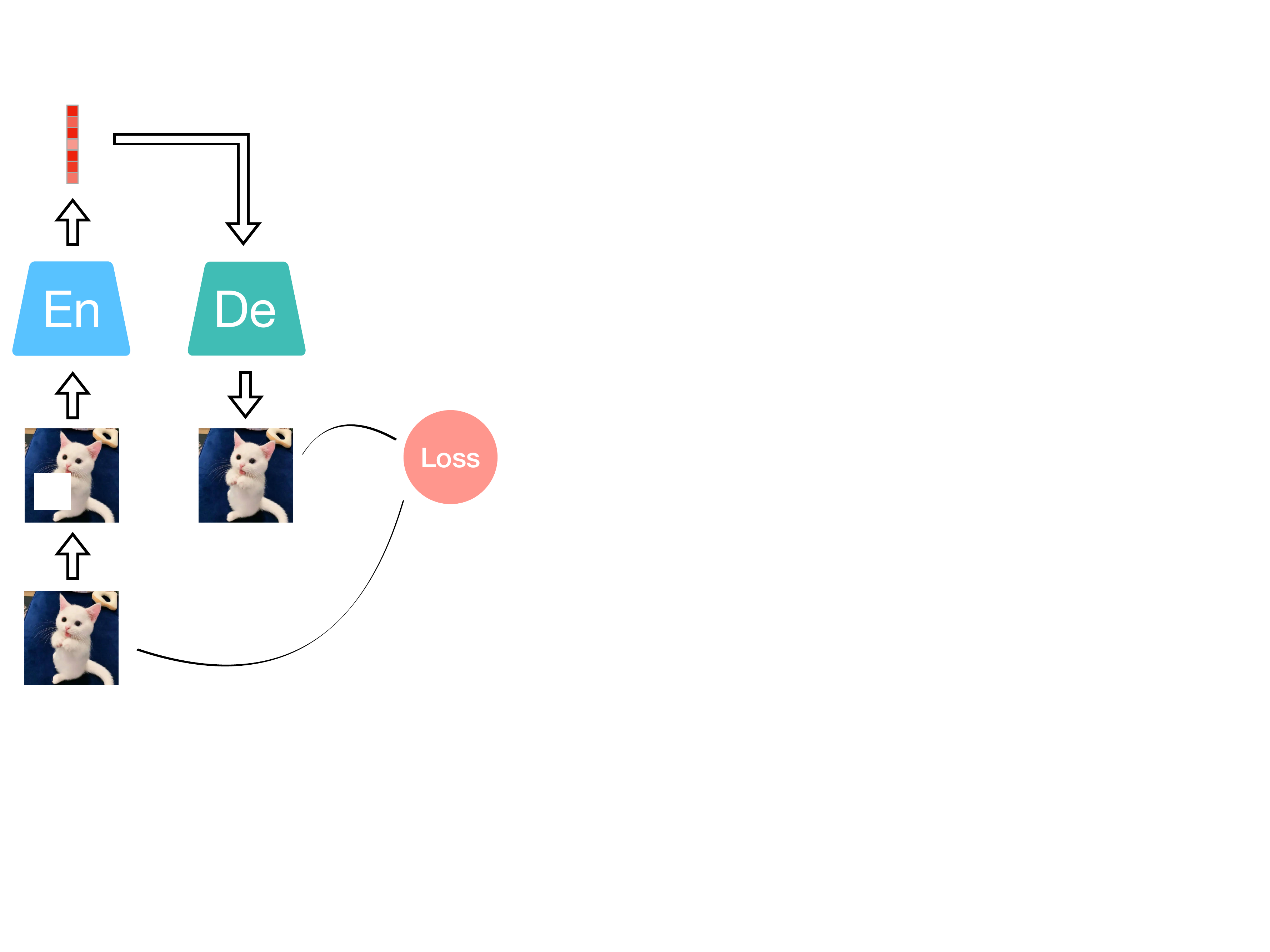}
        \vspace{-2cm}
        \caption{Auto-encoding}
        \label{fig:models(a)}
    \end{subfigure}%
    ~ 
    \hspace*{-0.2cm}
    \begin{subfigure}{0.25\textwidth}
        \centering
        \vspace{0.3cm}
        \hspace*{0.8cm}
        \includegraphics[height=2.4in]{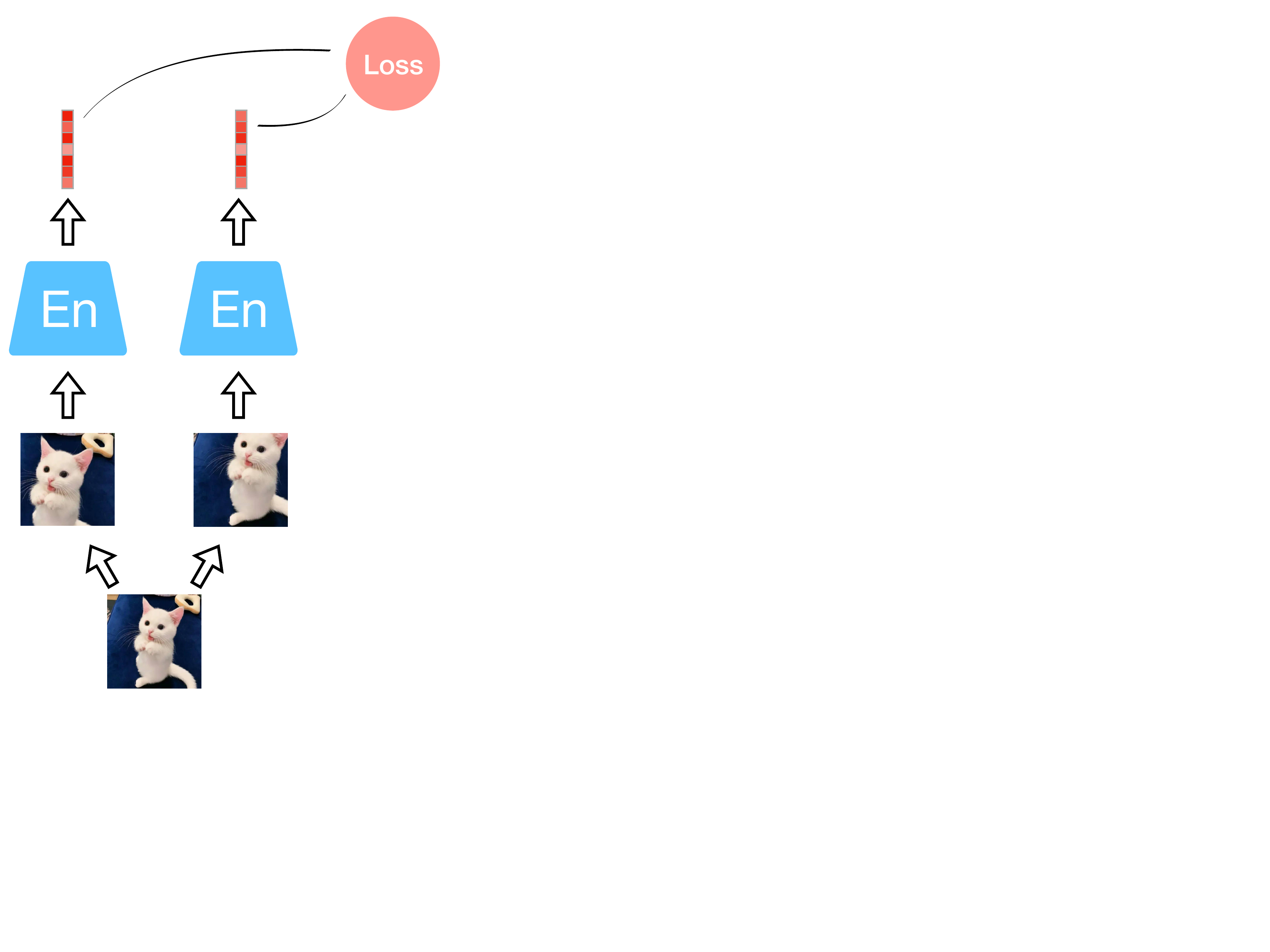}
        \vspace{-2cm}
        \caption{Siamese Network}
        \label{fig:models(b)}
    \end{subfigure}% 
    %\vspace{0.5cm}
    \hspace*{-0.8cm}
    \begin{subfigure}{0.25\textwidth}
        \centering
        \vspace*{-0.4cm}
        \hspace*{1cm}
        \includegraphics[height=2.4in]{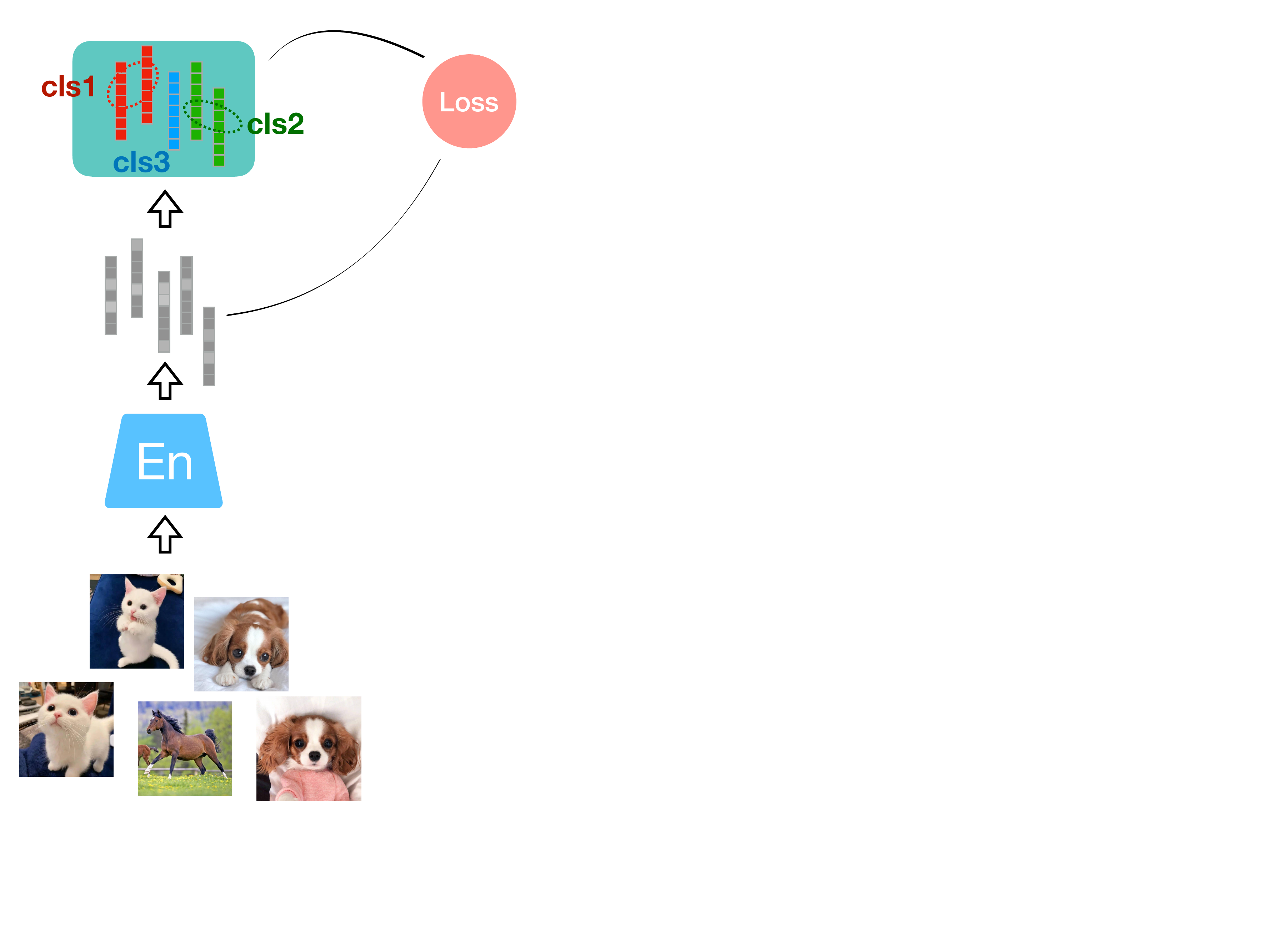}
        \vspace{-1.4cm}
        \caption{Clustering}
        \label{fig:models(c)}
    \end{subfigure}
    ~
    \hspace*{0.6cm}
    \begin{subfigure}{0.25\textwidth}
        \centering
        \hspace*{-0.5cm}
        % \vspace*{1.7cm}
        \includegraphics[height=2.4in]{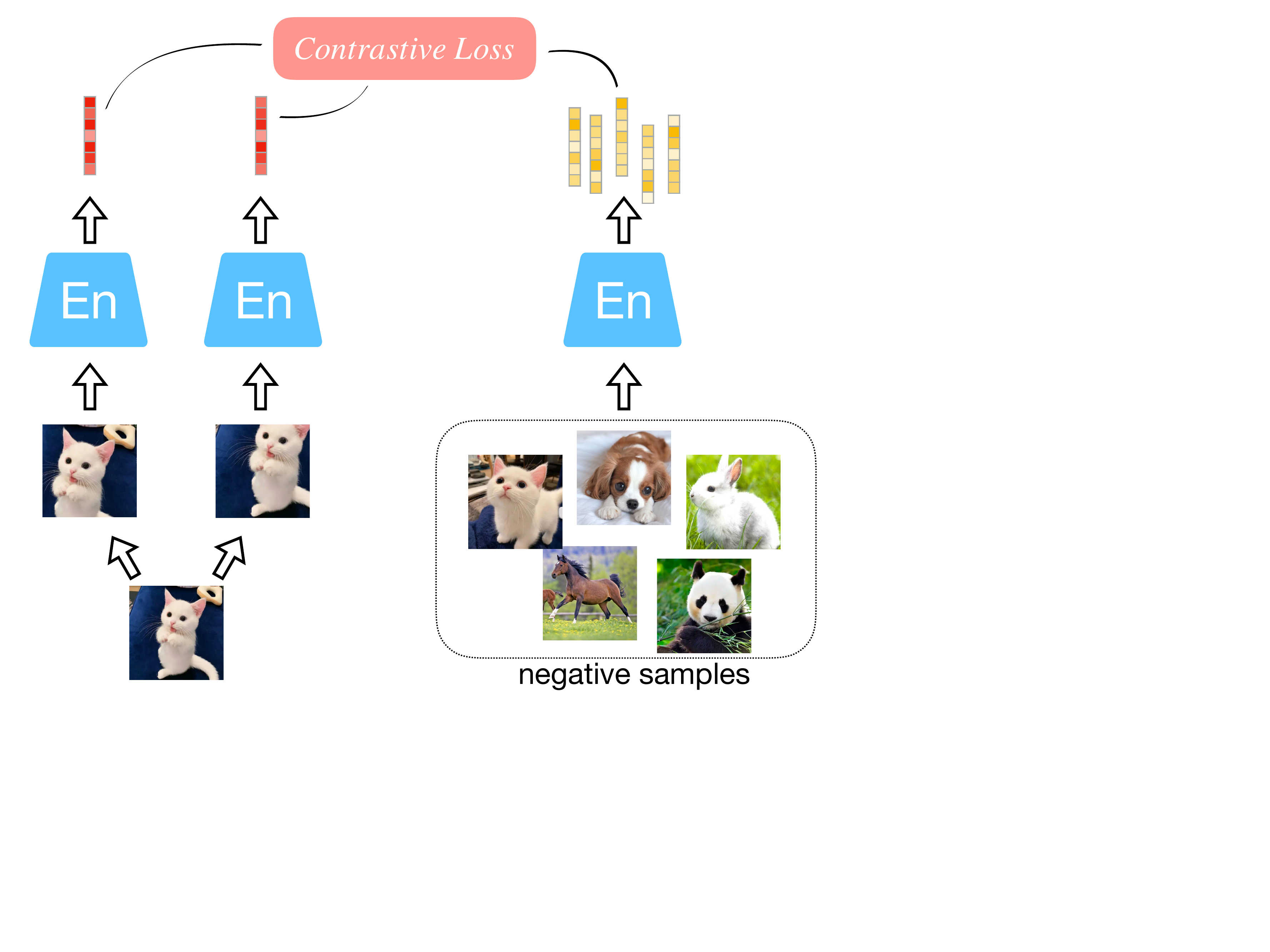}
        \vspace{-2.cm}
        \caption{Contrasitve SSL}
        \label{fig:models(d)}
    \end{subfigure}
  \caption{Predictive SSL frameworks (a-c) and contrastive SSL framework (d). For each framework, the diagram shows the components, including pseudo-labels that are used to construct training objectives. (a) Auto-encoding contains an encoder (En) and a decoder  (De). The encoder learns representations from a distorted signal input while the decoder aims at recovering the clean signal from the learnt representations; (b) a Siamese network processes two views of the same data point, hence the latent representation of one sub-network is seen as pseudo-label of the other sub-network; (c) clustering is applied for grouping the learnt representations -- the clustering centroids are used as pseudo-labels for training; (d) contrastive SSL constructs the contrastive loss through negative samples.}
  \label{fig:frameworks}
\end{figure*}

% \begin{figure}[t!]
%   \hspace*{-1cm}
%   \centering
%     \centering
%     \hspace*{1.2cm}
%     \includegraphics[height=3.5in]{Figures/contra.pdf}
%     \vspace{-3cm}
%     \caption{Contrastive SSL framework, which constructs the contrastive loss through negative samples.}
%     \label{fig:contrastive}
% \end{figure}

\section{Self-supervised Learning: A General Overview}
\label{sec:ssl}
\noindent
SSL aims at learning latent representations from large-scale data by solving designed pretext tasks, rather than using human annotations. 
%To this end, pseudo-labels are created from the data itself and subsequently used as training target.
%Each pseudo-label can be seen as another view of its corresponding data instance. Hence, both views of the same object, \ie instance and pseudo-label, present common attributes in a latent high-dimensional space. 
%SSL approaches exploit the natural correlations between the two views in order to learn a representation of the object which can be, to some extent, generalise. 
%Considering the difference between objects, comparing similar objects (or views) to dissimilar ones (defined as negative samples) in training objectives can improve the distinctiveness of the learnt representations. 
To this end,  different views of an object, which are of high natural correlation, are created. An SSL model is trained to generalise, to some extent,  its  representations in a latent high-dimensional space. 
By comparing the representations of the same object to other objects (defined as negative samples) in training, an SSL model is expected to produce representations that are of better distinctiveness. 
Dependent on whether negative samples are taken into the training process, SSL frameworks can be categorised into two classes: Predictive and contrastive (cf.\ \Cref{fig:frameworks}). %(cf.\ \Cref{fig:contrastive}). 
We will introduce these two basic frameworks and their variants in \Cref{subsec:pred} and \Cref{subsec:contrast}, respectively. 
In addition, as predictive models can also utilise contrastive loss as its training objectives, contrastive predictive coding (CPC), based on auto-regressive predictive codding (APC) and masked predictive coding (MPC), will be described in  \Cref{subsec:cpc}. 
We will also discuss the benefit of employing negative samples in training SSL models and the potential drawbacks.
As the process of generating the views of an object is the fundamental step for SSL, we will start by summarising the different methods  in the following.

\subsection{Views Generation}
\noindent
% In contrastive SSL, a positive pair is considered as  input for the model. This pair can be generated following several procedures described in the literature. 
%A typical way to get multiple views of the same context is to record it using multiple sensors, 
%EMI%
A typical way to get multiple views of the same context is  recording it by the use of multiple sensors, 
%EMI%
for example, using multiple cameras to shoot a scene from different angles \cite{sermanet2018time}. The %different 
views are not limited to be of the same modality but can also belong to a different one, 
such as an utterance with its text transcriptions \cite{chung2018unsupervised,sermanet2018time} or audio and image frames from the same video source \cite{zhao2018sound,alwassel2020self}. 
As using multi-modality and cross-modality is typical %applications 
in SSL, in \Cref{sec:mm}, the works specific to audio representation will be presented.
%that consider audio as one of the modalities . 
%\emi{I think you need to justify better why do you focus only on audio. Maybe saying something about the fact that the amount of works is so large that goes beyond the scope of this article, which will focus on revising audios, since underrepresented?}

%2) data transformation
An alternative way to create effective views in CV-related tasks is through data transformation, including image cropping \cite{simclr,he2020momentum}, rotation \cite{komodakis2018unsupervised}, colouring \cite{Larsson2017ColorizationAA}, and distortion\cite{simclr}, amongst others. 
In several audio works, after converting an audio waveform to its spectrogram or Mel representations, similar paradigms can be applied \cite{nandan2020,al2021clar}. In this case, data augmentation techniques such as warping and SpecAugment \cite{48482} can also be used for creating another view of the audio signal.  

% 4) sequential coherence and consistency (instance-instance)
Considering the spatial or temporal coherence and consistency in a signal, local features of different patches of an image or segments of an audio can be considered as multiple views of the same type of data \cite{dwibedi2018learning,sermanet2018time}. This approach is widely adopted for sequential input, such as video as a sequence of image frames \cite{sermanet2018time}, where two frames in a short temporal range are considered a positive pair, while frames that are far away in the same sequence or from other sequences are taken as negative samples. %
For an audio sequence, the positive pair can be segmented from the same recording, and the negative pair can be extracted from different recordings \cite{micro2019,saeed2021contrastive,fonseca2021unsupervised}.
%Besides, an image can also be decomposed into a sequence of pixels, in order to fit the framework. 

% 3) global-local mutual information (context-instance)
Another solution is to exploit the relationship between local features and global information, which aims at maximising their mutual information \cite{hjelm2018learning}. 
The global representation, often defined as context vector, aggregates the information of the entire context. An SSL model then learns to represent local features by capturing meaningful information relevant to the aggregated global representation. 
Deep InfoMax \cite{hjelm2018learning} codes an image into a global context vector, and contrasts  its distance to the spatial patches of the same image against the distance to spatial patches of different images. 
More often, the mechanism is used for a sequential data structure. As in \cite{oord2018representation}, a context vector collects the global information along the temporal direction in an auto-regressive manner. The context vector is then compared to the learnt representations of each local frame. Note that, in this work, contrastive loss (introduced in \Cref{subsec:contrast}) has been used in the framework of an auto-regressive coding model, leading to a contrastive predictive coding that can learn effective fixed-length audio representations from speech input. We leave its discussion for the next section, to which the reader is referred  for more details.    
Finally, \cite{49568} investigates the effect of the redundancy in two views of the positive pair, suggesting that the views with less mutual information  should be selected for training. The idea is to compress the redundancy in the embeddings of the two views that are not relevant to the labels \cite{49568}. Similarly, %Tschannen 
\cite{tschannen2019mutual} provides also empirical evidences about the fact that the contrastive loss is not only attributed to mutual information.

\subsection{SSL Frameworks}
\label{subsec:framework}

\noindent
% \am{Paragraph here briefly introducing what the reader will find in this section}

\subsubsection{Predictive Models}
\label{subsec:pred}
Predictive SSL, different from contrastive SSL, optimises the similarity or correlations between the representations of two views of the same object, without considering their similarity to that of negative samples in training objectives. Typical frameworks %such frames 
are auto-encoding and Siamese Networks. A clustering method requires no additional views generation, but explicitly groups the learnt representations based on the underlying similarity between each input. We describe the three typical predictive SSL frameworks in detail in the following.      

\begin{table*}[t!]
\centering
\caption{\shuo{An overview of the recent typical self-supervised learning methods. \textbf{FOS} abbreviates field of study. The type of frameworks refers to \Cref{fig:frameworks}. ``Other images'' in the \textbf{Source} column indicates other images of the mini-batch.}}
\label{tab:table1}
\centering
\setlength{\tabcolsep}{2mm}
\setlength{\arrayrulewidth}{0.3pt}
\renewcommand{\arraystretch}{1.2}
\begin{tabular}{c|c|c|c|c|c|c|c}
\hline
\multirow{2}{*}{\textbf{Model}} & \multirow{2}{*}{\textbf{FOS}} & \multirow{2}{*}{\textbf{Framework}} & \multirow{2}{*}{\textbf{Encoder}} & 
\multirow{2}{*}{\textbf{Pseudo-labels}} & \multirow{2}{*}{\textbf{Loss}} & \multicolumn{2}{c}{\textbf{Negative samples}} \\ 
%\cmidrule(lr){7-9}
&&&&&& \textbf{Source} & \textbf{Strategy} \\
\hline
\textbf{TCN embedding \cite{sermanet2018time}} & CV & (d) &  Inception network & Different but  & Triplet loss & Images of & End-to-end           \\
& & & + CNN & simultaneous viewpoint & & different time &\\
\hline
\textbf{SimCLR
\cite{simclr}} & CV & (d) &  ResNet & Data augmentation & NT-Xent loss & Other images& End-to-end\\
\hline
\textbf{SimCLR v2
\cite{chen2020}} & CV & (d) &  Variants of & Data augmentation & NT-Xent loss & Other images& End-to-end\\
(semi)& & &  ResNet & & & &\\
\hline
\textbf{MoCo
\cite{he2020momentum}} & CV & (d) &  ResNet & Data augmentation & InfoNCE loss & Other images & Momentum\\
\hline
\textbf{MoCo v2 \cite{chen2020improved}} & CV & (d) &  ResNet & Data augmentation & InfoNCE loss & Other images & Momentum\\
\hline
\textbf{MoCo v3 \cite{chen2021}} & CV & (d) &  Vision & Data augmentation & InfoNCE loss & Other images & End-to-end\\
& & &  Transformers & & & & \\
\hline
\textbf{RotNet
\cite{gidaris2018unsupervised}} & CV & (a) & ConvNet & Rotation directions & Prediction loss & - & -\\
\hline
\textbf{Colorization
\cite{Larsson2017ColorizationAA}} & CV & (a) & AlexNet, VGG-16, & Colour of missing patch & Regression loss & - & -\\
&&& ResNet-152 & & KL divergence & - & - \\
\hline
\textbf{DIM
\cite{hjelm2018learning}} & CV & (d) &  - & - & JSD, DV  & - & End-to-end\\
&&&&& or InfoNCE loss &&\\
% \hline
% \textbf{InfoMin
% \cite{49568}} & CV & (d) & ResNet-50 & - &  & - & end-to-end\\
\hline
\textbf{Word2Vec
\cite{word2vec}} & NLP & (a) & Auto-encoder & Context words & Prediction loss & - & -\\
\hline
\textbf{Speech2Vec
\cite{an2018}} & Audio & (a) & Auto-encoder & Context audio & Prediction loss & - & -\\
\hline
\textbf{Audio2Vec
\cite{audio2vec}} & Audio & (a) & Auto-encoder & Context audio & Prediction loss & - & -\\
\hline
\textbf{BERT
\cite{devlin2019}} & NLP & (a) & MPC & Masked words & Prediction loss & - & -\\
\hline
\textbf{ALBERT
\cite{lan2020self}} & NLP & (a) & MPC & Masked words & Prediction loss & - & -\\
& & & & Sentence order & & &\\
\hline
\textbf{NPC
\cite{liu21l_INTERSPEECH}} & Audio & (a) & MPC & Masked frames & Prediction loss & - & -\\
\hline
\textbf{BYOL
\cite{richemond2020byol}} & CV & (b) & ResNet & Data augmentation & MSE loss & - & -\\
\hline
\textbf{Barlow Twins
\cite{jure2021}} & CV & (b) & ResNet & Data augmentation & Eq. (3) & - & -\\
\hline
\textbf{SimSiam
\cite{chen2021exploring}} & CV & (b) & ResNet & Data augmentation & Negative & - & -\\
& & & & & cosine similarity & & \\
\hline
\textbf{DeepCluster
\cite{caron2018deep}} & CV & (c) & AlexNet, VGG-16 & Clustering centroids & Negative & - & -\\
& & & & & log-softmax loss & & \\
\hline
\textbf{Local Aggregation
\cite{zhuang2019local}} & CV & (c) & AlexNet, VGG-16 & Soft-clustering & Negative & - & -\\
& & & &centroids  & log-softmax loss & & \\
\hline
\textbf{SwAV
\cite{caron2020unsupervised}} & CV & (c) & Variants of & Online-clustering & Modified & - & End-to-end\\
& & & ResNet-50 &centroids  & cross-entropy & & \\
\hline
\textbf{CPC
\cite{oord2018representation}} & CV, audio & (d) & APC & - & InfoNCE loss & Other images & End-to-end\\
& NLP&&&&  &&\\
\hline
\textbf{CPC v2
\cite{henaff2020data}} & CV & (d) & APC & - & InfoNCE loss & Other images & End-to-end\\
\hline
\end{tabular}
\end{table*}

\textbf{Auto-encoding} 
%is a type of predictive models 
%is similar to an 
%originates from 
is based on the use of auto-encoders \cite{baldi2012autoencoders}, as %demonstrated 
depicted in \Cref{fig:models(a)}. An standard auto-encoder learns a compressed latent embedding that %can 
represents the input of the encoder and expects to reconstruct the original input from the latent representation, \ie  the decoder output. The dimensionality of the latent representation must be carefully designed, as it determines the representation reliability. When setting a too large latent dimensionality, an auto-encoder risks to learn an identity function, \ie maps the input directly to the output, and hence becomes useless. Various techniques  to prevent auto-encoders from learning an identity function do exist, \eg denoising auto-encoders \cite{lecun2015deep}, which partially corrupt the input data by randomly zeroing %out 
some input values and are trained to recover the original undistorted input. 
For the denoising to be successful, the model's ability to retrieve useful high-level representations becomes essential.
%Hence, the model needs to retrieve useful high-level representations in order to perform denoising well. 
The zero-out step can be replaced by other data augmentation techniques, such as geometric transformations including cropping, reordering, and colourisation, to name a few, which often appear in SSL studies.

\begin{figure}[t!]
  \centering
    \hspace*{-0.45cm}
    \begin{subfigure}{0.25\textwidth}
        \centering
        % \vspace{1.8cm}
        \hspace*{0.1cm}
        \includegraphics[height=2.8in]{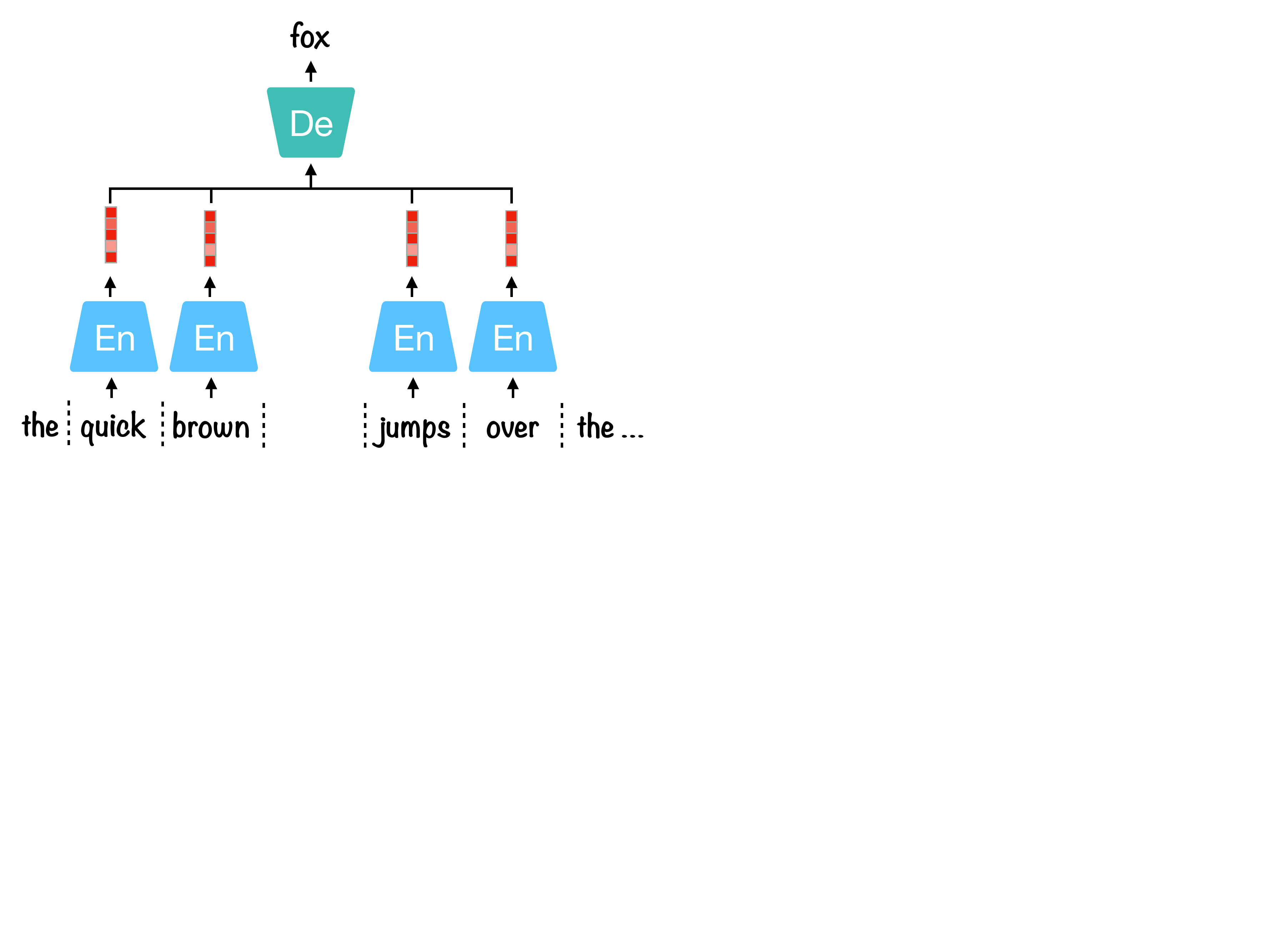}
        \vspace{-4.1cm}
        \caption{CBoW}
        \label{fig:CBoW}
    \end{subfigure}%
    ~ 
    \hspace*{-0.25cm}
    \begin{subfigure}{0.25\textwidth}
        \centering
        % \vspace{0.7cm}
        \hspace*{0.3cm}
        \includegraphics[height=2.8in]{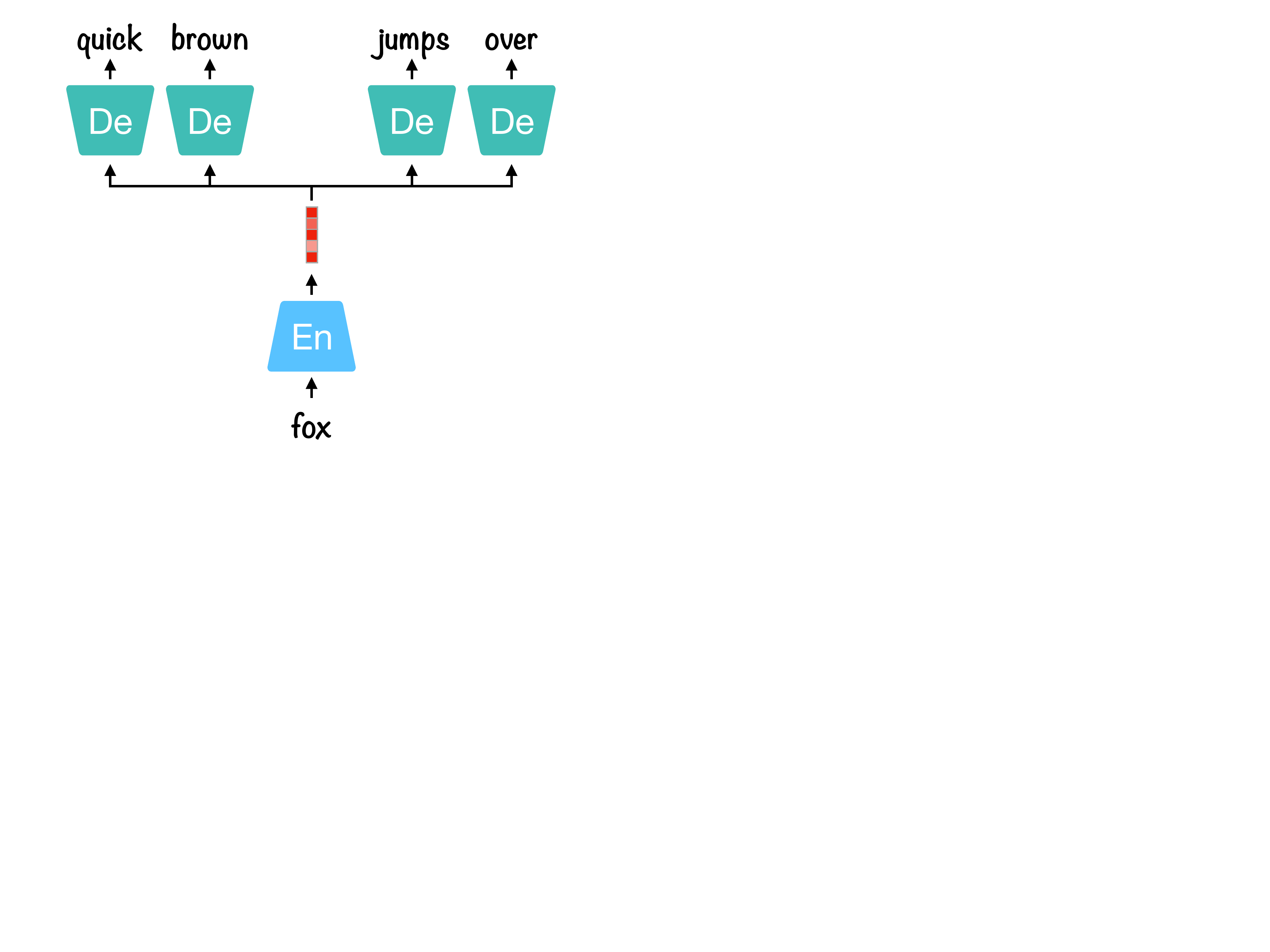}
        \vspace{-4.1cm}
        \caption{Skip-gram}
        \label{fig:Skip-gram}
    \end{subfigure}% 
  \caption{Two architectures of Word2Vec. (a) CBoW predicts a single word from the previous and future words. The context words are fed into an encoder (En) to aggregate a context vector, which is used to produce the target word using a decoder (De); (b) Skip-gram makes the opposite prediction from CBoW, \ie predicting previous and future words from a single centre word.}
  \label{fig:W2V}
\end{figure}

Nevertheless, such models do not always require to predict the entire original sample, \ie the prediction can be restricted to only recover the distorted part. This is typically the case %especially 
for sequential data as in Word2Vec \cite{mikolov2013distributed, word2vec} that is used to map one-hot representations of words to word embeddings. 
In Word2Vec, two formulations are used to learn underlying word representations:  Continuous Bag-of Words (CBoW) and Skip-gram, depicted in \Cref{fig:CBoW} and \ref{fig:Skip-gram}, respectively. CBoW is trained to predict a single word from its context words, whereas Skip-gram does the opposite, aiming at predicting the left and right context words of a single input word. 
CBoW performs better in learning syntactic relationships between words, however, it is prone to overfit frequent words.
Differently, Skip-grams are better at capturing semantic relationships and suffer less from overfitting, leading to a more effective solution in learning representations for general purposes. Both formulations have been considered for learning fixed-length vector representations of audio segments, such as in Audio2Vec \cite{audio2vec} and Speech2Vec \cite{an2018}. To this end, the input audio or speech waveform is transformed into its time and Mel-frequency representations, keeping a two-dimensional input format as for Word2Vec.

\begin{figure*}[t!]
  \centering
    \hspace*{0.1cm}
    \begin{subfigure}{0.5\textwidth}
        \centering
        % \vspace{1.8cm}
        \hspace*{0.4cm}
        \includegraphics[height=2.8in]{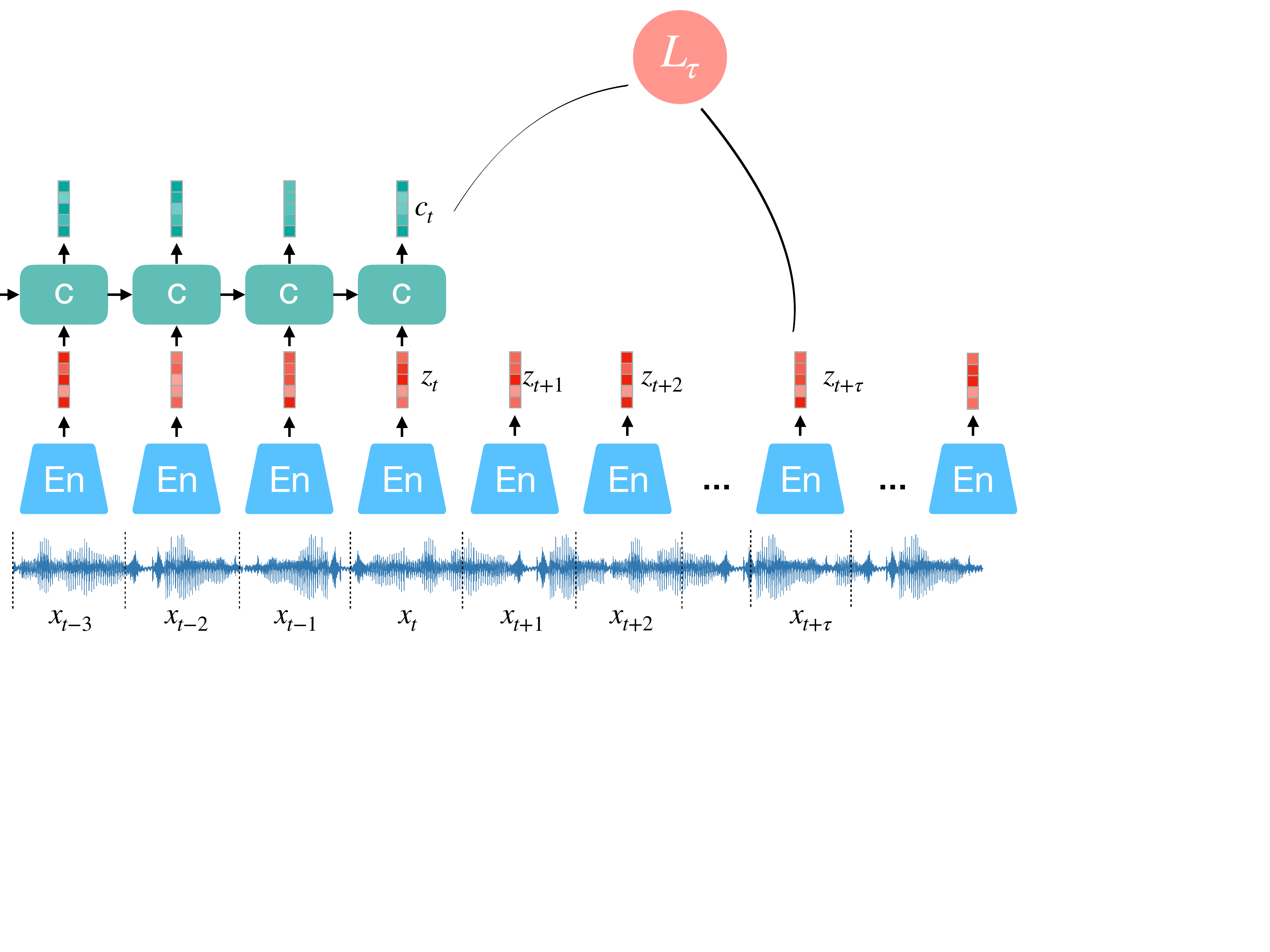}
        \vspace{-2.2cm}
        \caption{APC}
        \label{fig:APC}
    \end{subfigure}%
    ~
    \vspace*{0.2cm}
    \hspace*{0.2cm}
    \begin{subfigure}{0.5\textwidth}
        \centering
        % \vspace{0.7cm}
        % \hspace*{1cm}
        \includegraphics[height=2.8in]{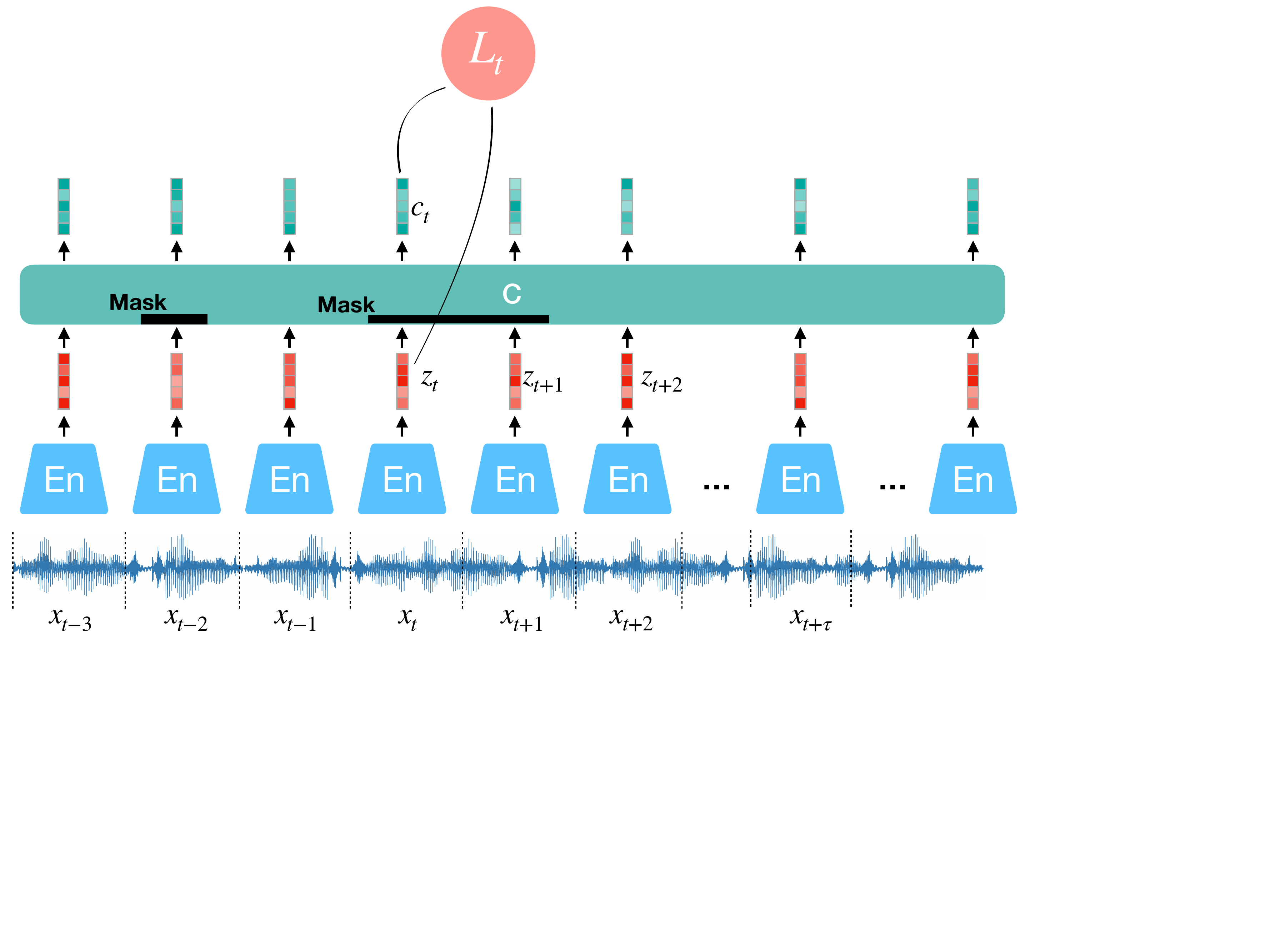}
        \vspace{-2.2cm}
        \caption{MPC}
        \label{fig:MPC}
    \end{subfigure}% 
  \caption{Diagrams of auto-regressive predictive coding (APC) and masked predictive coding (MPC).}
  \label{fig:PC}
\end{figure*}

%do not require entire sample as input, but rather learns to recover the whole, or parts of, or only some features of its input. 
%The input to the encoder can be a variant of a sample by using data augmentation techniques such as transformation, corruption and permutation etc.
%When the model aims at reconstructing its input or its transformation as its output, the framework turns into an auto-encoder architecture that optimises the reconstruction error., the optimisation targets can be used to predict some contents from its surrounding contexts, as for Masked Predictive Coding (MPC) \cite{liu2020mockingjay,liu2021tera} and Auto-regressive Predictive Coding (APC) \cite{chung2019}. 
The success of Word2Vec or alike is based on the consistency of the context surrounding the component to predict. 
Similarly, an auto-regressive model can be used to learn representations by making predictions of future information conditioning on the past context. Auto-regressive Predictive Coding (APC) \cite{chung19}   codes on wave samples (cf.\  \Cref{fig:APC}). An additional context network aggregates the resulting representations up to the current time step. Hence, the context network is usually a recurrent neural network (RNN) for modelling the temporal information. Its output context vector is then used to predict the next audio representations, for example, $\tau$ steps ahead of the current time step. APC is optimised by minimising L1 loss.    
The APC method takes into account only uni-directional information of a sequence and needs a combination of separately trained models for forward and backward directions in order to achieve a representation from both directions (past and future). 
%Inspired by BERT \cite{devlin2019}, Masked Predictive Coding (MPC)   directly trains a bidirectional architecture by first masking parts of the input signals and then predicting them through conditioning on context from both directions 
%EMI%
Inspired by BERT \cite{devlin2019}, Masked Predictive Coding (MPC)    trains directly a bidirectional architecture. This is made by  masking parts of the input signals, which are subsequently  predicted  by  conditioning on the context from both directions 
%EMI%
(cf.\  \Cref{fig:MPC}). Transformer encoders and bidirectional RNNs have been considered as context networks for realising MPC.
Similarly, the recently proposed  Non-autoregressive predictive coding (NPC) \cite{liu21l_INTERSPEECH} also applies a mask on its model input, but it learns representations based on local dependencies of an input sequence, rather than globally. The MPC approaches can learn effective representations of sequential data in a non-autoregressive way, and hence achieve considerable speed-up in training. 
%
% Predictive coding on wave samples [21] has a long and influential history in speech processing, 
%and its recent neural version [22] and variants, such as Contrastive Predictive Coding (CPC) [23], 
% \subsection*{Skip-grams and CBoW methods}
%- ranking permutation, position, puzzle， etc...
Finally, other auto-encoding predictive models for SSL also aim %include 
to predict the relative position of signal parts \cite{doersch2015unsupervised,gidaris2018unsupervised}, including solving a Jigsaw puzzle \cite{noroozi2016unsupervised,misra2020self}  or reordering the pieces of a shuffled %schuffled 
sequence input \cite{carr2021self,ryan2020,lan2020self}. 

\textbf{Siamese Models}
%Although both alignment and uniformity terms in contrastive loss is claimed to be important for achieving a good representation, recent proposed methods discard the use of negative samples and can achieve even better prediction results for CV tasks. 
%Another kind of predictive SSL models 
have a typical `two towers' architecture as shown in \Cref{fig:models(b)}. Each tower processes a view of a data sample. Considering the natural similarity between the two views of the same sample, the encoded representations in the high-dimensional latent space should be close to each other. Hence, during training, the representations from one tower can be seen as the training target, \ie  pseudo-labels, for the other tower. 
The neural encoders of both towers share the same or similar architecture -- their parameters can be shared or independent. 
When taking negative samples in the training objectives, \shuo{a contrastive loss is formulated} which pulls the representations of different views from the same data %but different views 
close together, while  pushing the one from  negative samples far away. The model can be optimised by applying standard back-propagation.   

However, without using negative samples, the Siamese model is prone to \textit{mode collapse}, \ie when the model's output is very similar (or even identical) for different inputs.  
Therefore, a specific training strategy is required for developing this kind of predictive model. 
Although the research of Siamese models starts from using contrastive learning as in SimCLR \cite{simclr} and MoCo \cite{he2020momentum}, recent works, such as  BYOL  \cite{grill2020bootstrap}, SimSiam \cite{chen2021exploring}, and Barlow Twins \cite{jure2021},  use only positive pairs without contrasting negative pairs. 
These approaches still maintain competitive (and even superior) SSL performance. 
Thus, in this part, we will focus on introducing the Siamese models trained without negative samples, 
%with the emphasis of their different ways to update the parameters, 
while the contrastive model will be discussed in \Cref{subsec:contrast}.

\begin{figure}[t!]
  \centering
    \hspace*{-1.9cm}
    \begin{subfigure}{0.33\textwidth}
        \centering
        % \vspace{1.8cm}
        \hspace*{1.9cm}
        \includegraphics[height=3.3in]{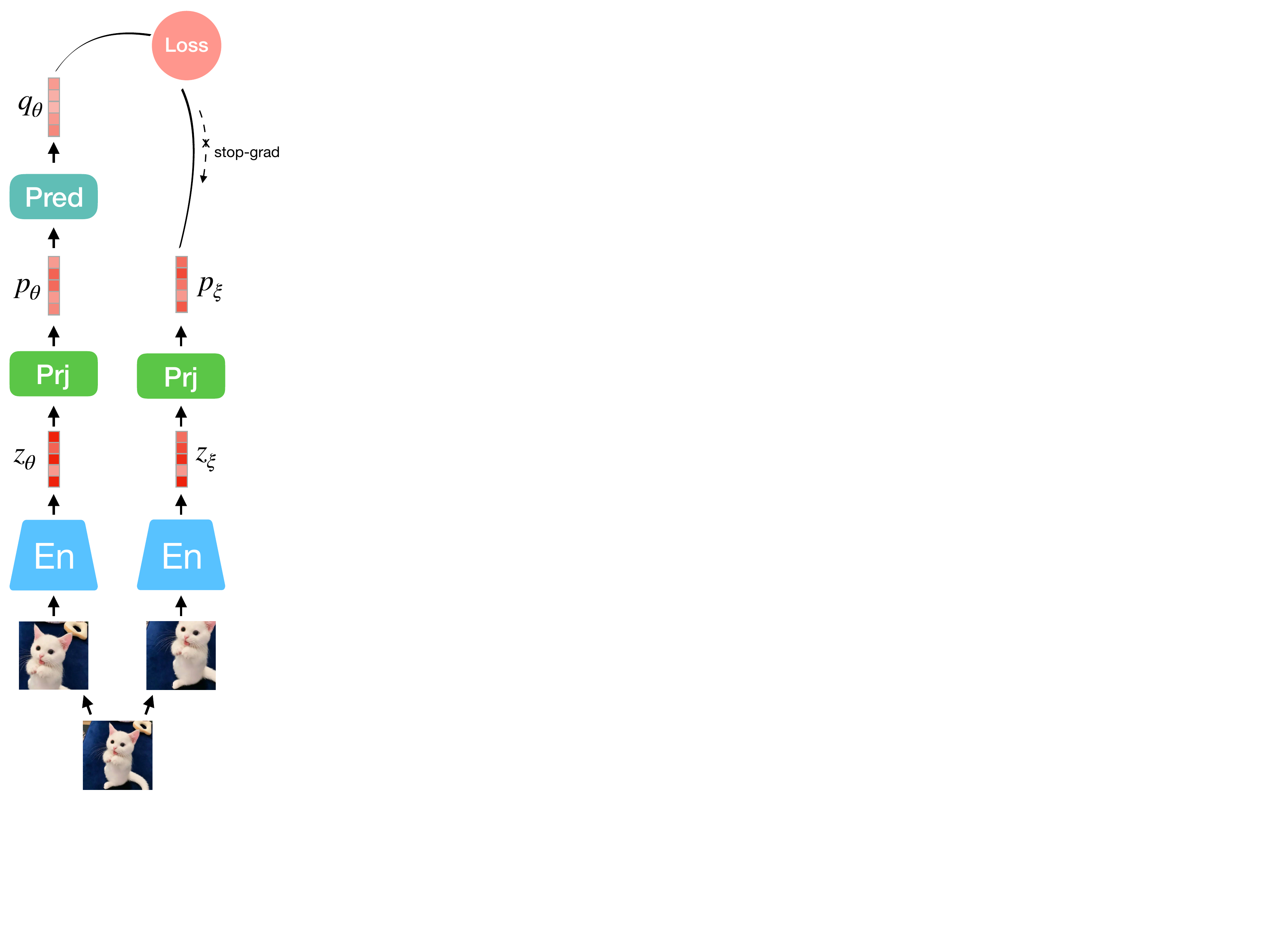}
        \vspace{-2cm}
        \caption{BYOL}
        \label{fig:byol}
    \end{subfigure}%
    ~ 
    \hspace*{-3cm}
    \begin{subfigure}{0.33\textwidth}
        \centering
        % \vspace{0.7cm}
        \hspace*{1.8cm}
        \includegraphics[height=3.3in]{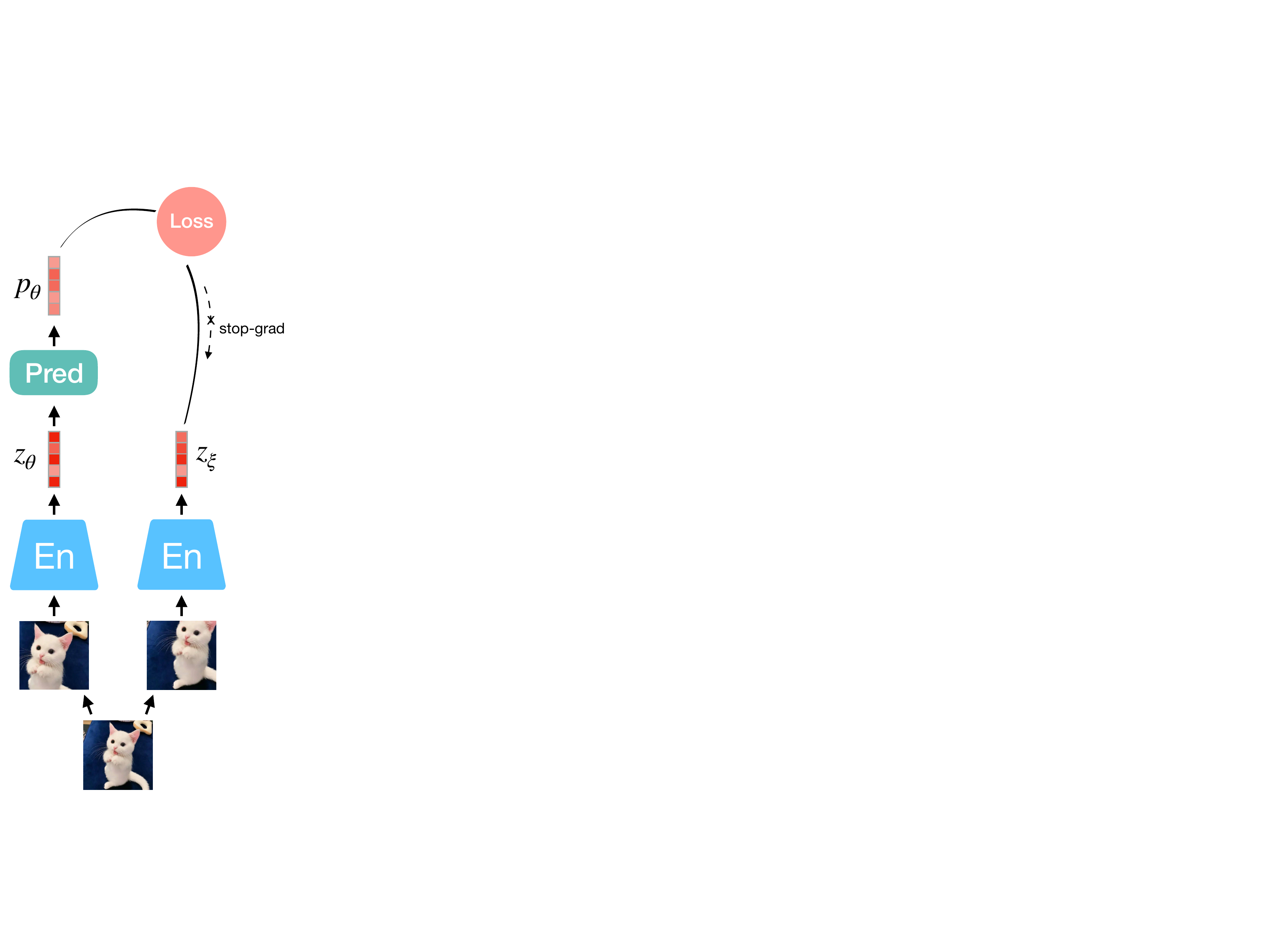}
        \vspace{-2cm}
        \caption{SimSiam}
        \label{fig:simsiam}
    \end{subfigure}% 
    ~
    \hspace*{-3cm}
    \begin{subfigure}{0.33\textwidth}
        \centering
        % \vspace{0.7cm}
        \hspace*{1.9cm}
        \includegraphics[height=3.3in]{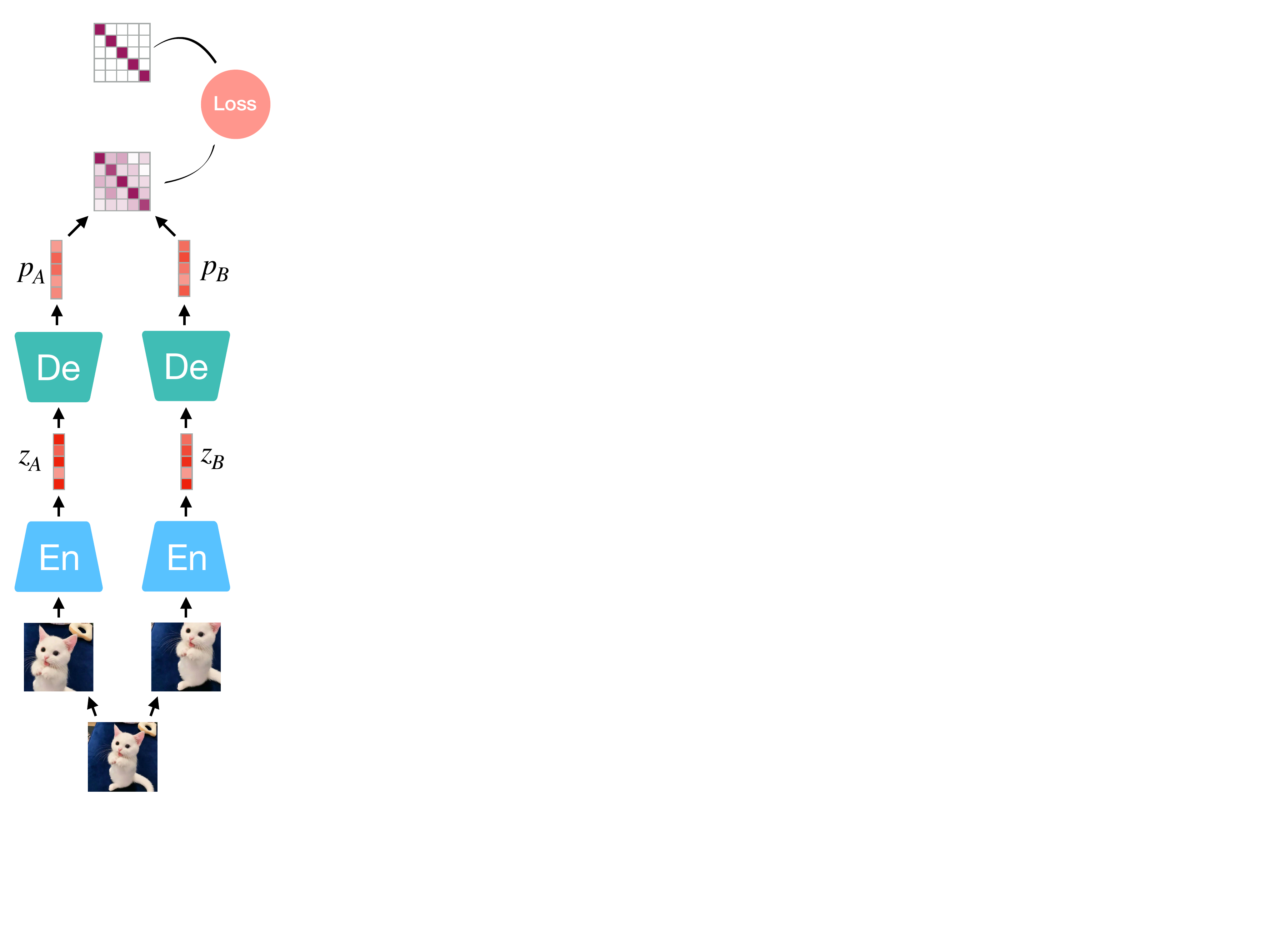}
        \vspace{-2cm}
        \caption{Barlow Twins}
        \label{fig:bt}
    \end{subfigure}% 
  \caption{Diagrams for Predictive Models using Siamese architectures.}
  \label{fig:siamese}
\end{figure}

BYOL \cite{grill2020bootstrap}, short for Boostrap Your Own Latent, trains two sub-networks separately denoted as online network and target network as shown in \Cref{fig:byol}. Both sub-networks contain an encoder $f$ and a projection layer $g$. The online network has an additional predictor layer $p$ build on MLP and is optimised to pull close the distance between the predicted representation and the projected embedding from the target network. 
To get rid of mode collapse, the two networks are asynchronously optimised in an iterative way.
The target network is randomly initialised and then, its parameters are updated using an \textit{exponential moving average} (EMA)  strategy during training, similar as presented in MoCo \cite{he2020momentum} and defined as:
\begin{equation}
    \xi \leftarrow \tau \xi + (1-\tau) \theta,
    \label{eq:moment}
\end{equation}
where $\theta$ and $\xi$ stand for the parameters of the online and target network, and $\tau \in [0,1]$ is a given decay rate for updating.
The online network follows the guidance of the slowly-updated target network, and is optimised by minimising the Mean Square Error (MSE) between the two network outputs:
\begin{equation}
\begin{aligned}
    L &= ||\bar{q_\theta} - \bar{p_\xi}||_2^2\\
      &= 2 - 2\frac{<q_\theta, p_\xi>}{||q_\theta||_2 \cdot ||z_\xi||_2},
\end{aligned}
\end{equation}
where $\bar{q_\theta}$ and $\bar{p_\xi}$ are l2-normalised $q_\theta$ and $p_\xi$, \ie $q_\theta/||q_\theta||_2$ and $p_\xi/||p_\xi||_2$. The two views are exchanged as the input of online and target networks once to create a symmetric loss, denoted as $\Tilde{L}$, leading to a complete training loss
of $L + \Tilde{L}$. 
The slow update of the target network progressively aggregates the parameters from the online network. This enables to produce more stable representations, which are used as the guidance to train the online network, by this, progressively yielding better representations. As updating the online parameters is a sensitive procedure that requires very careful fine-tuning, in order to avoid mode collapse, the authors additionally exploit LARS \cite{you2017large} as optimiser for training. Updating the  parameters in different layers and with different strength is expected to gradually lead to meaningful convergence.

SimSiam \cite{chen2021exploring} (shown in \Cref{fig:simsiam}) shares a similar structure as BYOL, but is removing the projection layers in both networks. 
%In addition, unlike BYOL, the target network is not trained via gradient descent \textcolor{red}{but through ????}. %as the same as for BYOL. 
%SimSiam directly employs a copy of the online target without using a momentum encoder that slowly follows the online network through EMA.  \emi{I am a bit lost here... is this the alternative to gradient descent used by SimSiam? I guess you should define EMA?} 
Its target network is also not optimised using back-propagation, but copied directly from the online network. Updating  the target network,  also found to be effective,  avoids the use of momentum encoding in BYOL. 

The extra learnable predictor and a stop-gradient operation are considered as the essential parts that prevent the model from collapsing into trivial representations \cite{pmlrv139tian21a}. The theoretical analysis and experimental study in \cite{pmlrv139tian21a} has also revealed that as long as the predictor is updated more often or has a larger learning rate, EMA is not necessary for successful convergence without
mode collapse. However, optimising the predictor too often or with too large a learning rate leads to its failure %fails to work 
for SimSiam  \cite{chen2021exploring}. 
%The online network is expect to grow along with the predictor, too often update of the predictor leads to its failure. 
The final predictor is expected to be optimal, \ie achieving minimal $l2$ error in predicting the output of the target network from the online network output. Moreover, weight decay has been shown to be very helpful in achieving stable convergence. 
Although the use of batch normalisation \cite{bn2015} was hypothesised to be crucial for preventing collapse in BYOL\footnote{https://generallyintelligent.ai/blog/2020-08-24-understanding-self-supervised-contrastive-learning/}, 
in previous work \cite{richemond2020byol}  batch normalisation has been  successfully replaced with group normalisation and weight standardisation, thus,  refuting the need of batch statistics for BYOL.  
% Another key component of BYOL, that leads to its successful convergence, is batch normalisation, used in the projection and prediction layers. 
% Batch normalisation calculates the global mean and standard deviation for normalising the layer features across a batch of samples. 
% Although BYOL requires no negative samples for contrastive learning, the batch normalisation identifys the common mode between examples in a mini-batch and removing it by using other represeantations in the min-batch as implicit negative examples. Therefore, batch normalisation works as a way of intrinsically implementing contrastive learning on embedded represenations. In another word, standard contrastive learning compare a sample with all other samples for maintaining good conformity, but BYOL compare the sample with the average of the other samples, which result in similar effects to some extent.  

%  suggests that BYOL does not even need the target encoder to be an exponential moving average of the online encoder
 
Barlow Twins (BT) \cite{jure2021}, is a neural network architecture inspired by the redundancy reduction principle described in the work of the neuroscientist H. Barlow \cite{barlow1961possible}. As depicted in \Cref{fig:bt}, BT contains two identical networks that process two distorted versions of a same sample to produce their representations. The model measures the cross-correlation matrix between the two learnt representations, which is expected to be close to the identity matrix. 
BT simplifies the training procedure with respect to BYOL and SimSiam, which requires asymmetric components, such as a predictor layer, as well as operations, including gradient stopping and EMA. 
A BT model benefits from very high-dimensional output vectors and its loss function is formulated as:
\begin{equation}
    L = \underbrace{\sum_i (1-C_{ii})^2}_{\text{invariance term}} + \lambda\underbrace{\sum_i \sum_{j \neq i}C_{ij}^2}_{\text{redundancy reduction term}},
\end{equation}
where the cross-correlation matrix computed between the outputs of the two networks along batch direction, is defined as:
\begin{equation}
C_{ij} = \frac{\sum_b p^A_{b, i}p^B_{b,j}}{\sqrt{\sum_b (p^A_{b,i})}\sqrt{\sum_b (p^B_{b,j})}}.    
\end{equation}
By minimising the training objective, the invariance term pushes the diagonal elements of the correlation matrix to $1$, which makes the learnt representations of the two distorted versions of a sample as close as possible. The redundancy reduction term compresses the correlations between the off-diagonal elements of the correlation matrix. This reduction of the redundancy between output elements in a representation vector results in representations of sufficient disentanglement.

\textbf{Clustering}, such as K-means clustering \cite{noroozi2018boosting, caron2018deep}, provides a way to yield pseudo-labels for SSL. 
Considering that different objects are naturally associated with 
%EMI%
distinct 
%EMI%
%different 
categories, each category should occupy a separate manifold in the representation space. 
Deep Cluster \cite{caron2018deep}, as shown in \Cref{fig:models(c)}, performs two steps iteratively. First, it exploits the K-means clustering method to group the encoded representations and produce pseudo-labels for each sample.  Then, with the created pseudo-labels assigned to each sample, the encoder network can be optimised by minimising the classification loss, such as by cross-entropy loss. 
Instead of the global clustering method of K-means clustering, Local Aggregation (LA)  \cite{zhuang2019local} allows to model more flexible statistical structures  by separately  identifying neighbours  for each example. Moreover, LA proposes an objective function that directly optimises a local soft-clustering metric, leading to better training efficiency.   
% VQ-VAE \cite{van2017neural}
% 
Another clustering method used in SSL is  SwAV \cite{caron2020unsupervised}, which introduces online clustering ideas into a Siamese architecture, by this avoiding the time consumption due to the two-steps training paradigm. The online clustering assignment provides pseudo-labels within mini-batches. In addition, the authors introduce multi-crop augmentation, aiming at solving a swapped prediction problem.  %For training, the created pseudo-labels from one branch can be used as the optimisation targets for the other branch, leading to a swapped loss formulised as the sum of the cross-entropy loss of each branch.

\subsubsection{Contrastive Models}
\label{subsec:contrast}
Considering negative samples when training an SSL model, such as a Siamese model, assists in achieving representations of better distinctiveness. The key idea behind is to pull the representations of two similar inputs (defined as a positive pair) close in the latent space, and to push those of dissimilar inputs (defined as a negative pair) far. Hence, extensive efforts are put on the designing of contrastive losses. Hereby, we summarise the most popular types of contrastive loss before going through the contastive SSL methods in the literature.

\textbf{Contrastive loss}
In early versions of contrastive loss, an anchor data is paired with only one positive and one negative sample, leading to a positive pair and a negative pair. Recent training objectives %include
%EMI%
take also into consideration  
%EMI%
multiple positive and negative pairs in one batch. 
Max margin contrastive loss, designed for deep metric learning \cite{chopra2005learning}, takes a pair of inputs and minimises the embedding distance when they are from the same class and maximises it,  otherwise.
More formally, given a data point $x$ in a batch of samples $\mathscr{X}$, CL learns an encoder $f$ that minimises

\begin{equation}
\begin{aligned}
L(x, x^+) &= \sum_{x\in \mathscr{X}} ||f(x) - f(x^+)||^2_2\\
L(x, x^-) &= \sum_{x\in \mathscr{X}} \mathrm{max}(0, \epsilon - ||f(x) - f(x^-)||_2)^2,
\label{eq:margin}
\end{aligned}
\end{equation}
where $x^+$ is a data point similar or congruent to $x$, and therefore is referred to as a positive sample. On the other hand, $x^-$ denotes a data point dissimilar to $x$, and hence is referred to as a negative sample. In addition, $\epsilon$ is a hyperparameter that defines the lower bound distance between representations of different samples.

Triplet loss, proposed for face recognition \cite{schroff2015facenet}, configures the offset distance between representations of positive and negative pairs: 
\begin{equation}
\begin{aligned}
\label{eq:triplet}
    L&(x, x^+, x^-) \\
    &=\sum_{x\in \mathscr{X}} \mathrm{max}(0, ||f(x) - f(x^+)||^2_2 - ||f(x) - f(x^-)||^2_2 + \epsilon).
\end{aligned}    
\end{equation}

Multi-Class N-pair loss \cite{NIPS2016_6b180037} generalises the triplet loss allowing joint comparisons among multiple negative samples. It is formulated similarly to the softmax loss:
\begin{equation}
\begin{aligned}
L &(x, x^+, x^-_{n \in [1, 2N-1]})\\ 
  &=log(1+\sum_{n=1}^{2N-1} e^{f(x)^Tf(x^-_{n})-f(x)^Tf(x^+)})\\
  &=-log\frac{e^{f(x)^Tf(x^+)}}{e^{f(x)^Tf(x^+)}+\sum_{n=1}^{2N-1}{e^{f(x)^Tf(x^-_{n})}}}.
\end{aligned}
\end{equation}

NT-Xent, short for normalised temperature-scaled cross-entropy loss, introduces an additional temperature parameter for controlling the penalty on the effect of negative samples, similarly as $\epsilon$ in \Cref{eq:margin} and \Cref{eq:triplet}: 
\begin{equation}
    L(x, x^+, x^-_{n \in [1, 2N-1]}) = -log\frac{e^{f(x)^Tf(x^+)/\tau}}{\sum_{n=1}^{2N-1}{e^{f(x)^Tf(x^-_{n})/\tau}}}.
\end{equation}

The loss function, also known as InfoNCE \cite{oord2018representation} objective, is  inspired by Noise Contrastive Estimation (NCE) \cite{gutmann2010noise}, which is proposed %for estimating 
%EMI%
in order to estimate the 
%EMI%
parameters of a statistical model. InfoNCE generalises by contrasting %to contrasts 
the distance of representations from a positive pair to their distance to $N-1$ negative pairs:
\begin{equation}
\begin{aligned}
\label{eq:infonce}
    L&(x, x^+, x^-_{n \in [1, N-1]}) \\
    &= \mathbb{E}[-log\frac{e^{f(x)^Tf(x^+)}}{e^{f(x)^Tf(x^+)}+\sum_{n=1}^{N-1}{e^{f(x)^Tf(x^-_{n})}}}].
\end{aligned}
\end{equation}
Its denominator terms contain one positive and $N-1$ negative samples. Hence, we can construct a softmax classifier which is optimised using cross-entropy loss for $N$ classes. The classifier  %that 
%assigns large values to the positive examples and small values to negative examples. 
%EMI%
assigns large and small values to the positive  and  negative examples, respectively. 
%EMI%
In this regard, InfoNCE can be seen as using categorical cross-entropy loss 
%to identify the positive sample amongst a set of unrelated noise samples.
%EMI%
for identifying  a positive sample within a set of (unrelated) noise samples. 
%EMI%

\textbf{Contrastive SSL for Siamese Models} A typical architecture that is trained by using contrastive loss is SimCLR \cite{simclr}. SimCLR %which 
exploits several different data augmentation techniques for transforming an input image, including random cropping, resizing, colour distortions, and Gaussian blur. The transformed images are then coded into representations using ResNet. After going through projection heads built on Dense-ReLU-Dense structure, NT-Xnet is used as objective function for self-supervised learning.  
The authors \shuo{of SimCLR}
emphasised the importance of `scaling up', \ie using a larger batch size, a deeper and wider network, as well as   training for longer epochs, in order to guarantee %for 
the success of the method. %, meaning larger batch size, deeper and wider network and training for longer epochs are needed. 

%SimCLR, using only one encoder $f$, while Montentum Contrast (MoCo) \cite{he2020momentum} exploits an additional momentum encoder $f_m$. 
Unlike SimCLR, which uses  only one encoder $f$,  Momentum Contrast (MoCo) \cite{he2020momentum} exploits an additional momentum encoder $f_m$. 
The encoder and momentum encoder,  sharing    %are of 
the same architecture and being identically initialised,  process two views of an image. The method also applies contrastive loss, where the negative samples are provided by previous batches. For this, representations of previous samples are stored into a queue during training. Representations from a new batch are pushed into the queue after training and %while
 old representations are excluded. 
The encoder is updated by applying back-propagation as in SimCLR, while the momentum encoder is updated by linear interpolation of the two encoders as introduced in \Cref{eq:moment}. The momentum parameter is set to $\xi=0.999$ by default, meaning the update of the momentum encoder is much slower.  However, the update mode of the momentum encoder avoids back-propagation, which hence can increase the amount of negative samples for training. The synchronise update of the encoder and momentum encoder also solves the problem of inconsistency of encoded representations happening in works using memory bank \cite{he2020momentum}. 
%Later, 
Posterior architectures such as MoCo v2 \cite{chen2020improved} integrate the effective components presented in SimCLR. % First, 
MoCo v2 incorporates stronger data augmentation techniques, \ie using an additional Gaussian Deblur method and a larger batch size. Moreover, its projection head layer is increased as a 2-layers MLP for both the encoder and momentum encoder. Similarly,  
SimCLR v2 \cite{chen2020} upgrades the system proposed in SimCLR by scaling up the model size from ResNet-50 to ResNet-152 and improving the depth of the projection head. In addition, the  authors leave one projection layer for fine-tuning on semi-supervised tasks, by this aiming at learning from few labelled examples while making best use of a large amount of unlabelled data. 
Furthermore, in order to efficiently provide a large size of negative samples for  training, the idea of memory mechanism used in MoCo v2 is employed in SimCLR v2 too. %in order to efficiently providing the training a large size of negative samples. 
Differently, the latest MoCo v3 \cite{chen2021}, %in the contrast, 
removes the memory queue with the cost of requiring a bigger batch size. In addition, it applies a prediction layer after the projection head, similarly as proposed in BYOL, which further %slightly 
improves the representation capability.

In constrastive SSL, given one %example sample, 
data point, the learnt representations of its positive and negative samples are seen as able to provide opposite pseudo-labels. 
A   positive sample provides an additional view of the same data.  Theoretical analysis has proven that 
%whenever the two views have redundant information about the label, linear functions of the learnt respresentations are nearly optimal on downstream prediction tasks %according to 
%EMI%
when two views provide redundant information of the label, applying linear projections to the learnt representations can guarantee the performance on downstream prediction tasks 
%EMI%
\cite{tosh2021contrastive}. This proof indicates that SSL can produce high-quality representations from the multi-views of a data point and guarantees prediction performance with simple downstream models.

Contrasting the distance of a data point  
to its positive samples with respect to the one to its negative samples avoids the model from falling into representational collapse. 
To analyse the effectiveness of contrastive loss, we can split it into two parts:
\begin{equation}
    \begin{aligned}
    L &= \mathbb{E}[-log\frac{e^{f(x)^Tf(x^+)/\tau}}{e^{f(x)^Tf(x^+)/\tau}+\sum_{n=1}^{N-1}{e^{f(x)^Tf(x^-_{n})/\tau}}}]\\
    &= \underbrace{\mathbb{E}[-f(x)^Tf(x^+)/\tau]}_{\text{alignment}}\\ & \;\;\;\; + \underbrace{\mathbb{E}[log(e^{f(x)^Tf(x^+)/\tau}+\sum_{n=1}^{N-1}{e^{f(x)^Tf(x^-_{n})/\tau}})]}_{\text{uniformity}},
    \label{eq:analysis}
\end{aligned}
\end{equation}
where the `alignment' term targets at maximising the similarity between the learnt embeddings of the positive pairs. Then, the `uniformity' term helps the contrastive learning to learn separable features by maximal-uniformly distributing the embeddings on a unit sphere given the normalisation condition. 
Both terms are crucial to the downstream tasks according to \cite{wang2020understanding}. Different from an instance discrimination objective, which pushes all different instances apart and considers no underlying relations between samples, the design of contrastive loss requires a proper temperature coefficient $\tau$ that finds a balance between learning separable features and at the same time provides some degree of tolerance to the closeness of semantically similar samples.
A too small $\tau$ loses the tolerance to group the similar input samples and hence may break the underlying semantic structure, by this harming the learnt features for its use in  %thus harm to the learn features for 
downstream tasks. The effect of the temperature parameter is similar as the margin value set in \Cref{eq:margin}, which has %also 
been investigated in detail by \cite{liu2021}.  

In \cite{wang2021understanding}, Wang suggests to adjust the alignment and uniformity loss to
\begin{equation}
\begin{aligned}
L_{align} &= \mathbb{E}[||f(x)-f(x^+)||_2^\alpha]\\
L_{uniform} &= \text{log}\mathbb{E}[e^{-t||f(x)-f(x^*)||_2^2}],
\end{aligned}
\end{equation}
%and 
indicating that both terms should be minimised simultaneously. Maintaining a good balance between these two losses has been found more effective than standard contrastive loss. 

% \begin{figure*}[t!]
%   \centering
%   \hspace*{-0.7cm}
%     \begin{subfigure}{0.5\textwidth}
%         \hspace*{-0.6cm}
%         \vspace*{0.5cm}
%         \centering
%         \vspace*{0.5cm}
%         \hspace*{0.6cm}
%         \includegraphics[height=2.7in]{Figures/CPC-A.pdf}
%         \vspace{-3.1cm}
%         \caption{CPC}
%         \label{fig:CPCA}
%     \end{subfigure}%
%     ~ 
%     \hspace*{0.9cm}
%     \begin{subfigure}{0.5\textwidth}
%         \centering
%         % \vspace{0.7cm}
%         % \hspace*{1cm}
%         \includegraphics[height=2.5in]{Figures/CPC-M.pdf}
%         \vspace{-1.2cm}
%         \caption{Contrastive MPC}
%         \label{fig:CPCM}
%     \end{subfigure}% 
%   \caption{Diagrams of contrastive predictive coding (CPC) based on the framework of APC and MPC.}
%   \label{fig:PC}
% \end{figure*}

\subsubsection{Contrastive Predictive Coding}
\label{subsec:cpc}
% - Constrastive Predictive Coding: pos + neg 
%    incorporating contrastive loss in predictive coding, including APC and MPC, to create CPC.
%    CPC - A
%    CPC - M
Contrastive Predictive Coding (CPC) \cite{oord2018representation}, %illustrated in \Cref{fig:CPCA}, 
exploits an auto-regressive predictive model, and is optimised to predict the correct future information based on the aggregated global context from the past frames. In addition, CPC gets use of negative samples to improve the representation discrimination in training objectives such as InfoNCE loss \Cref{eq:infonce}, replacing the query data point $x$ by a context vector $c_t$:
\begin{equation}
\label{eq:cpcinfo}
    L(c_t, z_{t+\tau}, z^-_{n \in [1, N-1]}) = \mathbb{E}[-log\frac{e^{c_t^Tz_{t+\tau}}}{e^{c_t^Tz_{t+\tau}}+\sum_{n=1}^{N-1}{e^{c_t^Tz^-_{n}}}}],
\end{equation}
where $z^-_n$ denotes a negative data point sampled from the proposal distribution of $z_{t+\tau}$, \ie randomly sampled from the sequence $z$. In CPC v2 \cite{henaff2020data}, the model is scaled up to achieve larger model capacity, and the batch normalisation is replaced by layer normalisation, as batch normalisation is found to harm downstream tasks for CPC frameworks \cite{henaff2020data}. Moreover, patch-based augmentation is introduced in order to add  %for introducing 
more diversity to the model's input.

Similarly, the contrastive loss has been used in masked predictive models, leading to constrastive MPC.% (\Cref{fig:CPCM}). 
The training objective can be formalised  as
\begin{equation}
\label{eq:cpcinfo}
    L(c_t, z_{t}, z^-_{n \in [1, N-1]}) = \mathbb{E}[-log\frac{e^{c_t^Tz_{t}}}{e^{c_t^Tz_{t}}+\sum_{n=1}^{N-1}{e^{c_t^Tz^-_{n}}}}].
\end{equation}
This loss has been modified and used in wav2vec 2.0 \cite{baevski2020wav2vec}, which will be introduced in details \Cref{sec:audio}. The negative samples are generated in the same way as for CPC. 

\subsection{Training with or without negative samples}
\noindent
Recent works have experimentally tested the importance of components for achieving effective SSL models \cite{kolesnikov2019revisiting}. %The training objectives are more crucial 
The training objectives have been shown to be more important 
than the network architecture, and the quality of the learnt representations can be improved by scaling up the model size and the representation size. %Furthermore, for SSL using contrastive learning, the quantity and quality of negative samples are important for the performance \cite{wu2017sampling}. 
Furthermore, the quantity and quality of the negative samples has also shown to be important for the performance of SSL using contrastive learning \cite{wu2017sampling}. 
% generalisation vs discrimination ability
The SSL models optimised without contrastive loss, such as BYOL \cite{grill2020bootstrap}, Barlow Twins \cite{jure2021}, SimSiam \cite{chen2021exploring}, and \shuo{the other previously introduced}
auto-encoder-like predictive methods, aim at reducing the distance between the latent representations of views from the same data point. 
Contrastive SSL aims at contrasting the distance between positive samples against the distance to negative ones. 
The advantages and disadvantages of these two training strategies can be traced back to the difference between generative and discriminative models in the wide field of machine learning methods. 
Training SSL without negative samples may be less effective in learning discriminative features between samples, but has more potential to code more complete information into representations. Differently, contrastive SSL approaches are expected to learn more discriminative features by comparing to negative samples at expenses of droping %however, it may drop 
the common attributes, which are salient to represent the sample itself but are not very informative for distinguishing between samples. 
Based on this, we infer that applying SSL without negative samples is more useful, %as a feature extractor without losing much salient information. 
as it would do a feature extractor without losing much salient information.
Contrastive SSL can perform better if a pretext task and the downstream task fit closer, for example, if the data for the pretext task contains reading speech and the  downstream task aims at speech recognition. 
However, as the parameters of a pre-trained model can be further fine-tuned for a downstream task in a specific domain, an SSL model optimised with negative samples can be improved to complement its generalisation ability to capture more complete representations. Furthermore,  a contrastive SSL model can also be further improved to reduce the redundant information in the representations for the downstream tasks. Still,  reducing redundancy in representations for downstream tasks seems to be harder than restoring the lost representation completeness, which is confirmed by the fact  
that non-contrastive SSL outperforms contrastive SSL, as shown in recent works
\cite{grill2020bootstrap,chen2021exploring}. 

\shuo{A summary table of the typical SSL methods is shown in \Cref{tab:table1}.}

\section{Audio SSL}
\label{sec:audio}
% audio2vec, speech2vec
% Predictive method: Lee Hunyee
% estimating missing parts - a combination of above both 
% PASE - extending to multiple output, multi-task reconstruction method
% Two towel methods: BYOL & CLAR 
% Wav2Vec
\noindent
Most of the above introduced SSL methods have been transferred from those aiming to solve audio tasks, especially  speech applications such as Automatic Speech Recognition (ASR) \cite{oord2018representation,schneider2019wav2vec,baevski2019vq,baevski2020wav2vec,chung2020generative,liu2020towards}. % Two towel methods: BYOL & CLAR 
Other methods, which differ in the used audio input formats, such as LIM \cite{micro2019}, COLA \cite{saeed2021contrastive}, CLAR \cite{al2021clar}, and the work by Fonseca et al.\  \cite{fonseca2021unsupervised}, expand the SimCLR approach for learning auditory representations. The LIM model \cite{micro2019}  processes directly  speech samples expecting to maximise local mutual information between the encoded representations of chunks of speech sampled from the same utterance. In COLA \cite{saeed2021contrastive} and the work by Fonseca et al.\   \cite{fonseca2021unsupervised}, the presented models take segments randomly extracted from time-frequency features along the temporal direction. 
%BS: it seems you might have taken images from others - see the guitar and piano pictures? If so, you HAVE TO write the image source in the caption. Obviously, it is MUCH BETTER to do your own images. Should be easy :) THIS INCLUDES repainting images of others in slightly different ways - then, still name the source, but write "adapted from" or "based on" or alike. Publishers don't want to ask for permissions for reusing images of others... And it is better to have your own ones.
Several stochastic data augmentations are adopted for the patches before feeding to the model in \cite{fonseca2021unsupervised}, such as random size cropping and Gaussian noise addition. In addition, the authors  proposed \textit{mix-back} for additional augmentation, which mixes the incoming patch with a background patch, by this ensuring that the incoming patch is dominant in their mixture.  
In CLAR \cite{al2021clar}, the paired views of the model's input are generated by applying data augmentations on raw audio signals and time-frequency audio features, for which effective compositions of data augmentation  are explored. In addition, the authors found that combining a contrastive loss, such as cross-entropy loss, in training objectives for supervised learning, %such as cross-entropy loss, 
while using substantially less labelled data, can provide significant improvements in terms of convergence speed and representation effectiveness, with respect to using SSL only.  
Similarly, Wang \cite{wang2020} also suggests to train audio SSL models with different formats of an audio sample. More precisely, the training objective is to maximise the agreement between the raw waveform and its spectral representation. The approach is shown effective for \shuo{downstream classification tasks on both AudioSet, and ESC-50 datasets.} 
BYOL has also been adopted in the audio domain, named as BYOL-A \cite{niizumi2021byol}, which learns representations from a single audio without using negative samples. 
%The authors first create two-dimensional input by converting a raw audio to its log Mel-spectrogram. The authors additionally apply data augmentation techniques including mixup and random resize crop and normalisations before and after the data augmentation which leads to performance gain.

% audio2vec & speech2vec
Other typical works include Audio2Vec \cite{audio2vec} and Speech2Vec \cite{an2018}, inspired by Word2Vec \cite{word2vec}, as introduced in \Cref{subsec:pred}. 
Both works learn audio representations using CBoW and skip-gram formulations. 
In the CBoW formulation, the task is to reconstruct a temporal spectrogram slice of pre-determined duration from a number of past and future slices. The method has also been shown effective for acoustic scene classification in \cite{gontier2021polyphonic}. 
%In the skip-gram formulation, the roles of the target and surrounding slices are reversed. 
%EMI%
Differently, the roles of the target and surrounding slices are reversed in the Skip-gram formulation.
%EMI%
Audio2Vec and Speech2Vec mainly differ in the following aspects: (i) Speech2Vec applies audio segmentation, by using an explicit forced alignment technique, in order to isolate audio slices corresponding to each word. The forced alignment segmentation may introduce supervision to some extent;  (ii) Audio2Vec requires no %any form of 
explicit assistance and hence completely removes the need of supervision; (iii) %Next, 
unlike neural network architectures, % are used, 
Speech2Vec is built based on an RNN encoder-decoder, and Audio2Vec is built of stacks of CNN blocks; 
(iv) as model input, Speech2Vec processes the Mel-spectrogram of an audio, while Audio2Vec operates on Mel-Frequency Cepstral Coefficients (MFCCs); and  
(v) in Audio2Vec, the TemporalGap formulation is additionally introduced, which requests %the model to estimate the absolute time distance between two slices sampled at random from the same audio clip. 
%EMI%
that the model  estimates the absolute time distance between two (randomly sampled) slices, taken  from the same audio clip. 
%EMI%

\begin{table*}[t!]
\centering
\caption{\shuo{An overview of the recent audio self-supervised learning methods. The ``speech'' column distinguishes whether a method addresses speech tasks or for general purpose audio representations. The ``framework'' type refers to \Cref{fig:frameworks}.}}
\label{tab:table2}
\centering
\setlength{\tabcolsep}{2mm}
\setlength{\arrayrulewidth}{0.3pt}
\renewcommand{\arraystretch}{1.2}
\begin{tabular}{c|c|c|c|c|c|c}
\hline
\multirow{1}{*}{\textbf{Model}} & \multirow{1}{*}{\textbf{Speech}} & \multirow{1}{*}{\textbf{Input format}} & \multirow{1}{*}{\textbf{Framework}} & \multirow{1}{*}{\textbf{Encoder}} & \multirow{1}{*}{\textbf{Loss}} &  \multirow{1}{*}{\textbf{\shuo{Inspired by}}} \\ 
%\cmidrule(lr){7-9}
\hline
\textbf{LIM\cite{micro2019}} & \CheckmarkBold & raw waveform & (d) & SincNet  & BCE, MINE or NCE loss & SimCLR \\
\hline
\textbf{COLA\cite{micro2019}} & \XSolidBrush & log mel-filterbanks & (d) & EfficientNet  & InfoNCE loss & SimCLR \\
\hline
\textbf{CLAR\cite{al2021clar}} & \XSolidBrush & raw waveform & (d) & 1D ResNet-18  & NT-Xent & SimCLR \\
(semi) &  & log mel-spectrogram &  & ResNet-18  &  + cross-entropy & \\
\hline
\textbf{Fonseca et al.\cite{micro2019}} & \XSolidBrush & log mel-spectrogram & (d) & ResNet, VGG, CRNN  & NT-Xent loss & SimCLR \\
\hline
\textbf{Wang et al.\cite{wang2020}} & \XSolidBrush & raw waveform & (d) & CNN  & NT-Xent loss & SimCLR \\
&  & + log mel-filterbanks &  & ResNet  &  + cross-entropy & \\
\hline
\textbf{BYOL-A\cite{niizumi2021byol}} & \XSolidBrush & log mel-filterbanks & (b) & CNN  & MSE loss & BYOL \\
\hline
\textbf{Speech2Vec\cite{an2018}} & \CheckmarkBold & mel-spectrogram & (a) & RNN  & MSE loss & Word2Vec \\
\hline
\textbf{Audio2Vec\cite{9060816}} & \CheckmarkBold \XSolidBrush & MFCCs & (a) & CNN  & MSE loss & Word2Vec \\
\hline
\textbf{Carr\cite{carr2021self}} & \CheckmarkBold  & MFCCs & (a) & Context-free network  & Fenchel-Young loss & - \\
\hline
\textbf{Ryan\cite{ryan2020}} & \XSolidBrush  & constant-Q transform & (a) & AlexNet  & Triplet loss & - \\
 &  & spectrogram & & & & -\\
\hline
\textbf{Mockingjay\cite{liu2020mockingjay}} & \CheckmarkBold  & mel-spectrogram & (a) & Transformer  & L1 loss & BERT \\
\hline
\textbf{TERA\cite{liu2021tera}} & \CheckmarkBold  & log mel-spectrogram & (a) & Transformer  & L1 loss & BERT \\
\hline
\textbf{Audio ALBERT\cite{chi2021audio}} & \CheckmarkBold  & log mel-spectrogram & (a) & Transformer  & L1 loss & BERT \\
\hline
\textbf{DAPC\cite{bai2021representation}} & \CheckmarkBold  & spectrogram & (a) & Transformer  & Modified MSE loss & BERT \\
 & & & & & + orthogonality penalty & \\
\hline
\textbf{PASE\cite{Pascual2019LearningPS}} & \CheckmarkBold  & raw waveform & (a) & SincNet + CNN  & L1, BCE loss& BERT \\
\hline
\textbf{PASE+\cite{Ravanelli2020MultiTaskSL}} & \CheckmarkBold  & raw waveform & (a) & SincNet + CNN + QRNN & MSE, BCE loss & BERT \\
\hline
\textbf{CPC\cite{oord2018representation}} & \CheckmarkBold  & raw waveform & (a) & ResNet + GRU  & InfoNCE loss& - \\
\hline
\textbf{CPC v2\cite{henaff2020data}} & \CheckmarkBold  & raw waveform & (a) & ResNet + Masked CNN  & InfoNCE loss& - \\
\hline
\textbf{CPC2\cite{kharitonov2021data}} & \CheckmarkBold  & raw waveform & (a) & ResNet + LSTM  & InfoNCE loss & - \\
\hline
\textbf{Wav2Vec\cite{schneider2019wav2vec}} & \CheckmarkBold  & raw waveform & (a) & 1D CNN & Contrastive loss & - \\
\hline
\textbf{VQ-Wav2Vec\cite{baevski2019vq}} & \CheckmarkBold  & raw waveform & (a) & 1D CNN + BERT & Contrastive loss & BERT \\
\hline
\textbf{Wav2Vec 2.0\cite{baevski2020wav2vec}} & \CheckmarkBold  & raw waveform & (a) & 1D CNN + Transformer & Contrastive loss & BERT \\
\hline
\textbf{HuBERT\cite{hsu2021hubert}} & \CheckmarkBold  & raw waveform & (c) & 1D CNN + Transformer & Contrastive loss & BERT \\
\hline
\end{tabular}
\end{table*}

Although TemporalGap fails to surpass the CBoW and Skip-gram approaches, it presents the idea of measuring the relative positions of audio components as a pretext task. Carr et al.\  \cite{carr2021self} proposed a training strategy based on permutations, \ie training  a model that can reorder shuffled patches of an audio spectrogram, as analogous to solving a jigsaw puzzle  \cite{noroozi2016unsupervised}. The method draws inspiration from `Shuffle and Learn'  \cite{misra2016shuffle} and has also been considered in another work for industrial audio classification \cite{ryan2020}. In \cite{carr2021self}, Carr et al.\ also leverage differentiable ranking to integrate permutation inversions into an end-to-end training, which enables %to solve the permutation inversion for the entire set of permutations, \ie casting the reordering task as classification and reducing the space of permutations that can be exploited. 
%EMI%
solving the permutation inversion for the whole set  of permutations, \ie  reducing the space of permutations that might  be exploited and performing the reordering  as a classification task. 
%EMI%

\begin{figure*}[t!]
  \centering
    \hspace*{-2.7cm}
    \begin{subfigure}{0.33\textwidth}
        \centering
        % \vspace{1.8cm}
        \hspace*{2.2cm}
        \includegraphics[height=4.6in]{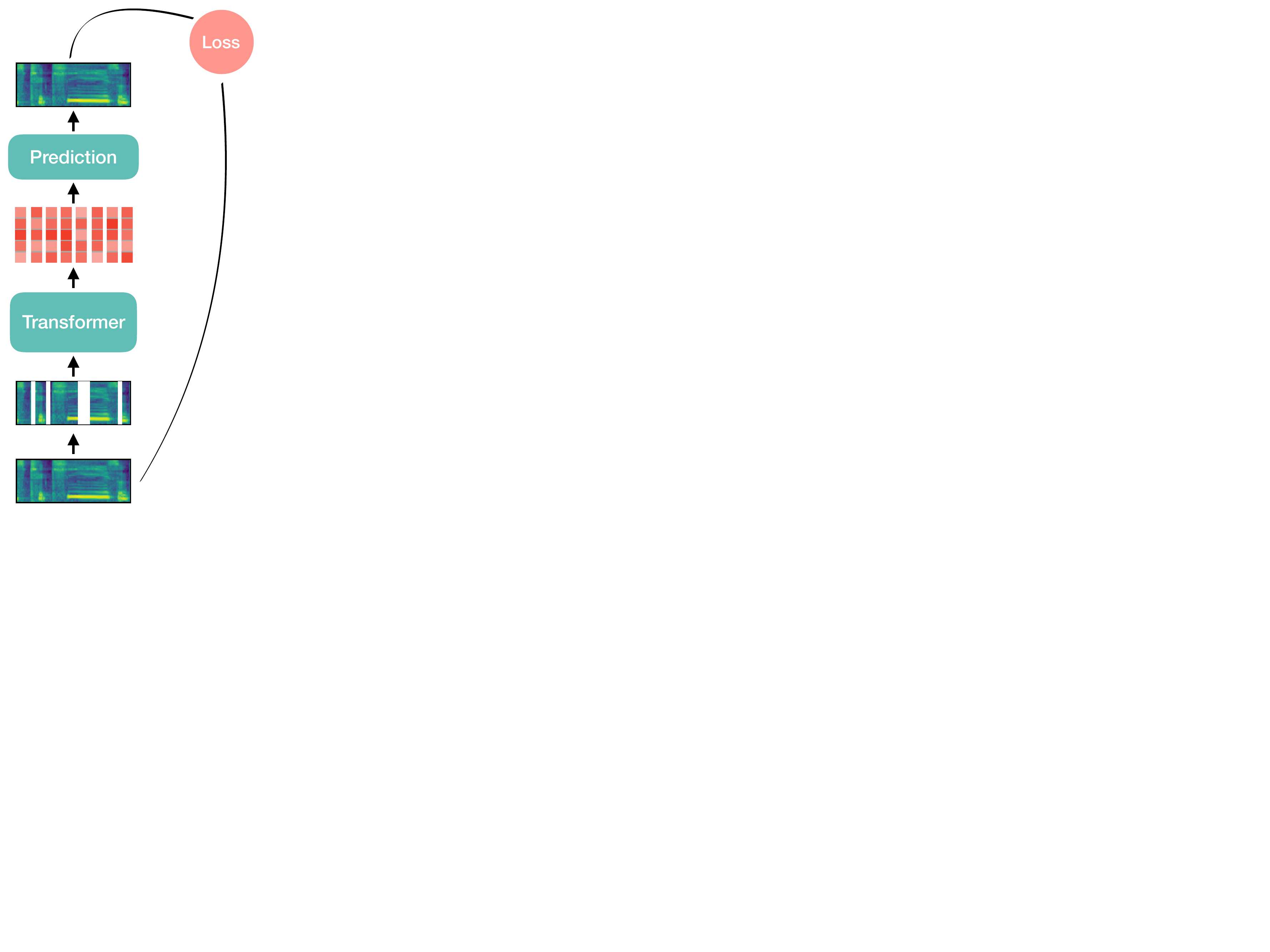}
        \vspace{-5.8cm}
        \caption{Mockingjay}
        \label{fig:mj}
    \end{subfigure}%
    ~ 
    \hspace*{-1.5cm}
    \begin{subfigure}{0.33\textwidth}
        \centering
        % \vspace{0.7cm}
        \hspace*{2.2cm}
        \includegraphics[height=4.6in]{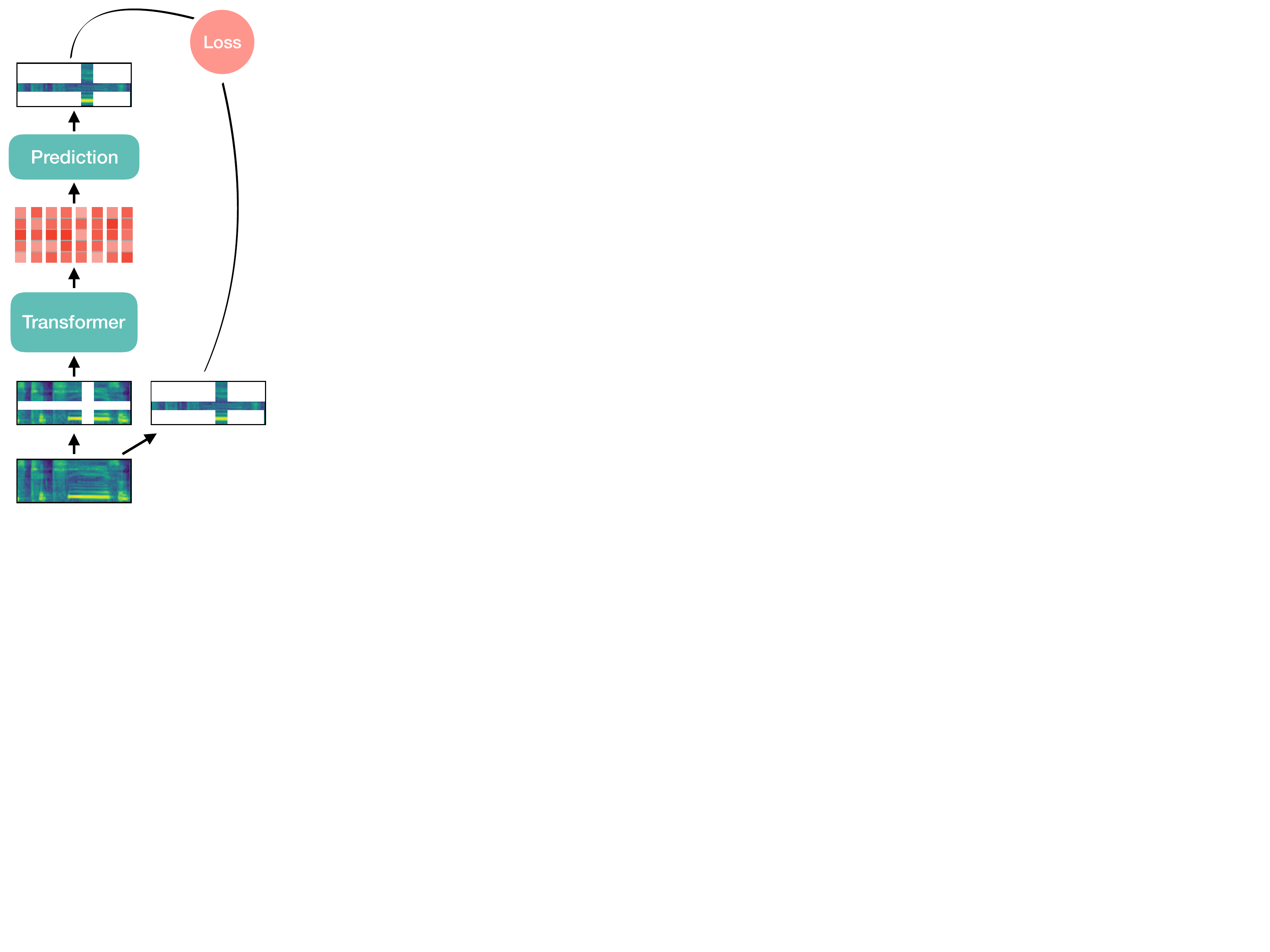}
        \vspace{-5.8cm}
        \caption{DAPC}
        \label{fig:dapc}
    \end{subfigure}% 
    ~
    \hspace*{1.5cm}
    \begin{subfigure}{0.33\textwidth}
        \centering
        % \vspace{0.7cm}
        \hspace*{-0.8cm}
        \includegraphics[height=4.6in]{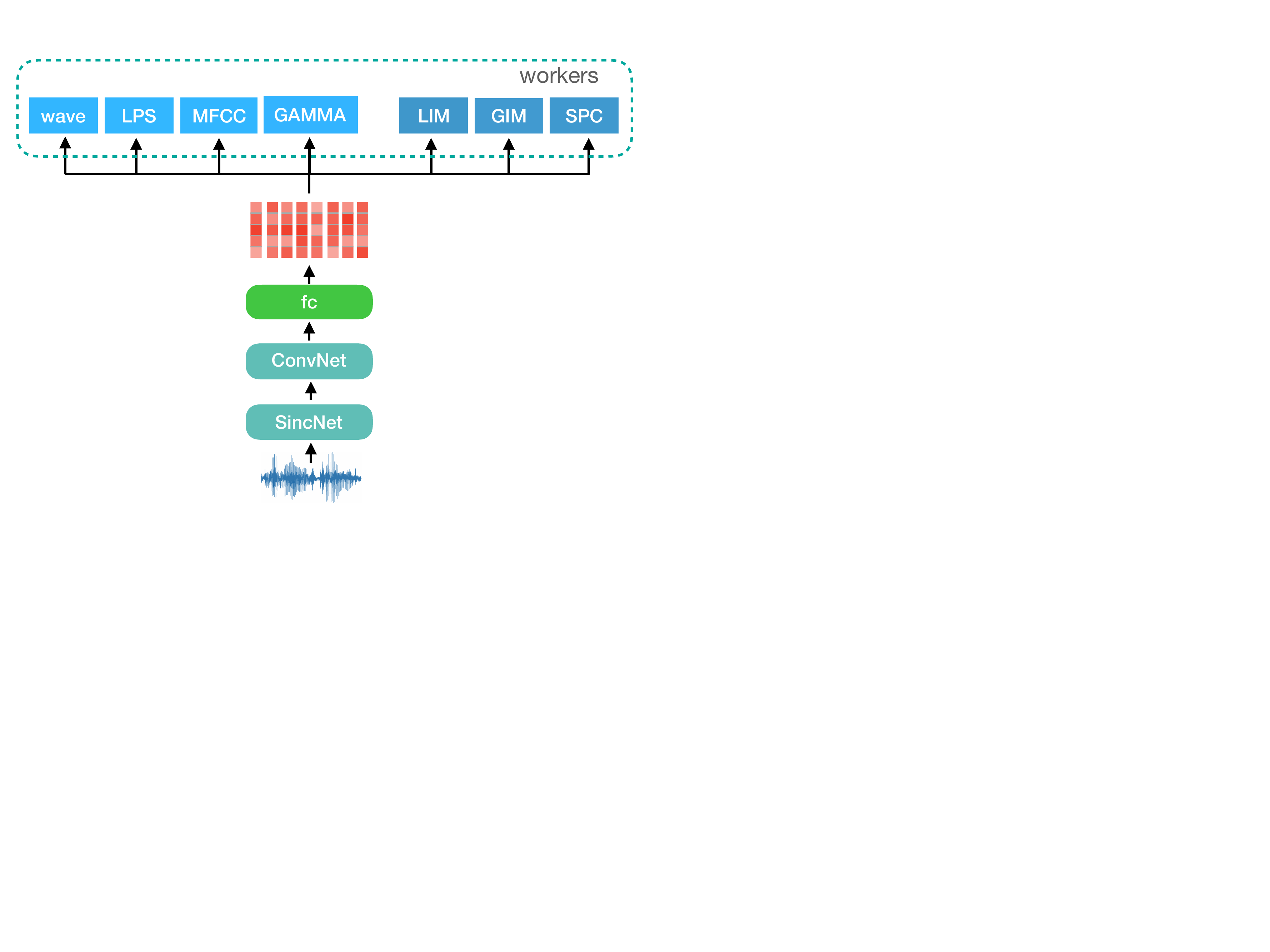}
        \vspace{-5.8cm}
        \caption{PASE}
        \label{fig:pase}
    \end{subfigure}% 
  \caption{Predictive models for audio SSL.}
  \label{fig:audiopredictive}
\end{figure*}

% Predictive method: Lee Hunyee
Another predictive model using an auto-encoder \cite{liu2020mockingjay,chi2021audio,liu2021tera}, as shown in \Cref{fig:MPC}, exploits a masked acoustic model (MAM) that masks some parts of an audio input and reconstructs the entire original input, essentially to fill the masked parts that are not known by MAM during training. The model is optimised by minimising the reconstruction error for learning general audio representations.  
Mockingjay \cite{liu2020mockingjay} (\Cref{fig:mj}) takes the Mel-spectrogram as input acoustic features and exploits transformers to code randomly masked frames into audio representations. The encoded representations are mapped to predict the complete frames using a projection head built of 2-layers MLP with layer normalisation. The transformer encoder and projection head are jointly optimised by minimising the L1 reconstruction loss. 
The effectiveness of self-attention in transformer encoders has been further explored in \cite{shuwen2020}; the authors also provide a visualisation tool for understanding the attention, based on which several attention refinement techniques are proposed to improve model performance. 
Audio ALBERT \cite{chi2021audio} has the same network architecture as Mockingjay, but the parameters are shared across all its transformer encoder layers, by this achieving a faster inference and increasing training speed without harming the performance of two evaluation downstream tasks, \ie speaker classification and phoneme classification. 
In TERA \cite{liu2021tera}, --short for Transformer Encoder Representations from Alteration -- the authors extend the used masking procedures,  %masking to more alterations, 
including replacing contiguous segments with randomness, masking along the channel axis, and applying Gaussian noise for pre-training the transformers. This resulted in a better representation performance than the one shown by Mockingjay and audio ALBERT, for the downstream tasks, phoneme classification, keyword spotting, and speaker classification \cite{liu2021tera}.
In addition, it shows also promising results for ASR tasks based on the Librispeech and TIMIT data sets.

%similar to 
% Missing leaving: do not estimate the entire output, but only missing part... this can also be seen as extending CBoW to an additional direction, frequency bins.
%Different from 
Unlike the works that predict the entire audio frames from their masked version, DAPC \cite{bai2021representation} (\Cref{fig:dapc}) proposes a method to only predict the missing components along the time- and frequency axes of an audio spectrogram by minimising a masked reconstruction loss.  
The method is also regarded as an extension of CBoW, for which the input masked spectrogram can be easily generated using SpecAugment \cite{48482}, and hence, the missing parts to be predicted are not only temporal frames, but also frequency bins.   
%Peng proposed MCVAE \cite{ma2019}, which extends a variational auto-encoder using RNN, and trained by a correspondence loss that encourages different intstances of the same word type to have similar latent embeddings. 
%
% PASE 
The problem agnostic speech encoder (PASE, \Cref{fig:pase}) \cite{Pascual2019LearningPS} is another approach that combines a CNN encoder with multiple neural decoders, defined as workers in the literature. The workers, %%are 
fed with learnt representations from the encoder, %and 
aim at solving regression or binary discrimination tasks. The regression tasks include, for instance,   recovering the raw audio waveform, the log power sepctrogram, MFCCs, and prosody. The binary discrimination tasks applies contrastive learning, by this maximising local and global mutual information  similar as in \cite{micro2019} and \cite{hjelm2018learning}, and optimises sequence predicting coding similar to \cite{van2017neural}.  
Each self-supervised task is expected to provide a different view of the speech signal; jointly solving self-supervised problems pushes the views into a unique representation that contains meaningful speech information such as speaker voice-print, phonemes, or emotions. 
In addition, to process the raw waveform as the encoder input, the SincNet \cite{Ravanelli2018SpeakerRF} model is used as the first stage of PASE, which performs a convolution with a set of parametrised Sinc functions that implement rectangular band-pass filters.
PASE+ \cite{Ravanelli2020MultiTaskSL} incorporates additional data augmentation techniques and more effective workers. The CNN encoder is combined with a Quasi-Recurrent Neural Network (QRNN) \cite{quasi2017} for capturing long-term dependencies in sequential data in a more efficient way.

\begin{figure}[t!]
  \centering
    % \hspace*{-1.5cm}
    \begin{subfigure}{0.4\textwidth}
        \centering
        \includegraphics[height=2.6in]{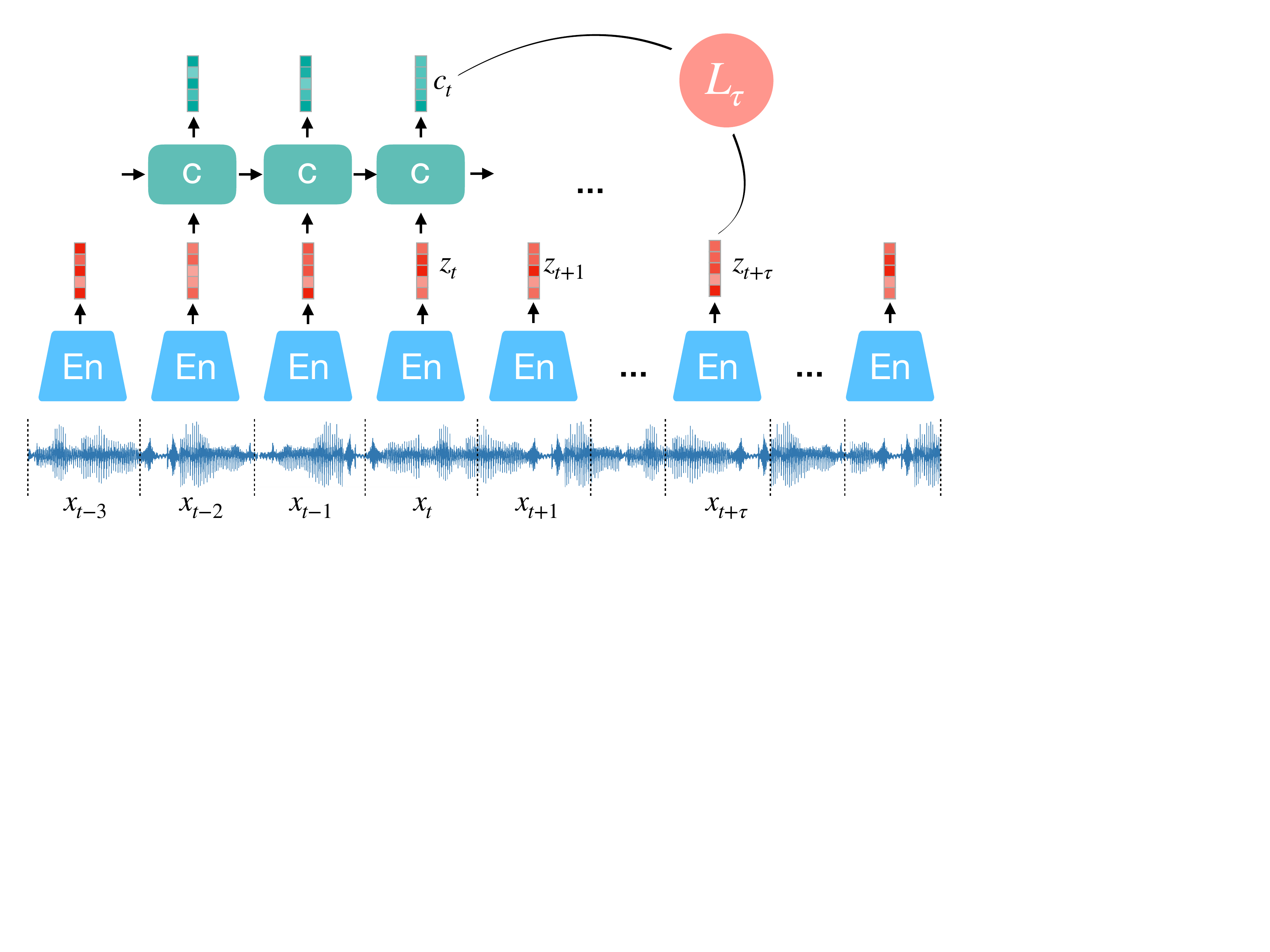}
        \vspace{-3cm}
        \caption{Wav2Vec}
        \label{fig:wv}
    \end{subfigure}%
    \\
    \vspace*{0.5cm}
    % \hspace*{-1.5cm}
    \begin{subfigure}{0.4\textwidth}
        \centering
        \includegraphics[height=2.6in]{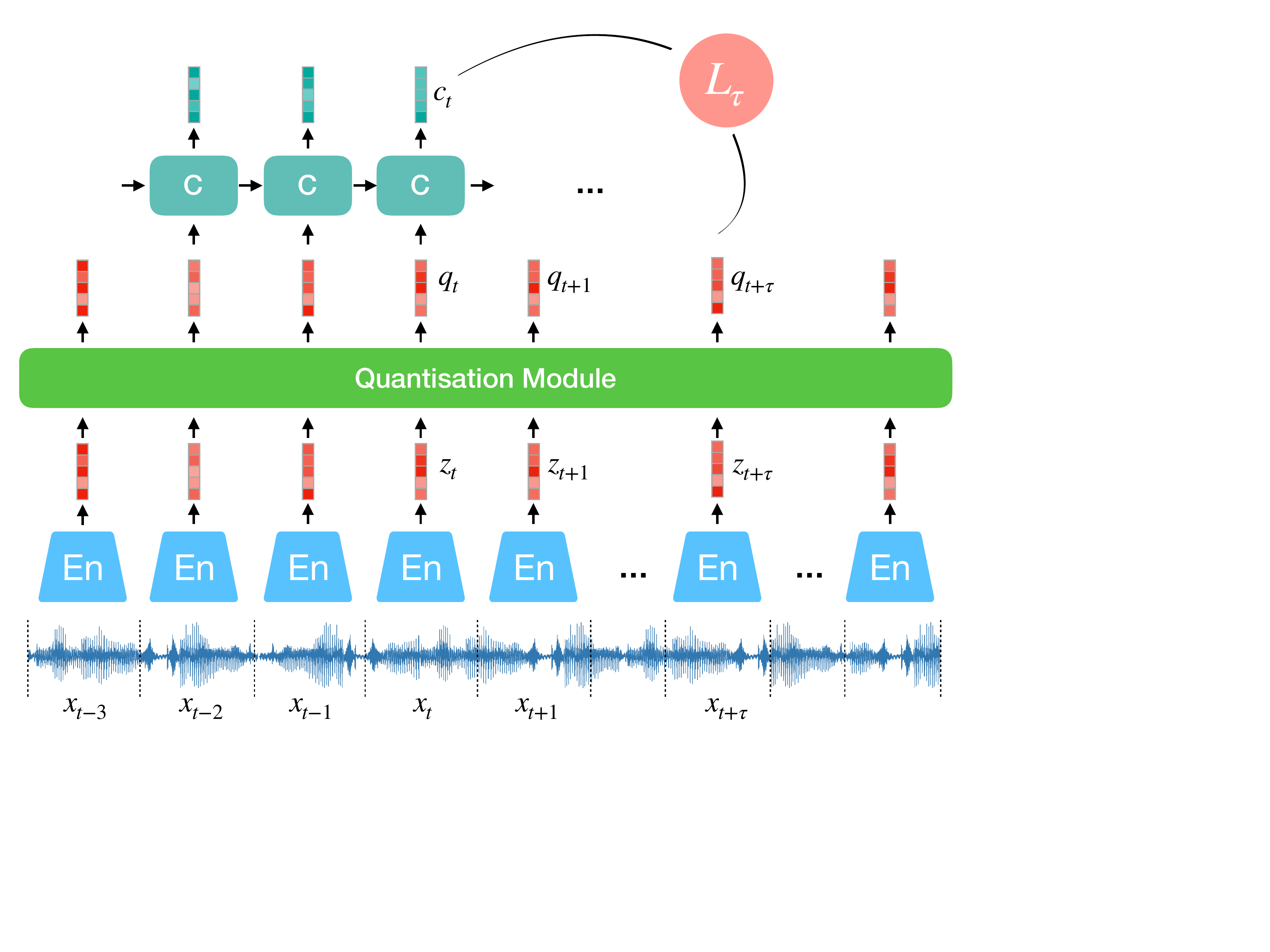}
        \vspace{-1.8cm}
        \caption{VQ-Wav2Vec}
        \label{fig:vqwv}
    \end{subfigure}% 
    \\
    % \hspace*{-1cm}
    \begin{subfigure}{0.5\textwidth}
        \centering
        % \vspace{0.7cm}
        \hspace*{0.7cm}
        \includegraphics[height=2.6in]{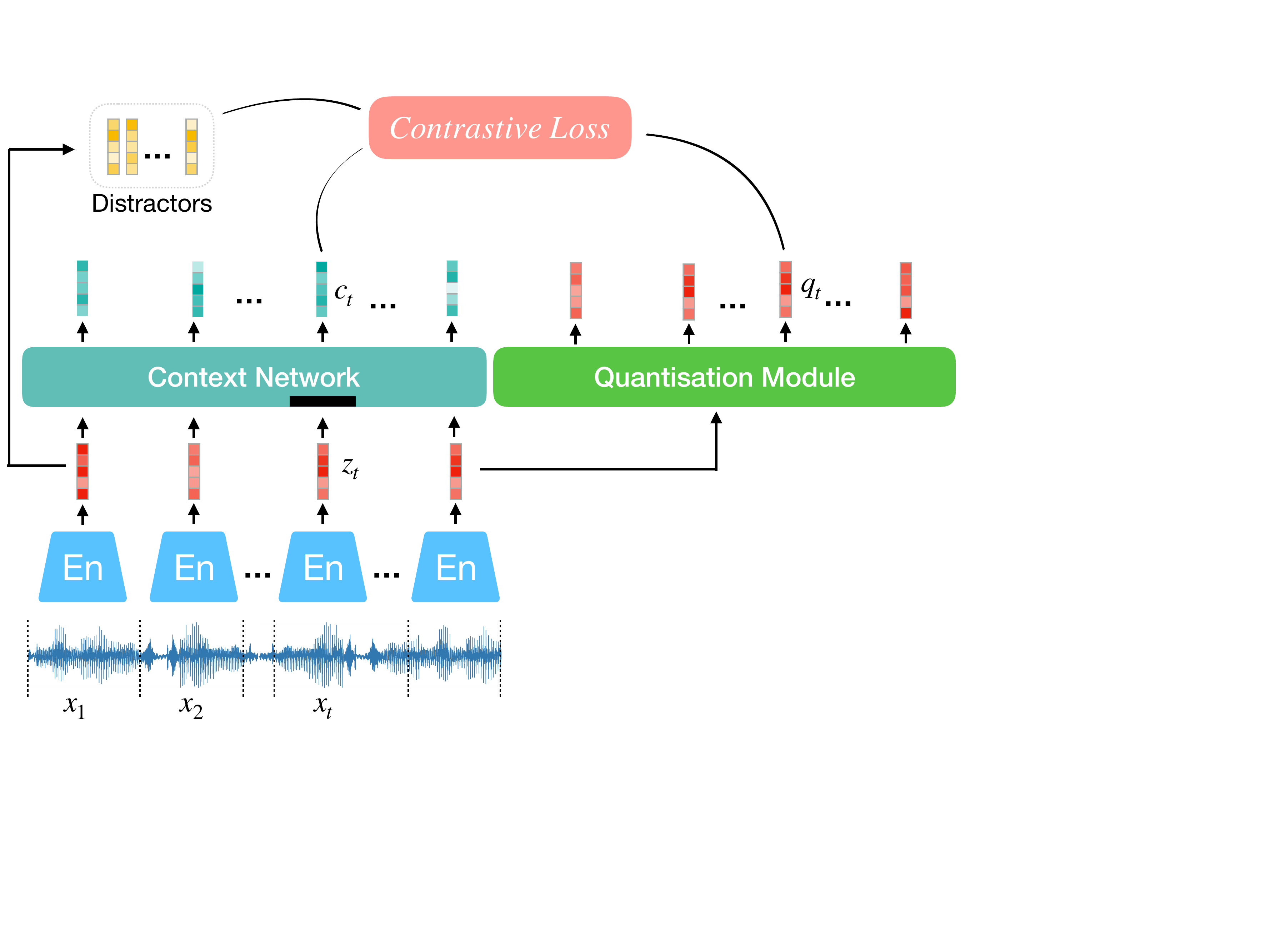}
        \vspace{-1.7cm}
        \caption{Wav2Vec 2.0}
        \label{fig:wv2}
    \end{subfigure}% 
  \caption{Predictive models for audio SSL.}
  \label{fig:w2v}
\end{figure}

% CPC
Van den Oord proposed CPC \cite{oord2018representation}, which can effectively learn representations by predicting the future in a latent space using an auto-regressive model, showing  very promising results for audio, images, text processing, and reinforcement learning.  
For audio, a strided convolutional network is used to encode raw audio to its latent representation. Then, a Gated Recurrent Unit (GRU) - RNN aggregates the information from all the past timesteps to form a context vector.  
More importantly, contrastive loss is applied to learn more discriminative representations by contrasting the true future to noise representations, given an aggregated context vector. 
Speech signals can be pre-processed by using a time-domain data augmentation library, such as  WavAugment \cite{kharitonov2021data}, in order to achieve more powerful representations by CPC. The library contains several DA techniques, including pitch modification, additive noise, reverberation, band reject filtering, or time masking, to name a few. In \cite{kharitonov2021data}, the authors define a CPC2 model, which replaces the GRU-RNN of CPC by a two-layers LSTM-RNN and replaces the linear prediction network by a single multi-head transformer layer, leading to better training efficiency without harming representation performance.

% wav2vec series
Wav2vec \cite{schneider2019wav2vec}, as show in \Cref{fig:wv}, adjusts the CPC structure to a fully convolutional architecture, enabling easy parallelisation over time on hardware. 
One CNN is used to produce a representation from audio, and the other captures global context information into a context vector for each time step. 
Besides moving beyond phoneme-based ASR, explored in \cite{oord2018representation}, it substantially improves a character-based ASR system. Specifically, the wav2vec approach is optimised by minimising contrastive loss for each step $k=1,...,K$:
\begin{equation}
\label{eq:wav2vec}
    L_k = - \sum_{i=1}^{T-k}(\text{log}\sigma(z_{i+k}^Th_k(c_i) + \lambda\mathbb{E}[\text{log}\sigma(-\Tilde{z}^Th_k(c_i)]),
\end{equation}
where $\sigma(x) = 1/(1+\exp(-x))$, and $\sigma(z_{i+k}^Th_k(c_i)$ stands for the probability of $z_{i+k}$ being the true future sample, applying an affine transformation, $h_k(c_i) = W_kc_i+b_k$. The total loss sums up considering $K$ steps;  $L=\sum_{k=1}^KL_k$ is minimised for training. 
After pre-training, the affine projection layer is removed for creating the learnt representations from the raw audio. 

\begin{figure}[t!]
  \centering
    % \hspace*{0.5cm}
    \begin{subfigure}{0.5\textwidth}
        \centering
        % \vspace{1.8cm}
        \hspace*{0.6cm}
        \includegraphics[height=3.5in]{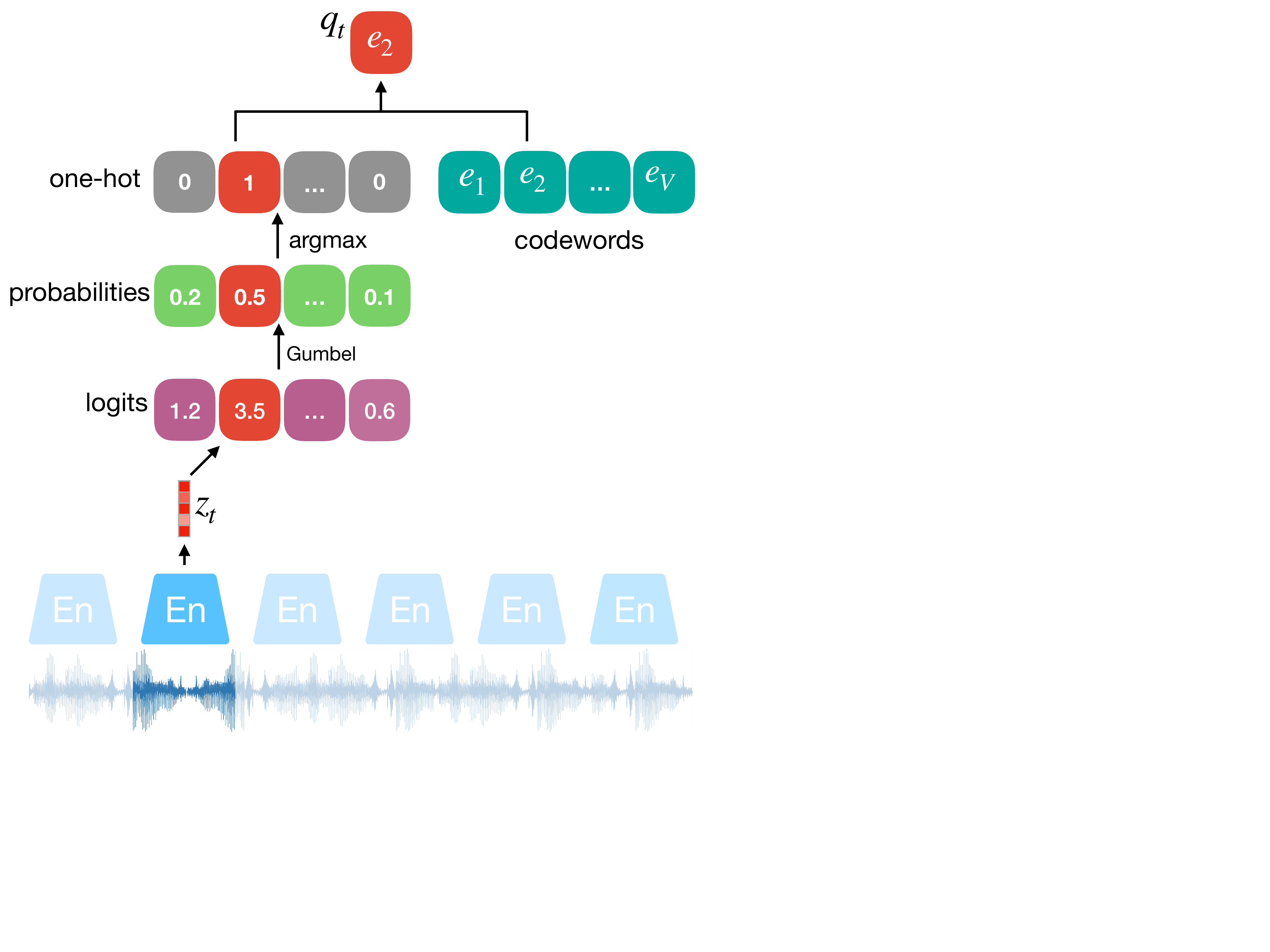}
        \vspace{-2.5cm}
        \caption{Gumbel softmax}
        \label{fig:gb}
    \end{subfigure}%
    \\
    \begin{subfigure}{0.5\textwidth}
        \centering
        \vspace*{0.5cm}
        \hspace*{0.8cm}
        \includegraphics[height=3.5in]{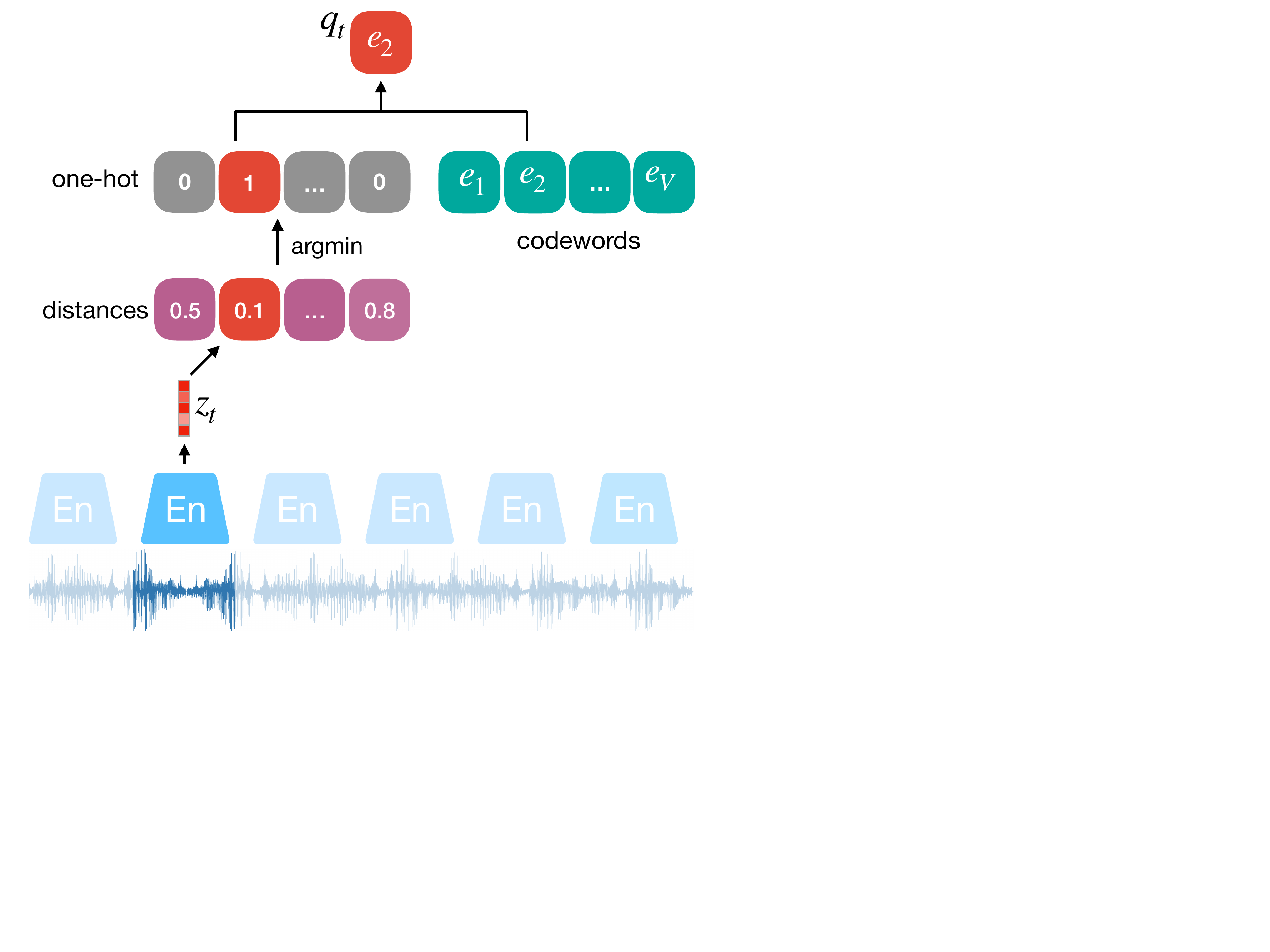}
        \vspace{-3.3cm}
        \caption{K-means clustering}
        \label{fig:clusteringcb}
    \end{subfigure}% 
  \caption{Diagrams illustrating the mechanisms to select codewords from a codebook.}
  \label{fig:PC}
\end{figure}

The works by Baevski et al.\  \cite{baevski2019vq,baevski2020effectiveness} (\Cref{fig:vqwv}) exploit a vector quantisation module after the wav2vec encoder in order to discretise the audio representations. This aims to find, for each representation, the closest embedding from a fixed size codebook $e \in \mathbb{R}^{V \times d}$ containing $V$ representations of size $d$.
The discrete representations are fed into the context network and then are optimised in the same way as for wav2vec. To solve the discontinuity caused by the argmax operation of the vector quantisation, Gumbel-Softmax \cite{jang2019} (refer to \Cref{fig:gb}) as a differentiable approximation of the argmax for computing one-hot representations as well as online k-means clustering (refer to \Cref{fig:clusteringcb}) are alternative solutions. This is similar to a vector-quantised variantional auto-encoder (VQ-VAE) \cite{van2017neural} and to vector-quantised autoregressive predictive codding \cite{chung2020vector}. 

Using a single codebook for coding representations %is prone to mode collapse where
%EMI%
tends to mode collapse in the cases in which   
%EMI%
only some of the codewords are actually used. To solve this issue, multiple codebooks are used as in product quantisation \cite{jegou2010product}. Specifically, product quantisation is equivalent 
to choosing quantised representations from multiple codebooks and concatenating them. Given $G$ codebooks with $V$ entries $e \in \mathbb{R}^{Vxd/G}$, one entry from each codebook is selected. A linear transformation is applied after concatenating the selected codewords.
The probabilities for choosing the $v$-th codebook entry for group $g$ are 
\begin{equation}
p_{g,v} = \frac{e^{(l_{g,v}+n_v)/\tau}}{\sum_{k=1}^V e^{(l_{g,k}+n_k)/\tau}},    
\end{equation}
where $l \in \mathbb{R}^{G\times V}$ represent the logits from projecting the encoded dense representation, $n=-\text{log}(-\text{log}(u))$ and $u$ are uniform samples from $U(0, 1)$, and $\tau$ is a non-negative temperature parameter. The codeword $i$ in group $g$ is chosen by $argmax_i p_{g,i}$.

K-means clustering can also be used for differentiable vector quantisation. A codeword is selected as long as it has the closest distance to the dense representations $z$. For training the model, in this case, additional terms are added in the wav2vec objective function, leading to 
\begin{equation}
\label{eq:vq}
    L = \sum_k L_k + (||\mathrm{sg}(z)-q||^2 + \gamma ||z-\mathrm{sg}(q)||^2),
\end{equation}
where $\mathrm{sg}$ is the stop gradient operator and $\gamma$ is a hyper-parameter. By miminising the loss, the term $||\mathrm{sg}(z)-q||^2$ freezes the encoder output $z$ and forces the codewords $Q$ to be closer to the encoder output. The other additional term $||z-\mathrm{sg}(q)||^2$ drives each encoder output to be close to one codeword, which is one centroid of the K-means clustering. 

The discretised representations are then used for training a BERT model in \cite{baevski2019vq}, aiming to predict masked input tokens based on encoding the surrounding context. The resulting representations are then fed into an acoustic model for producing transcriptions.

The optimisation of Wav2Vec and VQ-Wav2Vec are motivated by CPC, processing audio input for only one forward direction. Instead, Wav2vec 2.0 exploits a bidirectional MPC model, optimised by using contrastive loss, such as InfoNCE \cite{oord2018representation}. 
%instead of a conventional CPC model using a auto-regressive network architecture. 
The raw audio is encoded using multiple 1D-CNN layers, and the resulting representations are partly masked before sending to a transformer network to learn contextualised representations. The networks are jointly trained to contrast the true representations from distractors, given the contextualised representations. Similarly to VQ-Wav2Vec,  Wav2vec 2.0  applies product quantisation too; however, the quantised vector $q_t$ for each time step is not fed into a context network, but only used in the objective function:
\begin{equation}
    L = \mathbb{E}[-log \frac{e^{c^T_t q_t/\tau}}{\sum_{\Tilde{q} \sim Q_t}e^{c^T_t \Tilde{q}/\tau}}],
    \label{eq:w2v2}
\end{equation}
where $\Tilde{q} \sim Q_t$ includes $q_t$ and $K$ distractors.
In addition to InfoNCE, the training loss is regularised by a diversity loss $L_d$ to encourage the model to use $V$ codebook entries equally often. The diversity loss is formulated as
\begin{equation}
\label{eq:ld}
    L_d = \frac{1}{GV}\sum_{g=1}^G-H(\bar{p}_g) = \frac{1}{GV}\sum_{g=1}^G\sum_{v=1}^V\bar{p}_{g,v}log\bar{p}_{g,v}.
\end{equation}
Wav2Vec 2.0 shows the state-of-the-art results for ASR tasks evaluating on both Librispeech \cite{panayotov2015librispeech} and TIMIT \cite{garofolo1993darpa} data sets.

The method has been explored from the perspective of domain shift in \cite{hsu2021robust}, where the data for pre-training, fine-tuning, and evaluation originates from different domains. The authors conclude that the matching conditions between data of pre-training and testing are very important in order to achieve satisfying speech recognition results. Moreover, pre-training on multiple domains can improve the generalisation ability of the learnt representations. 

Based on Wav2vec 2.0, models for cross-lingual speech representation learning have been explored separately in \cite{riviere2020unsupervised} and \cite{babu2021xlsr}. The task of learning multilingual speech representations has also been undertaken in \cite{kawakami2020learning} but using a bidirectional CPC model.  
Sadhu et al.\  \cite{sadhu2021wav2vec} proposed a way to integreate Wav2vec 2.0 and VQ-VAE \cite{van2017neural} in a single model, defined as Wav2vec-C. The use of VQ-VAE, reconstructing the discrete codes to the original input features, can be seen as providing an additional regularisation when creating discrete speech representations by Wav2vec 2.0. The optimisation target combines the reconstruction error from VQ-VAE and  the loss function of Wav2vec 2.0. This enforces the learnt latent representations, which explicitly carry essential %crucial 
information for recovering the input features.  This is  expected to alleviate the codebook utilisation difficulties observed for Wav2vec 2.0. The method achieved some additional improvement in ASR on real-world far-field noisy data, compared to the original Wav2vec 2.0 \cite{baevski2020wav2vec}.

% clustering
In general, all the introduced Wav2vec audio SSL models learn latent representations without considering specific tasks for pre-training. After pre-training, they are fine-tuned for downstream tasks in an additional step. 
Baevski et al.\ present Wav2vec-U \cite{baevski2021}, short for Wav2vec Unsupervised, which learns a map from audio representations to phonemes directly without supervision. The method is a generative adversarial network, where the generator uses Wav2vec 2.0 to extract speech representations and generate phoneme sequence based on it using a clustering method; the generated phoneme tries to cheat a discriminator that is conditioned on a real phoneme sequence from unlabelled text. 

The idea of grouping quantised audio representations into phoneme sequences is named as phonetic clustering in SeqRA-AE \cite{liu2020towards}. In this work, the discrete representation is learnt in an auto-encoder architecture with vector quantisation. Moreover, the consecutive repeated quantised representations are further grouped to form phonetic units. Each phoneme can therefore correspond to several repeated codewords, which is similar to the format of Connectionist Temporal Classification (CTC) \cite{graves2006connectionist}.
Differently, Hidden unit BERT (HuBERT) \cite{hsu2021hubert} does not apply contrastive learning for training the same MPC model and avoids vector quantisation. Instead, each of the learnt audio representation is paired with a pseudo-label provided by applying K-means to MFCCs of the input audio. The method benefits from cluster ensembles, as the K-means clustering can be of different numbers of clustering centres, creating targets of different granularity.  

% ** Idea of audio codebook used in wav2vec and hubert**

We noticed that the classic formulation of several front-end audio processing tasks that have been explored for a long history are essentially using the framework audio SSL. For example, methods to solve the task of speech enhancement (SE) \cite{wang2018supervised, choi2018phase} share similar structure with the  auto-encoding predictive model, as shown in \Cref{fig:models(a)}. An SE model processes a noisy audio input and outputs clean speech. For generating the noisy input, clean speech is typically mixed with a noise recording. This is exactly the same as the processing of input to an auto-encoding predictive model, while the noise addition is seen as a step for data augmentation. Hence, the latent features in the middle layers of an SE model are seen as a kind of audio representation of the clean speech. The formulation is not limited to speech enhancement, but is applicable to all the pre-processing tasks that aim at predicting audio of interest from additive and multiplicative noise or interference, such as, source separation, de-reverberation, and echo cancellation. %, etc.

%Similar frameworks for representation learning have been explored for audio tasks such as speech enhancement, source separation, de-reverbration, echo cancellation. 
%For these tasks, deep learning models take noisy audio as input, and expect to output clean audio. For training these models, the noisy audio is usually synthesised by adding noise or multiplying room effect to clean audio. This is similar way like how an audio is transformed or augmented. Later netowrks should be able to recover clean speech from the augmented audio. SEGANs and so on....
%However, these tasks are limited to supervised learning as they have a groundtruth in training. 
% Therefore, authors believe most standard GAN framework belongs to supervised learning. 
% One exception is Cycle-GAN?

In some very recent works, audio SSL approaches have been  specifically chosen to solve typically challenging tasks, such as 
% to address more front-end audio tasks, such as 
speech enhancement  \cite{wang2020self,aswin2020,qiu21_interspeech} and source separation \cite{huang2021}.  
In \cite{wang2020self},  a pair of variational auto-encoders, named  clean auto-encoder (CAE) and mixture autoencoder (MAE), were  exploited. A CAE is trained to learn representations of clean speech by minimising the reconstruction error of its input spectrogram. 
An MAE encodes a noisy utterance and forces the encoded representation into the same latent space of the CAE by using a cycle-consistency loss terms. This paradigm leans a mapping from the domain of mixtures to the domain of clean sounds without using paired training examples. %\emi{I do not get this sentence. Can you rephrase?}
MixIT, which is short for Mixture Invariant Training, is proposed in \cite{scott2020} for solving unsupervised sound separation. In MixIT, a separation network takes a mixture of multiple single-channel acoustic mixtures (MOM) as model input, where each of the acoustic mixtures is comprised of several speech sources. The separation network decomposes the MOM into separate audio sources, which are then selected to be re-mixed up to approximate each acoustic mixture of the MOM. Similarly as for the  Permutation Invariant Training (PIT)  \cite{yu2017permutation}, the remix matrix is optimised  by choosing the best match between the separated sources and the acoustic mixtures. The method shows improvements for reverberant speech separation, universal sound separation, and is effective for speech enhancement, too.  
Finally, using denoising pre-training is an alternative solution to solve the permutation switching problem of source separation \cite{huang2021}. In this work, the authors use speech denoising as a self-supervised pre-training task to learn the structure information of speech from large-scale data. The model is subsequently  fine-tuned with the normal training objective of source separation. As knowledge about the speech structure has been captured in the pre-trained model, it relaxes the permutation problem.

To develop an SE system specialised in a particular person (PSE), %in the case of few-shot learning scenario where access to cleaning recordings of a speaker is very limited, 
%EMI%
%EMI: here is complicated... I do not fully understand this sentece. Can we simply skip it?
%**%
%%in the case of few-shot learning scenario where access to cleaning recordings of a speaker is very limited, 
%**%
%EMI%
\cite{aswin2020} presents two SSL algorithms, pseudo speech enhancement (PseudoSE) and contrastive mixtures (CM), for extracting speaker-specific discriminative features. 
A PseudoSE model is trained to recover a premixture signal (\ie a clean speech contaminated by noise) from a pseudo-source (\ie a mixup of the premixture signal and additional noise). 
% \emi{Do not get this.}\sure{The premixture signal is clean + noise. An additional noise is added to premixture signal, premixture + noise. A model is trained to recover the premixture signal from premixture+noise. So the premixture signal is used as target for denoising.}  
The CM method generalises the training via contrastive learning, for which a positive pair shares the same premixture signal (but deformed with different additional noises), while a negative pair stems from two different premixture sources mixed with the same additional noise. The trained model, using either  contrastive  or non-contrastive SSL, is trained to recover premixture sources rather than clean speech, and hence, it requires fine-tuning for the downstream task. Data purification (DP) \cite{sivarama2021} is later introduced in the pseudo speech enhancement training. Specifically, a separate model is trained to estimate the segmental SNR of the premixture signals, measuring the different importance of the audio frames. Injecting the importance measurement in pseudo SE training enables the model to benefit from segments of higher quality, and hence,  enables to derive more meaningful speaker-specific features.  

\shuo{A summary table of the typical audio SSL methods is shown in \Cref{tab:table2}.}

\begin{figure*}[t!]
\vspace{-0.8cm}
  \centering
    \hspace*{-0.5cm}
    \begin{subfigure}{0.33\textwidth}
        \centering
        % \vspace{1.8cm}
        \hspace*{1.3cm}
        \includegraphics[height=3.5in]{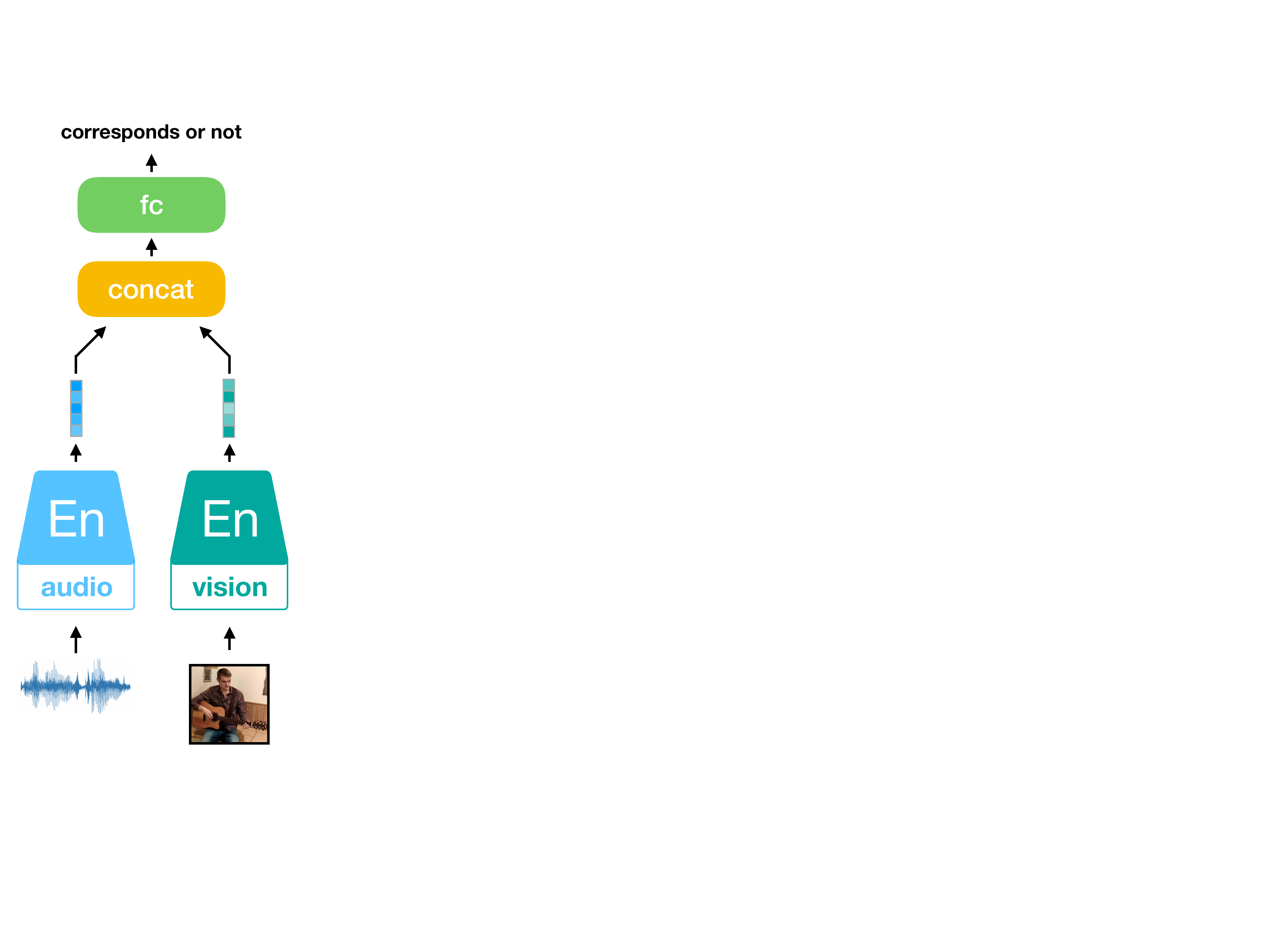}
        \vspace{-2.4cm}
        \caption{L$^3$-Net}
        \label{fig:l3}
    \end{subfigure}%
    ~ 
    \hspace*{-1.5cm}
    \begin{subfigure}{0.33\textwidth}
        \centering
        % \vspace{0.7cm}
        \hspace*{1.3cm}
        \includegraphics[height=3.5in]{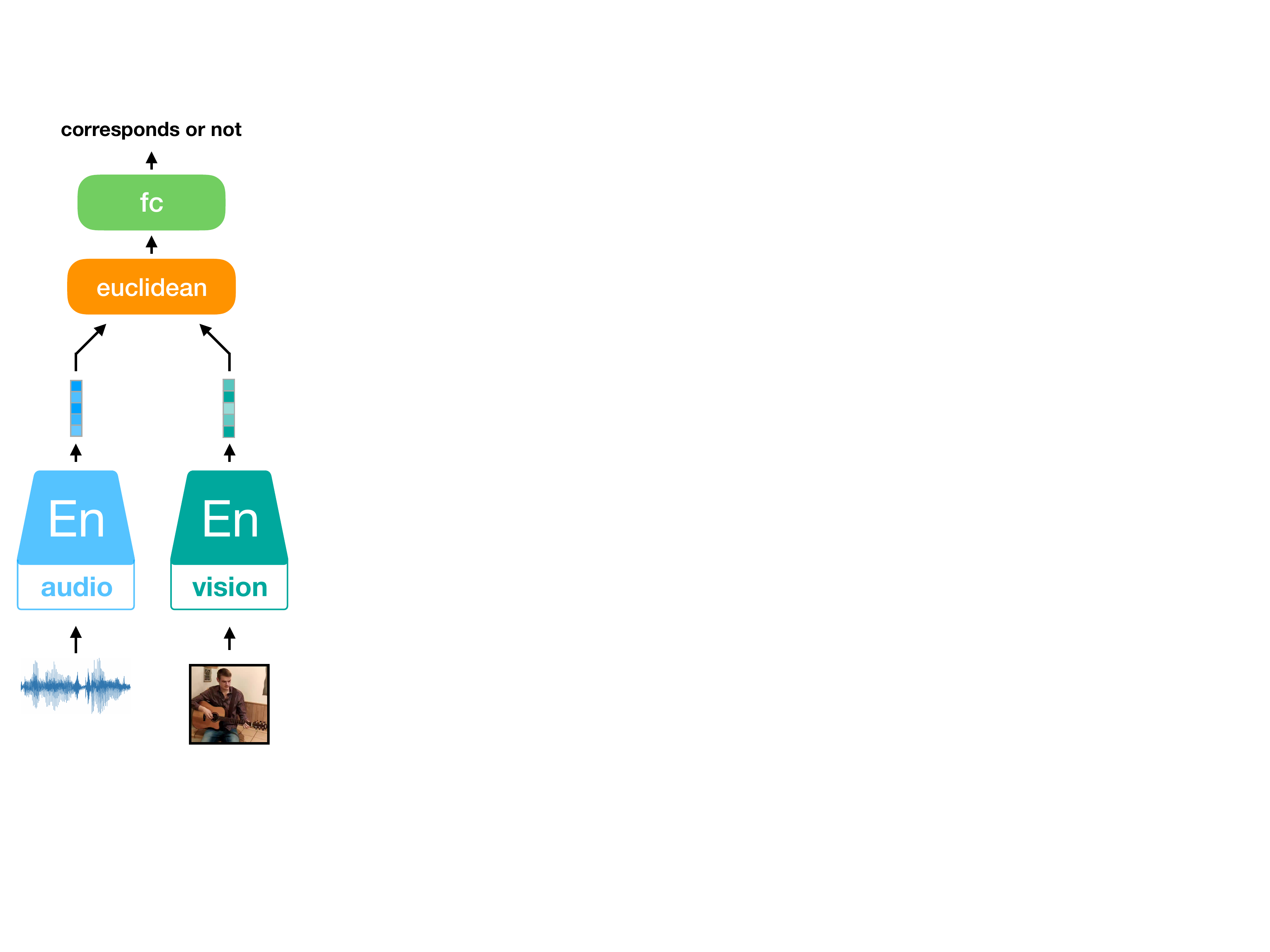}
        \vspace{-2.4cm}
        \caption{AVE-Net}
        \label{fig:ave}
    \end{subfigure}% 
    ~
    \hspace*{-2cm}
    \begin{subfigure}{0.33\textwidth}
        \centering
        % \vspace{0.7cm}
        \hspace*{1.3cm}
        \includegraphics[height=3.5in]{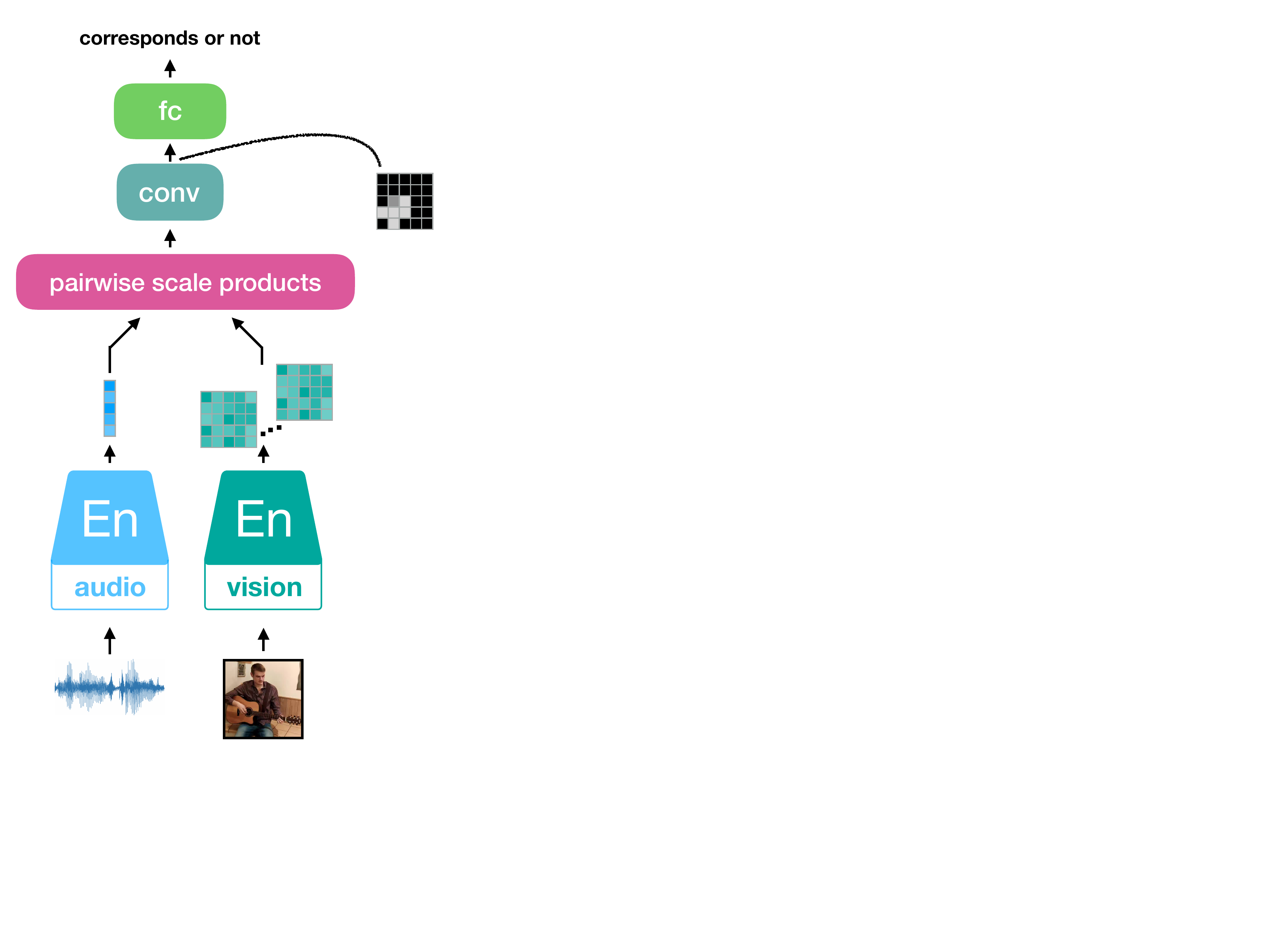}
        \vspace{-2.4cm}
        \caption{AVOL-Net}
        \label{fig:avol}
    \end{subfigure}% 
  \caption{Diagrams for L$^3$, AVE-Net, and its modified version -- AVOL-Net -- for visual object localisation.}
  \label{fig:Twotowers}
\end{figure*}

\section{Multi-modal audio representation}
\label{sec:mm}
\noindent
The successful adoption of SSL has spread over many academic and industrial fields, including, but not limited to, CV, audio processing, and NLP, to cite a few. 
Moreover, exploiting multi-modal SSL has also been explored for a variety of applications whereby the representation learning of different modalities can be performed simultaneously. The mutual complementarity between different modalities, treated as different views, representing one unique object, is beneficial for  %that represent one thing benefits 
the representation learning of each considered modality. 
In this section, we discuss multi-modal SSL approaches that use audio as one modality. Most of these works are based on audio-visual processing which aim, for example,  at %by 
determining the correspondence between video frames and its audio sequence.  %\shuo{for example}. 
Other visual-audio methods, similarly as the SSL works previously described, %same as before SSL works 
make use of the synchronisation of the visual and audio streams of a video, taking the two views as input of a Siamese network. 
In this case, each modality of the two can be seen as the supervisory signal for the other. Audio representation can also be learnt during a task of video generation, where the representation of each segment of an utterance is expected to carry adequate speech information in order to transfer the knowledge into video frames. 
Some more interesting approaches are motivated by classic tasks in CV, \ie  object segmentation and localisation, and audio processing, \ie  source separation.   
On the other hand, text is also considered as one modality that assists in learning speech representations, regarding the similarity between speech and text in a linguistic structure.

%BS: All images using images from others MUST HAVE a reference to the source IN THE CAPTION!!! I still see, e.g., Fig 8 using guitar images from the net w/o reference - this is NOT POSSIBLE! Also, it's always better to avoid taking others' pictures - please replace where possible!

\subsection{Audio \& Visual}

\noindent
% \am{Introductory paragraph}

\subsubsection{Visual-Audio Correspondence Decision}
By splitting a video into visual and audio streams, both L$^3$-Net \cite{arandjelovic2017look} (\Cref{fig:l3}) and AVE-Net \cite{arandjelovic2018objects} (cf.\ \Cref{fig:ave})  exploit two convolutional networks, named as vision sub-network and audio sub-network, to separately encode the two streams into a common space for cross-modal retrieval. 
%Using audio-visual correspondence (AVC) of an video as the training objective, L$^3$-Net \cite{arandjelovic2017look} and AVE-Net \cite{arandjelovic2018objects} exploits two networks that can embed audio and visual inputs into a common space for cross-modal retrieval. 
Specifically, based on the alignment between both streams, one second of audio segment and the corresponding centre video frame are fed into these two networks. The model is required to decide whether the two inputs are in correspondence or not. 
For a video clip, the audio segment and video frame at the same time step are considered as a positive pair for model input, while a negative input pair is the audio segment paired with a video frame from another different video clip. 
In L$^3$-Net \cite{arandjelovic2017look}, the audio and visual representations are concatenated before sending into fully-connected layers to predict the correspondence score. In contrast, AVE-Net \cite{arandjelovic2018objects} measures the correspondence degree by computing the Euclidean distance between audio and visual representations that are designed to be of the same dimensionality. 
Moreover, both L$^3$-Net and AVE-Net are especially designed for recognising where the sound is generated in the video frame, for example, the location of specific instruments in a band. %guitars, keyboards, etc. 
The vision sub-network of L$^3$-Net has the intrinsic ability to recognise semantic entities that make sound, while AVE-Net needs additional modifications on its model architecture, incorporating a comparison mechanism to the audio representation with each spatial grid of the 3D visual representations (\Cref{fig:avol}). 
%the vision subnetwork to construct AVOL, which computes the correspondence scores between the audio embedding and a grid of region-level image descriptors. 
The method encourages at least one region to respond highly for a corresponding audio and video frame, and hence enabling the localisation of the object that sounds. 
As these two AVC works formulate the task as a binary classification problem, the models can be optimised by minimising a logistic loss. 

\begin{figure*}[t!]
  \centering
    %\hspace*{-2cm}
    \centering
    % \vspace{1.8cm}
    \includegraphics[height=4.5in]{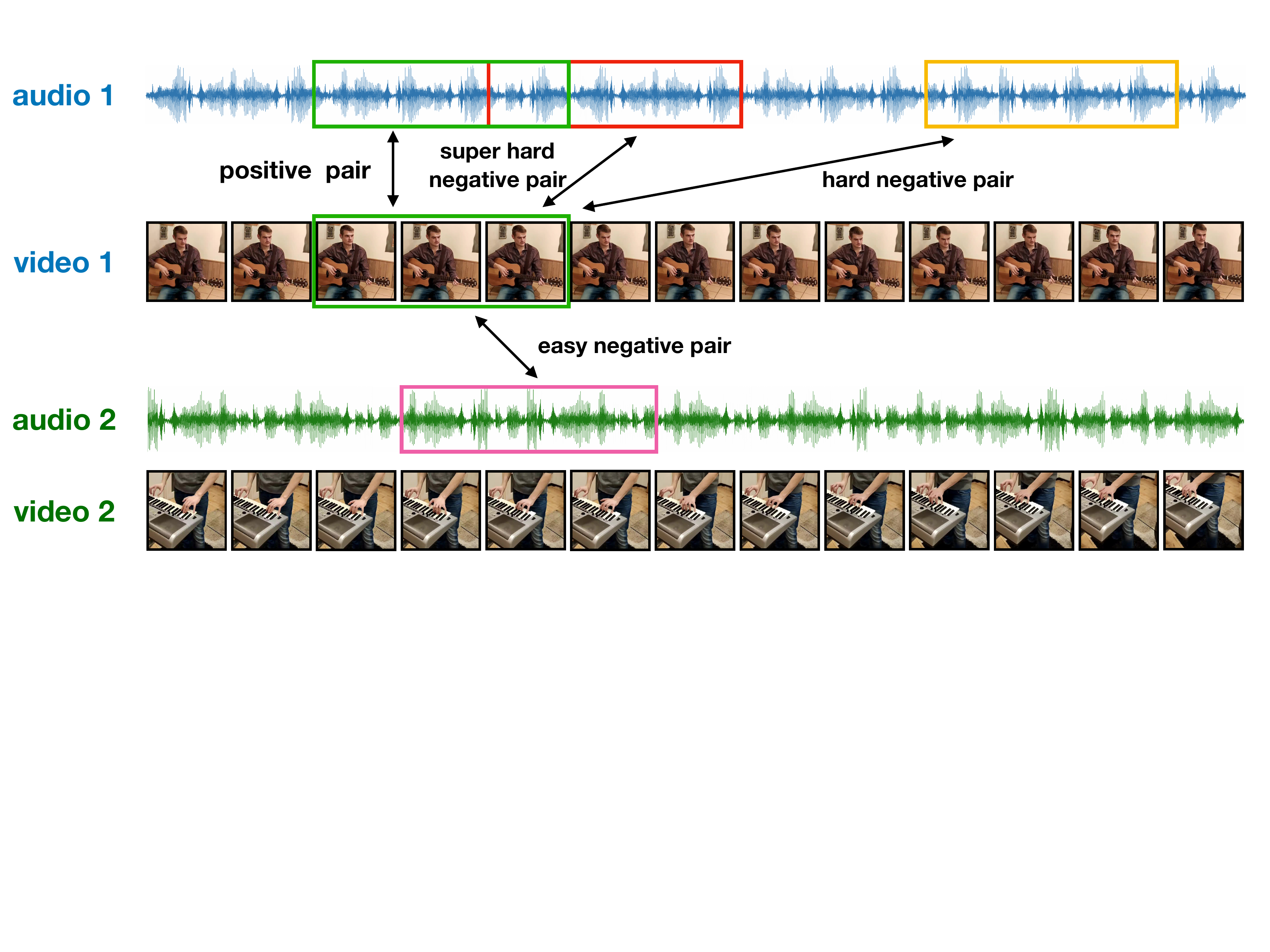}
    \vspace{-4cm}
  \caption{Diagrams for demonstrating a negative sampling strategy \shuo{introduced in \cite{korbar2018cooperative}}.}
  \label{fig:samples}
\end{figure*}

In \cite{owens2018audio}, a model to predict whether the visual and an audio stream are synchronised is trained,  with  contrastive %training 
objectives, %, with a slight difference in visual data format, \ie  
using a sequence of video frames instead of a single frame.
Nagrani et al.\ \cite{nagrani2018learnable} suggest to apply curriculum learning, \ie starting the training with relative easier negative and positive pairs for good model initialisation, and gradually increase the difficulty of input pairs for easier model convergence. The approach has shown promising results in learning cross-modal embeddings for the recognition of a person's identity. 
However, an incoming problem is that the model tends to rapidly learn to differentiate easy negative pairs from positive pairs while harder input pairs have very limited effect on learning discriminative representations. 
An opposite curriculum schedule has been shown to be effective for training a model that learns the cross-modal embeddings for ultrasound \cite{jiao2020self}. The visual and audio streams used in this work are medical ultrasound video and voice from a sonographer during the video recording. Due to the sparse correlation between the two inputs, hard positive and negative input pairs are first used in order to force the model to learn more strongly correlated representations. 
%To enhance the association between the two streams, the authors of \cite{nagrani2018learnable} suggests to apply curriculum learning, meaning to start training the model with harder negative and positive pairs, forcing the model to learn more strongly correlated representations. 
%Similar solution has been employed in \cite{nagrani2018learnable} for learning cross-modal representation for personal identity. 
In \cite{zhang2021enhancing}, Zhang introduced a two-stage curriculum learning solution based on %using a 
teacher-student training and identified it as % manner, named  
self-supervised curriculum learning, for audio-visual representation learning. Before joint training of vision and audio sub-networks, one of the sub-networks is updated using the representations from the other sub-network as teacher, which is frozen for updating. 
The two sub-networks exchange the role for training with the other sub-network.     
In addition, to enlarge the number of negative samples for training, a memory bank is applied, resulting in considerable improvements in a visual task of action recognition and an acoustic task of audio sound recognition. 

Differently, in \cite{korbar2018cooperative}, the authors make use of margin loss in order to contrast positive and negative pairs that are of equal proportion. 
The negative examples of different hardness difficulty are considered, including easy negatives, hard negatives, and super-hard negatives (shown in \Cref{fig:samples}). The easy negatives are video frames and audio segments from different videos, while hard negatives are those pairs taken from the same video but that are at least half a second distant from each other. The super-hard negatives are the pairs that overlap for a certain temporal extent. 
The authors also confirmed the need of starting training the model with easy negatives and then gradually adding harder negatives, which has shown to be effective for learning high-quality representations. 
Similarly, treating negative pairs of different specialities, \ie different difficulty levels,  
is  investigated in \cite{ding2020self} for audio-visual speaker diarisation.  In this work,  the margin value used in a triplet loss is controlled by the shifted range between audio and visual streams, by this representing a different difficulty degree of negatives. 
Nagrani et al.\  \cite{nagrani2020disentangled} optimised a model to learn audio-visual representations by formulating negative samples from the same video and different videos in content loss and identity loss.   Additionally, in order to encourage explicit separation of representations, they  used a disentanglement loss which is implemented as confusion loss in \cite{alvi2018turning}.

Harwath et al.\ \cite{harwath2017unsupervised,harwath2018jointly,harwath2019learning} proposed another interesting pretext task by associating spoken audio captions with their corresponding image for learning audio-visual representations. Two networks are used to process the audio and image as input. In \cite{harwath2017unsupervised}, the dot product of a pair of visual and audio representations is calculated as their similarity score. Similar to the AVOL-Net, similarity between audio representation and the visual embedding of each pixel can be computed to construct spatial activation maps, leading to a solution for object localisation \cite{harwath2018jointly}. In a different way, in \cite{harwath2019learning}, the generated audio and visual representations are pushed into a common latent space using triplet loss as well as   by   contrasting the positive pair to the negative pairs that contain either an unmatched caption or an unmatched image.  
Hsu et al.\ \cite{Hsu2021TextFreeIS} solve the same task by building a system based on ResDAVEnet-VQ %proposed in 
\cite{harwath2019learning} and an   Image-to-Unit Model %from 
\cite{xu2015show}. Although  each of these models is  used to process an input stream, the two representations are pushed into the same latent space. Instead of using contrastive training objectives to reproduce the audio input, %, %but in another way. 
the learnt discrete linguistic units,  learnt through  ResDAVEnet-VQ from the input utterances, are fed to Tacotron2 \cite{shen2018natural}, \ie  a text-to-speech (TTS) model for speech synthesis.  %, to reproduce the audio input. 
The visual sub-network, \ie ResDAVEnet-VQ, is then expected to learn representations that are close to the discrete linguistic units, by this enabling the representation learning to retrieve information from both modalities.     

\begin{figure*}[t!]
  \hspace*{-1cm}
  \centering
    \begin{subfigure}{0.5\textwidth}
        \centering
        % \vspace{1.8cm}
        \hspace*{2.2cm}
        \includegraphics[height=3.2in]{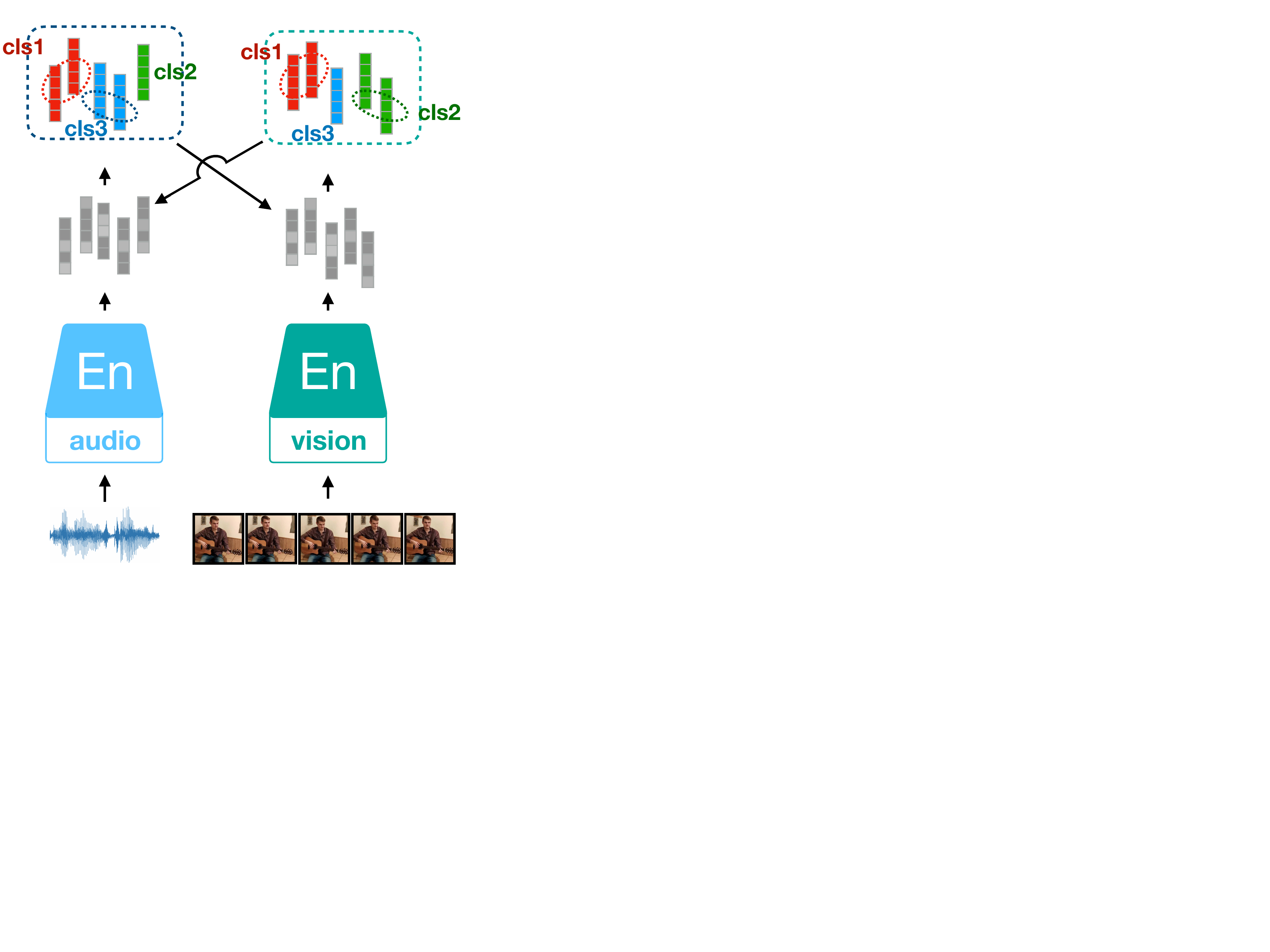}
        \vspace{-3.2cm}
        \caption{Clustering}
        \label{fig:ms_clustering}
    \end{subfigure}%
    ~ 
    \hspace*{-0.8cm}
    \begin{subfigure}{0.5\textwidth}
        \centering
        \hspace*{0.8cm}
        \includegraphics[height=3.2in]{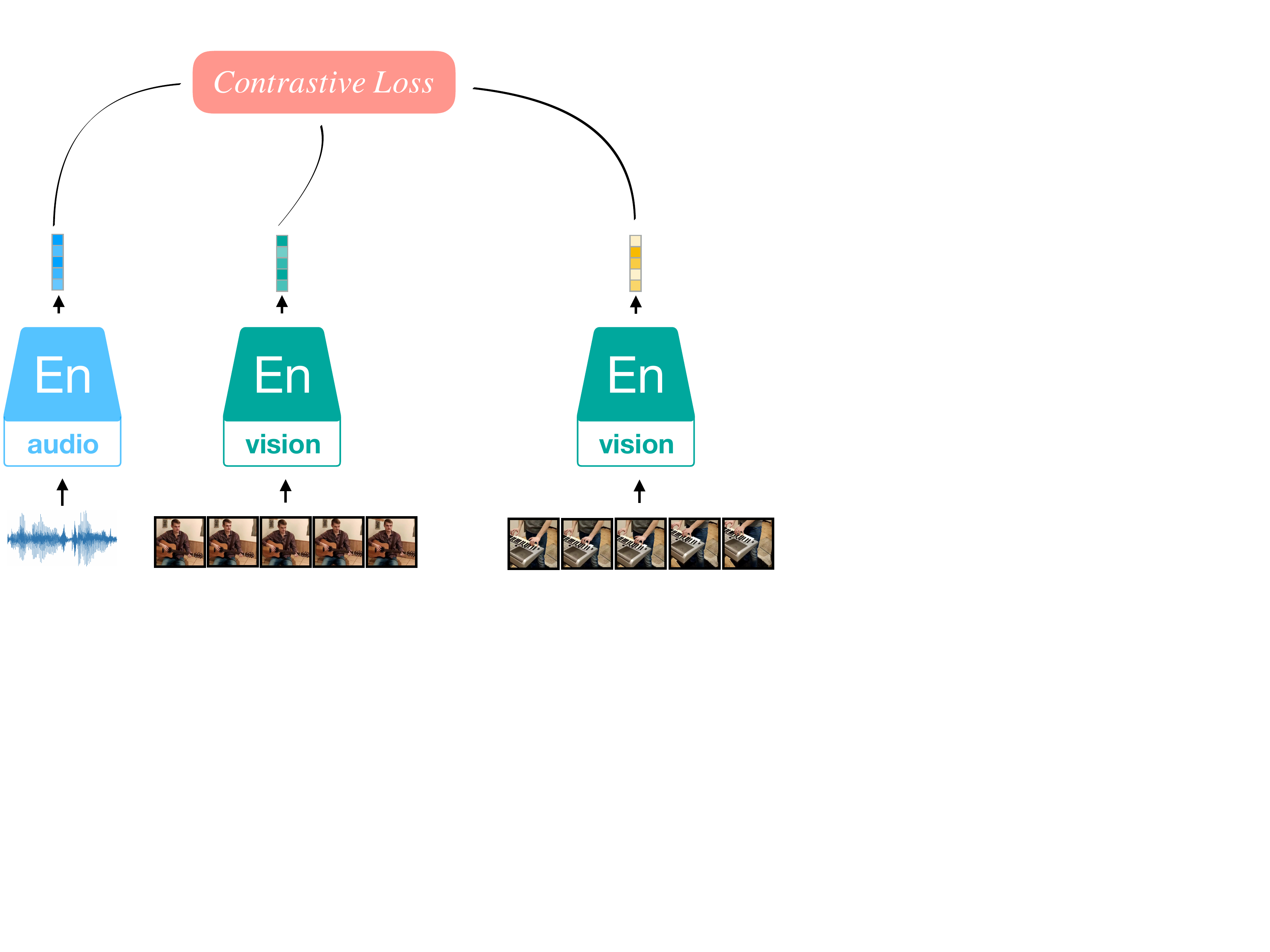}
        \vspace{-3.2cm}
        \caption{Contrastive Learning}
        \label{fig:ms_cont}
    \end{subfigure}% 
  \caption{Methods for audio-visual mutual supervision.}
  \label{fig:PC}
\end{figure*}

% Mutual Supervision
% The learning of audio and visual representations from video clips
 Based on their mutual correspondence,  the audio and visual streams of a video clip can be seen (to each other) as the supervisory signal  for representation learning. % based on their correspondence. 
An earlier study has shown the success of using synchronously recorded ambient sounds as supervision for visual learning \cite{owens2018}. Later on, Alwassel et al.\ \cite{alwassel2020self} and Morgado et al.\ \cite{morgado2021audio} empirically verified that exploiting the representation of one modality to create pseudo-labels for training the encoder network of the other modality  outperforms not only SSL on a single modality, but also SSL based on pseudo-labels of both modalities.  %\emi{mmm you need to simplify this. Maybe "multi-modal SSL" or "SSL based on pseudo-labels" (i do not know how to sumarise it...) but something  that sounds like a name, not an additional sentence.} 
In \cite{alwassel2020self}, the pseudo-labels are generated using a deep clustering method  (cf.\ \Cref{fig:ms_clustering}). Differently,   \cite{morgado2021audio} aggregates `memory features' by computing the slow exponential moving average (EMA), and subsequently applies contrastive learning  (cf.\ \Cref{fig:ms_cont}), similarly as    \cite{grill2020bootstrap}.  
In \cite{morgado2021audio}, \textit{Cross-Modal Agreement} (CMA) is additionally introduced to enhance the interactions between instances, specifically to calibrate within-modal similarities between positive pairs. 
Both methods, \ie clustering- and contrastive learning-based modelling, learn effective audio-visual representations, evaluated on a downstream task of action recognition based on video. 

\subsubsection{Audio-Visual Source Separation}
%PixelPlayer \cite{zhao2018sound} extends the AVC methods for localising sound sources in a video and separating the audio into its components without supervision. The method adopts the Mix-and-Separate framework, which first generates a synthetic sound separation training set by mixing the audio signals from two different videos and then trains a neural network to separate the audio mixture conditioned on the visual input corresponding to one of the audio signals. The framework consists of three components: an image analysis network, an audio analysis network and an audio synthesizer network. 

% \begin{figure*}[ht!]
%   \centering
%     \hspace*{4.4cm}
%     \centering
%     % \vspace{1.8cm}
%     \includegraphics[height=4.5in]{Figures/PlayPixel.pdf}
%     \vspace{-4.5cm}
%     \label{fig:pp}
%   \caption{PixelPlayer}
%   \label{fig:Twotowers}
% \end{figure*}

\begin{figure*}[t!]
  \centering
    \hspace*{0.7cm}
    \begin{subfigure}{0.3\textwidth}
        \centering
        % \vspace{1.8cm}
        %\hspace*{-0.4cm}
        \includegraphics[height=2.8in]{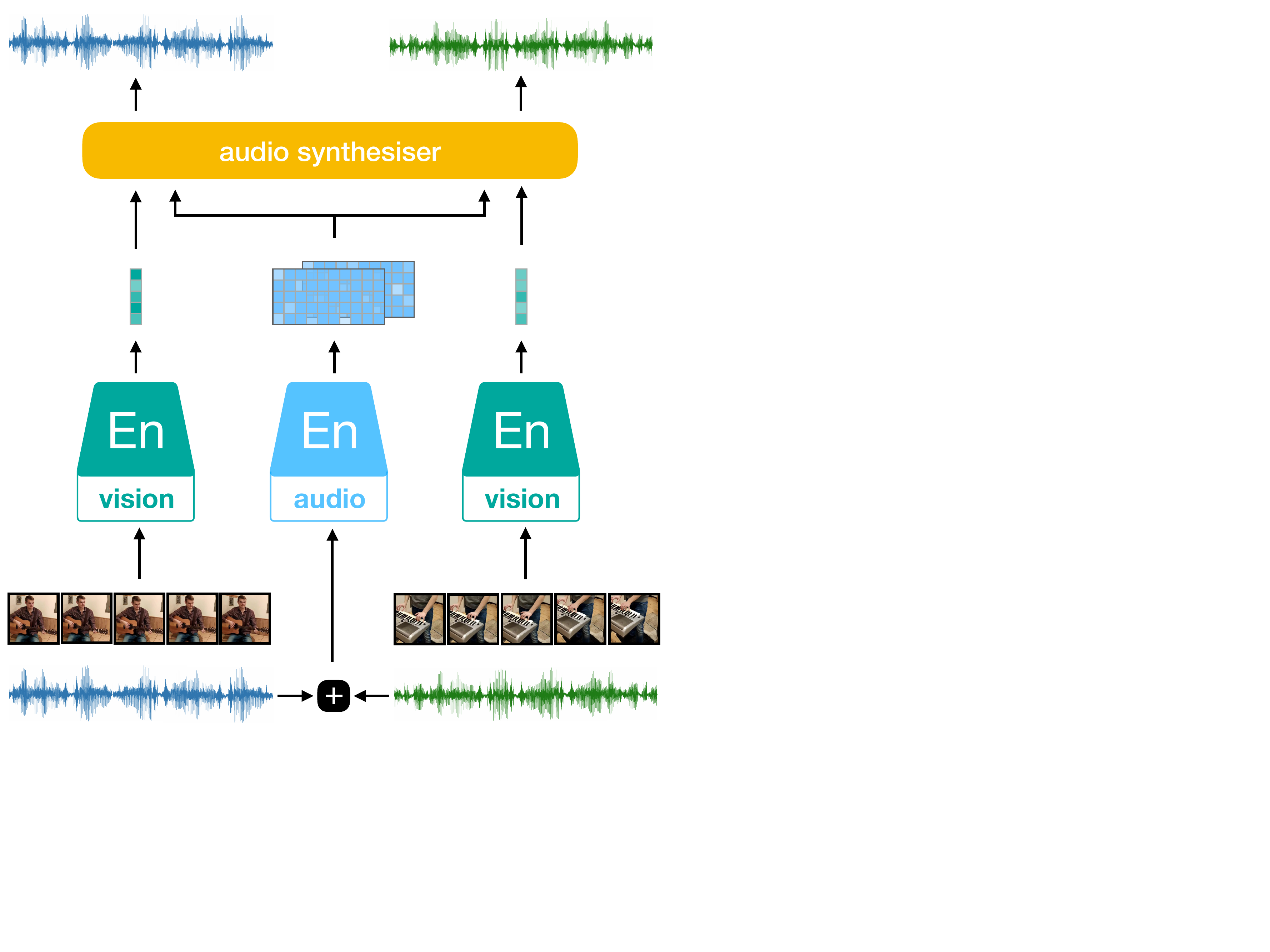}
        \vspace{-2cm}
        \caption{PixelPlayer}
        \label{fig:pixelplayer}
    \end{subfigure}%
    ~ 
    \hspace*{0.3cm}
    \begin{subfigure}{0.3\textwidth}
        \centering
        % \vspace{0.7cm}
        %\hspace*{-1.8cm}
        \includegraphics[height=2.8in]{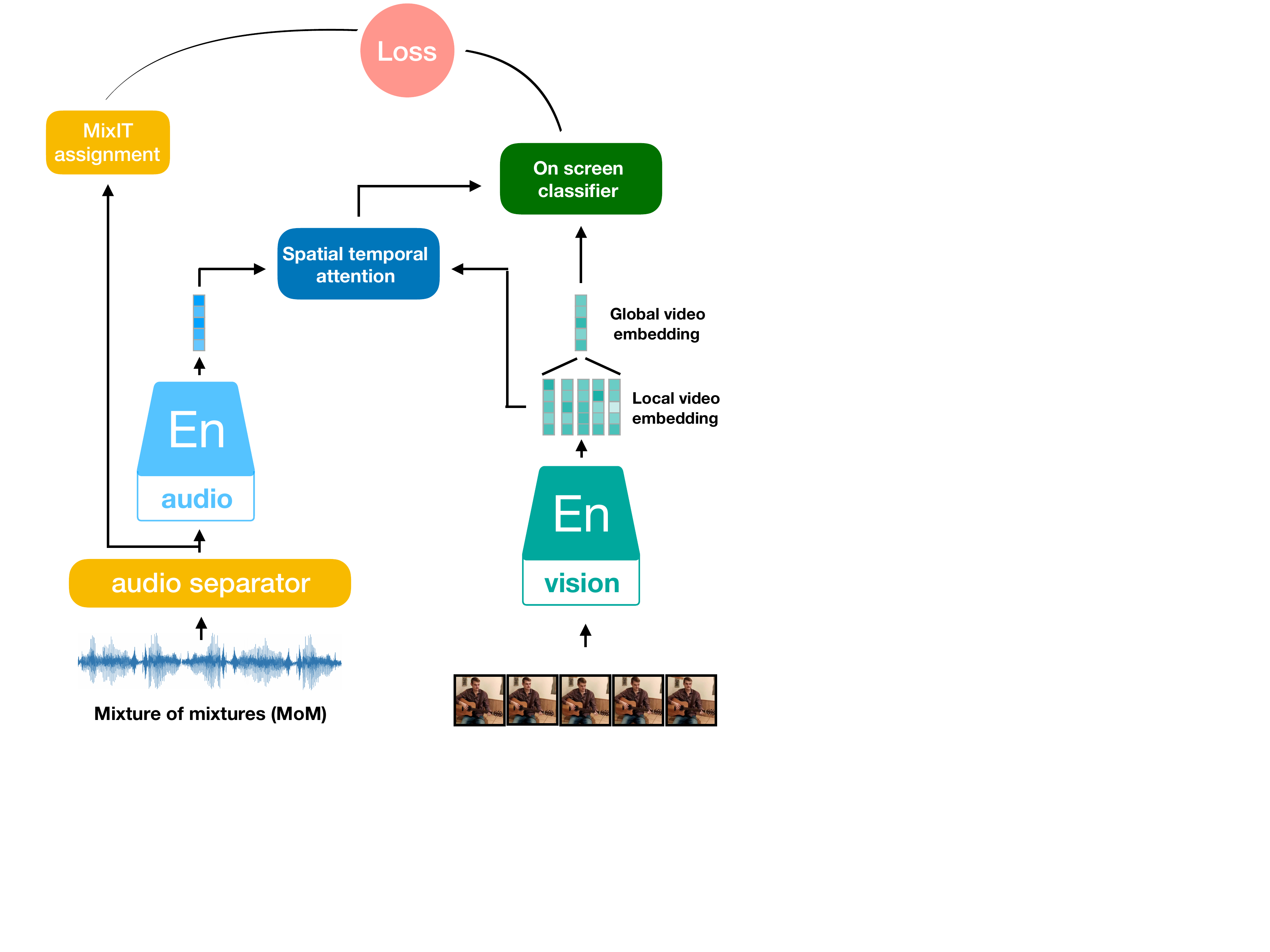}
        \vspace{-2cm}
        \caption{AudioScope}
        \label{fig:audioscope}
    \end{subfigure}% 
    ~
    \vspace*{0.5cm}
    \hspace*{0.1cm}
    \begin{subfigure}{0.3\textwidth}
    \centering
    \hspace*{1.2cm}
    \includegraphics[height=3.2in]{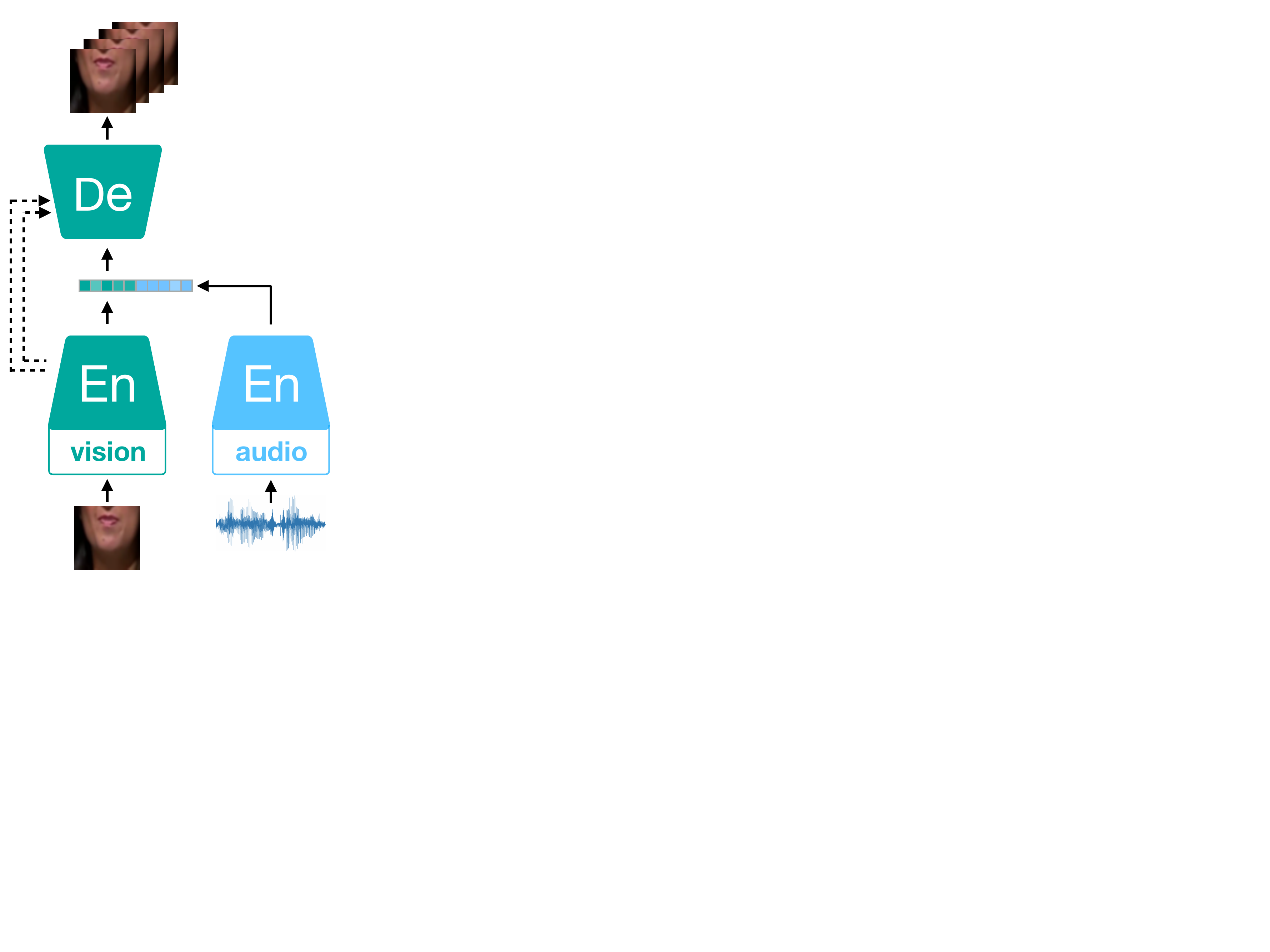}
    \vspace{-3.2cm}
    \caption{Video generation}
    \label{fig:vg}
    \end{subfigure}%
  \caption{Diagrams for PixelPlayer, AudioScope and the framework for video generation.}
  \label{fig:av}
\end{figure*}

The PixelPlayer \cite{zhao2018sound} effort proposed a Mix-and-Separate framework that solves visual object segmentation and audio source separation together. The framework consists of three networks: a video analysis network, an audio analysis network, and an audio synthesiser network; as shown in \Cref{fig:pixelplayer}.  
The video analysis network extracts visual features from a sequence of video frames while the audio analysis network processes the mixture sound from two different video clips. The audio synthesiser network aims to separate the audio sources based on the learnt audio representations of the mixture, conditioned on the corresponding video frames. In this way, the model can learn a better semantic visual representation that is highly associated with its own audio, but less relevant to the audio of the other video clips. 
Although the learnt audio representations   enable the model to retrieve  information from the mixture sound,  it cannot  separate audio from each video. 
%However, the learnt audio representations in this model retrieve the information of the mixture audio but not the separate audio from each video. 
% features of diffrent components of the input sound. 
%The audio synthesiser network finally predicts the sound by pixel-level visual feature and audio feature. A mask that could separate the sound of the pixel from the input is estimated and then used to reconstruct the spectrogram of each audio source. 
%Later, the authors
In a later work, the same authors also indicate that having   %\hl{release the requirement} \emi{what do you mean?} of 
synchronised audio and visual data is a requirement to disentangle %for inference by this disentangling that 
the learnt audio and visual representations before feeding them into the audio synthesiser network \cite{rouditchenko2019self}. By doing this, the learnt audio and visual representations can be used independently.

AudioScope \cite{audioscope} expands the conditioning audio separation approach, and exploits an additional audio embedding network to process the separated audios. An audio representation then aggregates the global information from each resulting audio embedding using temporal pooling. Subsequently, attention is used to retrieve the mutual information between the local spatial-temporal video embedding (learnt with the video embedding network) and the global audio representations. This allows  to generate an audio-visual representation that combines the audio and visual information. Finally, the audio-visual representations, \ie  the global video embedding and the global audio embedding, are concatenated together. By this, based on the separated audios, it is expected  to create the MixIT assignment  \cite{scott2020}. % based on the separated audios. 

%Gao \cite{gao2018learning} solves a similar problem by using multi-instance multi-label learning framework to disentangle the audio and frequency bases, extracted using NMF, that map to individual visual objects. They verified the recovered disentangled bases can be used to guide audio source separation to obtain better-separated, object-level sounds. (not pure SSL method)

LWTNet \cite{triantafyllos2020} designs a model that can ingest a video and transform it into a set of discrete audio-visual objects using SSL. Similarly, an audio network and a video network encode the audio and video frames; then, a fine-grained audio visual attention map is computed by solving a synchronisation task, \ie  measuring the similarity between the audio and the visual features at every spatial location. The model can detect and track multiple audio-visual objects and extract an embedding for each of them.  Given negative audio samples from shifted audio segments of the same video clip, contrastive loss is applied to maximise the similarity between a video frame and its true audio track.  
This, which is made in the form of an attention map, also  minimises the similarity of the misaligned versions of the audio. 

\subsubsection{Video Generation}
Given a starting video frame of a speaker, previous work has shown that  a model can be trained to generate the subsequent video sequences based on the corresponding speech utterance  \cite{shukla2020learning,shukla2020visually,shukla2021does}.
%and the entire speech sequence, machine can generate a sequence of image frames that starts with the given frame and the following frames match to the input speech. 
In these works, a U-Net is used
%contains an identity encoder 
to code the starting video frame into a latent representation while  an audio encoder is used to learn a representation of the speech utterance. 
In \cite{shukla2020learning}, the visual and audio representations are concatenated and then fed into the decoder of the U-Net to generate the full sequence of video frames. 
Using neural decoders, the raw waveform, the log Mel-spectrogram, and the MFCC-spectrogram of the audio input are expected to be constructed from the learnt audio representation. 
L1 reconstruction errors between the output of the  decoders (both audio and video) and the ground-truth, are minimised. 
Differently, 
% The length of input audio is reduced to $0.2$\,s 
in \cite{shukla2020visually} %, and additionally, 
%the work avoids 
the use of audio decoders is avoided and the length of the input audio is reduced to $0.2$\,s. In this work, 
a vector randomly created using Gaussian noise is appended to the audio-visual representation, by this injecting randomness in the procedure of face synthesis. %These two works 
In  \cite{shukla2020learning,shukla2020visually},  Shukla et al.\ show the efficacy of these two methods for spoken word classification and lip reading tasks. A later approach by the same authors is also  developed  for  %is further investigated for the task of 
emotion recognition  \cite{shukla2021does}.

\subsection{Audio \& Text}
\noindent
As already introduced in \Cref{sec:audio}, Wav2vec-U uses 
 %self-supervised speech representations to segment unlabelled audio and learn a mapping from these representations to phonemes via adversarial training. 
%EMI%
self-supervised speech representations in order to fragment unlabelled samples and subsequently map them  to phonemes through adversarial training.
%EMI%
Similarly, Chung et al.\  \cite{chung2018unsupervised} proposed to learn the individual speech and text embedding spaces, by aligning the two spaces via adversarial training and subsequently applying a refinement procedure. %attempting to align the two spaces via adversarial training, followed by a refinement procedure.
%Specifically, 
Under the assumption that  embedding spaces from  two modalities share a similar structure,  in this work, the audio and text embeddings are first learnt using Speech2Vec and Word2Vec, respectively;  then,  adversarial training is used 
%to learn a linear mapping from the speech embedding space to the text embedding space. 
%EMI%
in order to learn a linear mapping  between the speech embedding space and the text embedding space. 
%EMI%
%with the assumption that the embedding spaces of the two modalities share a similar structure. 
The approach has shown promising results for the task of ASR and speech-to-text translation systems for low- or zero-resource languages such as German, which has little audio-text pairs for training. 

\begin{figure*}[t!]
  \centering
    \hspace*{-2cm}
    \begin{subfigure}{0.33\textwidth}
        \centering
        % \vspace{1.8cm}
        \hspace*{1.3cm}
        \includegraphics[height=2.5in]{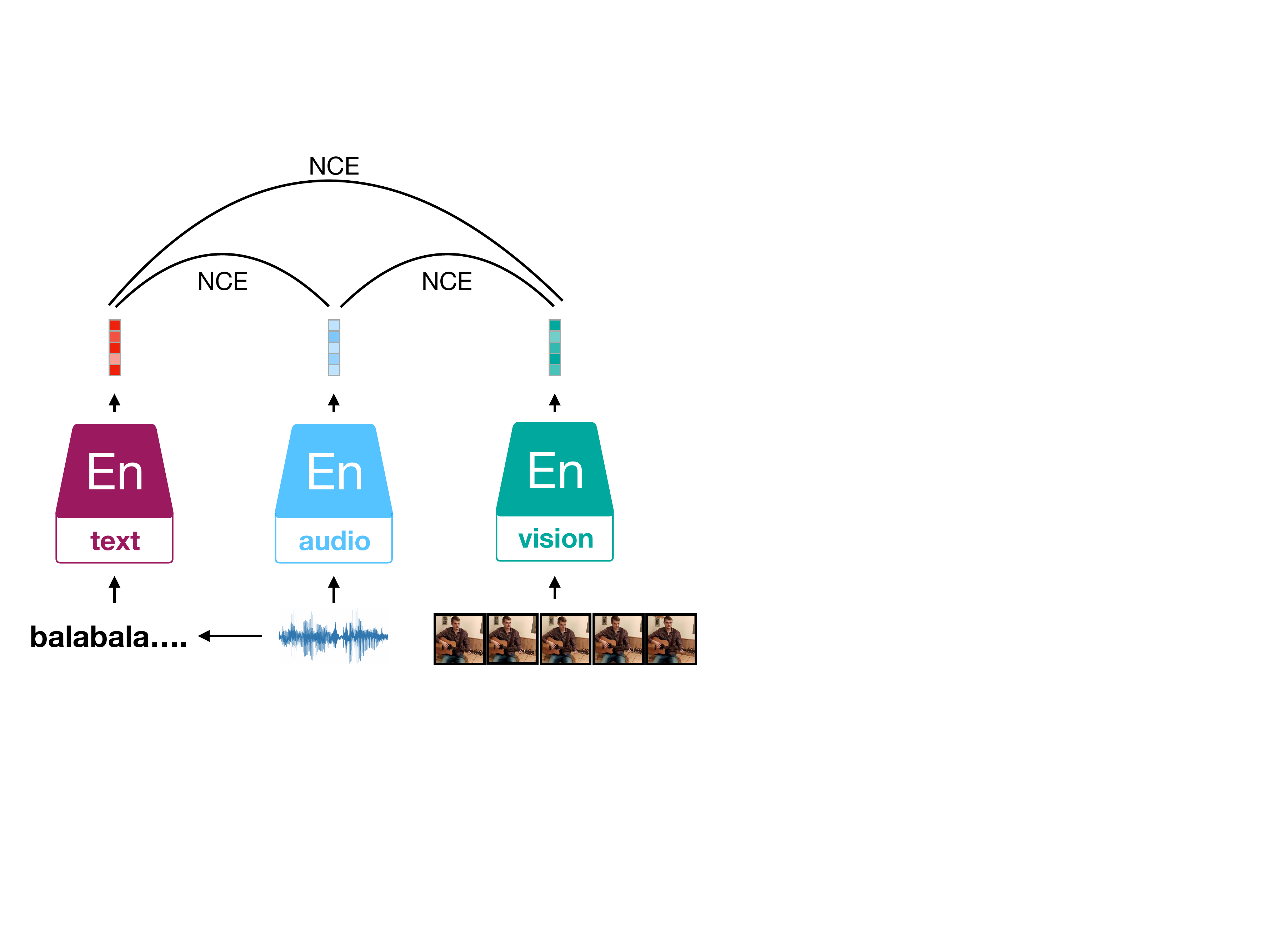}
        \vspace{-2.5cm}
        \caption{Shared}
        \label{fig:shared}
    \end{subfigure}%
    ~ 
    \hspace*{-0.5cm}
    \begin{subfigure}{0.33\textwidth}
        \centering
        % \vspace{0.7cm}
        \hspace*{1.3cm}
        \includegraphics[height=2.5in]{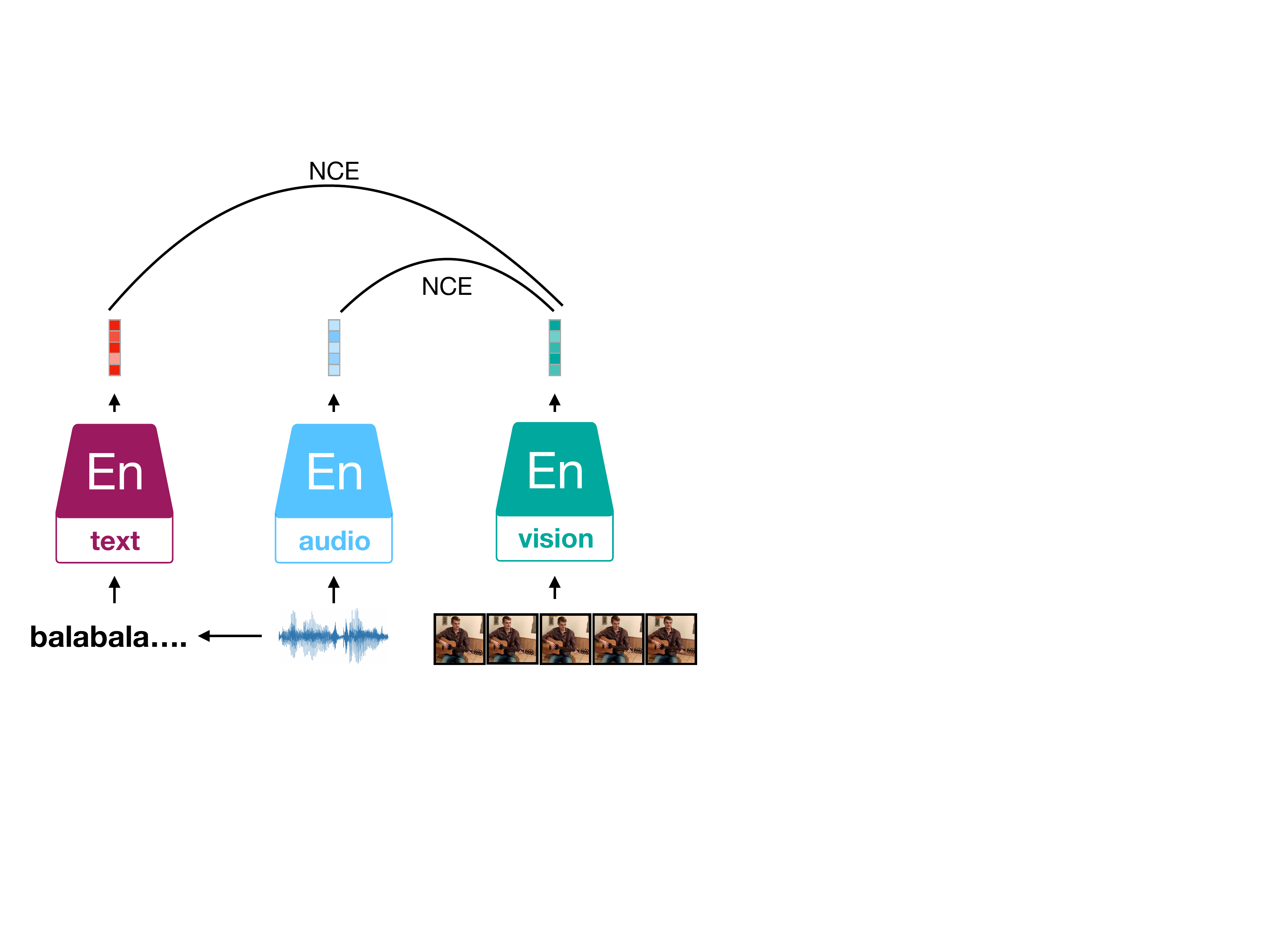}
        \vspace{-2.5cm}
        \caption{Disjoint}
        \label{fig:disjoint}
    \end{subfigure}% 
    ~
    \hspace*{-0.5cm}
    \begin{subfigure}{0.33\textwidth}
        \centering
        % \vspace{0.7cm}
        \hspace*{1.3cm}
        \includegraphics[height=2.5in]{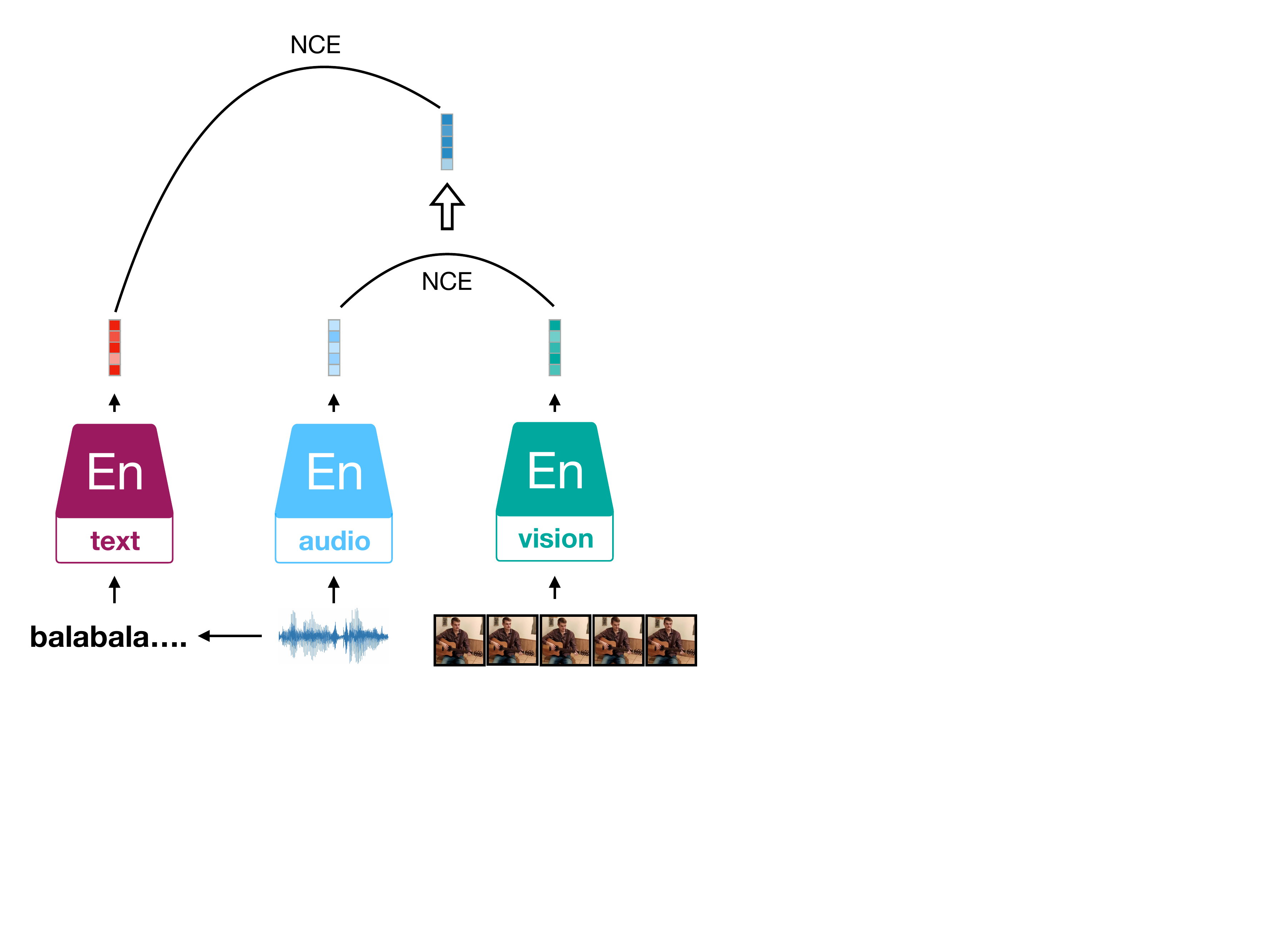}
        \vspace{-2.5cm}
        \caption{FAC}
        \label{fig:fac}
    \end{subfigure}% 
  \caption{Diagrams for three modes in a multi-model versatile network \shuo{(based on \cite{alayrac2020self})}.}
  \label{fig:152}
\end{figure*}

The co-alignment of audio and text can also be done within an auto-encoder (AE) architecture. 
COALA, presented in \cite{coala2020}, applies two AEs to process an audio spectrogram and the audio tag.   
Both AEs are optimised to reconstruct its input, resulting in semantic features of the audio and the text. The paired semantic features are pulled closer and the unpaired semantic features are pushed far using using a contrastive loss. %The whole system is jointly optimised by minimising the reconstruction errors of each modality and contrastive loss \hl{between} \emi{What does "between" refer to?} the representations of these two modalities in a multi-task learning problem. 
The whole system is jointly optimised by minimising the two reconstruction errors and the contrastive loss as a multi-task learning problem.  
%\sure{minimising three losses: 1. the reconstruction error of audio, 2. the reconstruction error of text and 3. the contrastive loss is applied to their latent representations.}
Affine transformations are applied to the two learnt representations, reducing the difficulty in maximising their agreement. 
An auto-encoder is also used in \cite{haque2019audio} for encoding audio spectrograms. As in COALA  \cite{coala2020},  the latent representations are also expected to be able to reconstruct the spectrogram and predict the linguistic features simultaneously. 
In these two works, one modality, either text or audio, is used to learn an embedding that is used to predict its paired input (in the other modality). 
%is considered as a representation of one of both streams.  
%As the embedding representation is then used to} %and 
%predict its paired stream, %and hence, 
Hence, it can be assumed that the learnt embedding contains the information from both streams.  
The reconstruction can be seen as a regularisation term that enables the embedding to reconstruct the input stream, by this ensuring that the learnt latent contains the salient features of the input stream.  
Similarly,
CSTNet \cite{khurana2020cstnet} is trained for speech-translation, but speech utterances are in English while the text translations are in any other language from French, German, Spanish, Portuguese, or Italian.  
The experimental results obtained from CSTNet indicate that the speech representation learnt using this framework can achieve comparable results for two downstream tasks, \ie a Minimal-Pair ABX task and phone recognition.

\subsection{Audio \& Text \& Video}
\noindent
To conclude this section, we will introduce some SSL works that %can 
learn audio representations through the use of three modalities: video, audio, and text.  
In these works, texts are commonly achieved by using off-the-shelf ASR systems from audio.  For instance, 
in \cite{sun2019learning}, the authors  presented the use of keyword localisation as the pretext task. The authors also  compare the performance obtained by  separately using text or images as supervisory signal.  
They conclude that the visually supervised model performs worse than a text supervised model based on %realised by using 
BoW. Indeed, although the visually trained model can sometimes locate semantically related words, this  phenomenon is not consistently observed.

A multi-modal versatile network is presented in \cite{alayrac2020self}, a  study that aims to find 
the best combination of the modalities. 
Learning a shared space of the three modalities, as well as two separate disjoint spaces for video-audio and video-text (considering that text originates from the audio), are investigated. Fine and Coarse (FAC) spaces are additionally proposed due to the fact 
%that the visual and audio domains are different from the language domain in terms of their granularities. In FAC, vision and audio are compared in the fine-grained space, while text is compared with audio and vision in a lower dimensional coarse-grained space. 
%EMI%
that the visual and audio domains  differ (in terms of  granularity) with respect to  the language domain. In FAC,  vision and audio  are compared in a fine-grained space while  text,  audio, and vision are compared in a lower dimensional coarse-grained space. 
%EMI%
%The procedure is seen as 
For this, the visual representation is first mapped into  common latent spaces of audio and video, and sequentially projected into the common latent spaces of text and audio-visual common spaces. The authors  consider no direct link between audio and text. %The training objective for FAC is a combination of NCE loss between audio and visual inputs, and a MIL-NCE variant that is tailored to account for the misalignment (weak correspondence) between video and text. 
Similar to the FAC approach, VATT \cite{akbari2021} also presents a two-stage multi-modal projection.
In VATT,  audio and video are compared first using NCE loss. Subsequently, through the use of MIL-NEC loss \cite{alayrac2020self} for optimisation purposes, the text is included in order to learn  common latent spaces for the three modalities. Moreover, the authors suggest to use transformers for encoding all three modalities, which leads to a more uniform but efficient architecture.

\section{Downstream Audio Tasks \& Benchmarks}
\label{sec:ds}
\noindent
%Although most audio SSL papers presented above evaluate algorithms performance using ASR task, which provides relative fair comparisons for language models, 
After solving pretext tasks, an audio SSL model is expected to produce high-quality audio representations that are of sufficient generalisation and discrimination, by this, guaranteeing a good  performance on downstream tasks. %abilities for performing well on downstream tasks. 
Several different downstream audio tasks have been considered for empirically measuring the audio representation quality. For example, ASR is used for evaluating all Wav2vec based methods \cite{baevski2019vq,baevski2020wav2vec,schneider2019wav2vec}. Other tasks include 
%Self-supervised learning has been shown to be effective forhas also been specifically adopted for several other audio applications, including 
speaker identification \cite{micro2019,Ravanelli2018SpeakerRF,chen2021}, 
speech emotion recognition \cite{jiang2019improving,neumann2019improving,nandan2020,shamane2020}, 
speech machine translation \cite{ha2020}, pitch detection \cite{engel2020},  and acoustic scene classification  \cite{gontier2021polyphonic}, amongst others.

%\hl{The rapid increase in scientific activity on representation learning has been accompanied and nourished by a remarkable string of empirical successes both in academia and in industry.} \emi{You need to be super careful with copy-paste.  This sentence can be found  here:  \url{https://www.gabormelli.com/RKB/2013_RepresentationLearningAReviewan} You need to paraphrase it.} 
%\emi{I think when writing a review this might easily happen. Could we do a plagiarism check before submitting? Do you know any free suitable tool? (Maybe the university has one) I think this would be a good idea, to be sure that we have paraphrased everything.}
Hereby, in the following we will describe  some publicly available benchmarks that enable a %provide the convenience for
fair comparisons between different audio SSL algorithms.   %\cite{tu2020,yang2021,EVAIN2020,kahn2020libri}.

The Zero resource Speech challenge (ZeroSpeech) \footnote{https://zerospeech.com} \cite{tu2020} started their first challenge in 2015 with the task of unsupervised discovery of linguistic units from raw speech in an unknown language. The tasks are split into two tracks, \ie unsupervised sub-word modelling and spoken term discovery, each focusing on a different level of linguistic structure. 
The first track aims at constructing a representation of speech sounds that is robust to within- and between-speaker variation and supports word identification.  The second targets at unsupervised discovery of `words', taking raw speech as input.
%Solving these two key tasks can help machines to equip with functionalities like keyword spotting, audio document classification or retrieval, language documentation, etc. 
In 2017, the organisers extend the study for the variants in language and speaker, considering the topic of cross-language generalisation and speaker adaptation. 
The task of ZeroSpeech 2019 was to address the problem of a speech synthesiser without any text or phonetic labels. Participants were expected to discover sub-word units in an unsupervised way given raw audio. Subsequently, they were supposed to  align them to the voice recording (as good as possible) %in a way that works best 
for the purpose of synthesising utterances of target speakers. 
The latest challenge, launched in 2021, provided several tasks for spoken language modelling, based on speech only as well as visually-grounded. Speech-based language modelling consists in learning language models directly from raw audio in an unknown language.   Visually-grounded language modelling aims at learning language models incorporating the visual information. 
%providing speech features that highlight linguistically relevant properties of the speech signal (phoneme structure) and downplay the linguistically irrelevant ones (speaker ID, emotion, channel, etc). 

The Speech processing Universal PERformance Benchmark (SuperB) \footnote{https://superbbenchmark.org/} \cite{yang2021}  aims to present %offer the community  
a standard and comprehensive testbed for evaluation which can be  generally applied  to  pre-trained models on various tasks. Taking into account the available datasets  and  conventional evaluation protocols, ten downstream tasks are provided:   
%to benchmark the capability of SSL pre-trained models on speech processing. 
%These tasks 
Phoneme Recognition, Automatic Speech Recognition, Keyword Spotting, Query by Example Spoken Term Detection, Speaker Identification, Automatic Speaker Verification, Speaker Diarisation, Intent Classification, Slot Filling, and Emotion Recognition.

LeBenchmark \footnote{http://lebenchmark.com/} \cite{EVAIN2020} is another reproducible and multi-faceted benchmark for evaluating speech SSL models for the French language. It is based on four tasks  which aim to assess the speech representations:  Speech Recognition (ASR), Spoken Language Understanding (SLU), Speech Translation (AST), and Emotion Recognition (AER).  %, for measuring the speech representations. 
For reproducibility, the LeBenchmark organisers provided pre-trained SSL models learnt on different subsets of a large and heterogeneous collection of read, prepared, and spontaneous speech utterances in French.

Libri-Light \cite{kahn2020libri} is a benchmark specifically designed for the task of ASR with limited or no supervision. Libri-Light is based on  spoken English audio collected from open-source audio books of the LibriVox project.  

HEAR \footnote{https://neuralaudio.ai/hear2021-holistic-evaluation-of-audio-representations.html}, short for Holistic Evaluation of Audio Representations, extends a benchmark suite for both speech and non-speech tasks. The challenge requests participants to create an audio representation that is as holistic as the human ear. In HEAR 2021, three main tasks are used including word classification, pitch detection, and sound event detection. 

\section{Discussion}
\noindent
In this section, we first clarify the differences and similarities  %correlations 
between SSL methods and other confusing machine learning mechanisms. Next, we discuss the the common problems and difficulties met during the development of SSL models. We further point out some additional concerns regarding audio SSL, considering the difference in data processing and augmentations, negative sample generation, and network construction, compared to the SSL approaches for other modalities.  
%1. self-supervised learning is in between supervised and unsupervised-learning.

\subsection{Difference from Other Confusing Learning Mechanisms}
\noindent
Generally speaking, representation learning aims to capture the posterior distribution of the underlying explanatory factors from the observed input data. A good representation should be of sufficient generalisation and distinctiveness, so that it carries complete salient information of the data that is useful as input for supervised tasks, such as classification. 
Representation 
SSL is  a representation learning approach that  trains a model in order to produce representations. This is achieved by solving specially defined pretext tasks based on, usually very large-scale, data without human annotations. 
This  is different to the classic learning mechanisms of transfer learning and domain adaptation, which %as the latter ones 
learn to generate representations in supervised frameworks, \ie using labelled data. SSL is commonly regarded as an unsupervised leaning method, as the ones using data without human annotations. However, it is also different from classic unsupervised learning, such as clustering, because these kinds of unsupervised learning concentrate on grouping inputs that have similar data patterns,  whereas SSL learns representations with supervision of some automatically created training targets, such as pseudo-labels. 
Likewise, it is considered unsupervised in the sense of no labels from the target task are involved.

Contrastive SSL is highly related to distance metric learning (or simply, metric learning)  \cite{suarez2021tutorial}. Given an anchor paired with positive samples and negative samples, a weakly-supervised metric learning constructs a distance metric that puts positives close together and negatives far away in a  latent space. Hence, contrastive SSL can be seen as a metric learning where the positive pairs are created from the same data source through procedures such as  data augmentation.
Contrastive SSL is also similar to instance discrimination \cite{wu2018unsupervised}. Instead of processing positive and negative pairs, instance discrimination takes each data sample as from a separate class, and learns feature representation that discriminates among individual instances. According to our analysis of \Cref{eq:analysis} given in \Cref{subsec:contrast}, when the temperature parameter is set too small, the InfoNCE loss tends to take the two inputs of a positive pair as the different instances, and therefore optimises the SSL model in a way of the same effect of instance discrimination.

Besides, Generative Adversarial Networks (GANs) are also seen as a kind of SSL framework in some works \cite{jing2020self,liu2021self,le2020contrastive}. For instance,  the generator creates data from a random vector by  taking the real data as training targets. Then, the discriminator network aims  to measure the similarity between  generated and real data. It is worth noticing that the similarity measure  changes, as the discriminator is updated. Such a kind of generative contrastive model has been successfully investigated for NLP tasks, such as in ELETRA \cite{clark2020electra}, but rarely been explored for audio SSL.
Hence, we did not introduce it as an audio SSL form in the literature review, though it should be naturally considered for future works.

\subsection{Difficulties and Problems for SSL Optimisation} 
\noindent
The representation quality using SSL is determined by the efficacy of pretext tasks, of which the key component is the design of training targets or objectives. The training objectives for both predictive and contrastive SSL concentrate on the correlations between representations of observed data. 
Both methods concentrate on maximising the similarity between the representations of the two views of one unique data sample. Additionally,  contrastive methods  contrast the similarity against the distance to other data samples. 

%- mode collapse
Representational collapse often appears when training a predictive SSL model, such as using a Siamese network architecture. To tackle this issue, the pair of networks is usually designed to be of asymmetric architecture and is updated asynchronously. Data augmentation techniques, used to generate different views of an input data, are used to additionally force the Siamese network to process asymmetric input.
Contrastive SSL alleviates the problem of mode collapse by driving the representations of samples, including positives and negatives, to maximal-uniformly distributed appearance on a unit sphere. 
Minimising a contrastive loss, such as InfoNCE, is found to be approximately equivalent to maximise the mutual information between representations. With the rise of the number of negative samples, a lower bound on mutual information is raised up \cite{oord2018representation,le2020contrastive}. Therefore, better representations that carry more correlation information between representations can be obtained by enlarging the amount of negative samples, for instance, as shown in \cite{simclr} and \cite{he2020momentum}. 
Similar considerations have led to the success of contrastive audio SSL \cite{oord2018representation,baevski2020wav2vec,schneider2019wav2vec}. 
% - sampling efficiency
For this to happen, however, the requirement on memory dramatically boosts. Therefore, negative sampling of better efficiency needs to be further explored. 
On the other hand, according to the theoretical analysis in \cite{saunshi2019theoretical}, a too large number of negative samples may not be profitable for training contrastive SSL models. So far, no research has been done to suggest a golden standard rule for setting a proper number of negatives. Moreover,   the setting should  potentially be  considered differently for different tasks and applications.
%- early degeneration....
Another issue that can hamper contrastive SSL is early degeneration, which means that the SSL model over-fits to the discriminative pretext task in very early training steps, and therefore, the representations do not present a  sufficient generalisation ability. Solutions that can relax this early degeneration issue should also be addressed in future work.  

%tends to get trapped into embedding spaces
%\subsection*{The effect of more negative samples}  
% 
%Constrastive training objectives such as InfoNCE, that are commonly used for audio SSL incorporates negatives in training. 
 %For audio processing in , this is to maximise the mutual information between local representation and a global contextualised representations.

\subsection{Additional adjustments on SSL for Audio}
\noindent
As introduced above, SSL approaches that have been well explored for CV and NLP tasks, are being transferred to the audio domain. For this, some works process 1D audio data into a 2D format, in order to match the formulations of these SSL frameworks. 
For example, time-frequency representations of audio and the advanced transformations based on it, such as a spectrogram, Mel-spectrogram, and MFCC can be used as `image' to some SSL models designed for CV tasks  \cite{saeed2021contrastive,niizumi2021byol,liu2021tera}. For this case, data augmentation techniques widely used in the CV domain have also been considered, which are essential for achieving high-quality representations.
Taking the features as sequential frames, we can process them in  similar ways as we would do  for NLP tasks \cite{an2018,audio2vec}.

An alternative way is to directly process the 1D waveform using deep learning encoders, such as 1D convolution, able to convert the 1D signal into higher-dimensional features for further processing. This solution has been successfully used in  \cite{oord2018representation,baevski2020wav2vec,Pascual2019LearningPS}.
The focus of this paper was not assessing %This paper did not address too much to 
network architectures, but rather concentrating on the framework and formulations of SSL approaches. As the network architecture is found to be less important than formulations in achieving good representations, ResNet is typically used in most visual SSL works. However, the importance of neural network architectures are not that clear for audio SSL. Researchers tend to use the network architectures that were designed to respect the speech or audio structure, which can achieve more promising results in the context of audio SSL. Still, more research evaluating  the effect of network components, such as assessing the effect of the attention mechanism used in transformers \cite{shuwen2020}, should be carried out.

\subsection{Fitness and mismatch between pretext and downstream tasks}
\noindent
In general, we expect that by  training a model with SSL, it is possible to learn general representations that are effective for downstream tasks. Although this is slightly different from classic transfer learning which performs pretext tasks in a supervised framework,
the gap between the source data in pretext tasks and target data for downstream tasks is expected to be matched. 

Comparing speech and other audio signals, such as acoustic scene recordings, the speech signal is more variable from a temporal and frequency perspective, while the acoustic scene recordings are usually more stationary along the temporal axis. Hence, an SSL model that is trained  on a speech signal may still cover the ability to learn discriminative  representations  to be used in  acoustic scene classification. However, in the reverse way, an SSL model trained  on scene recordings may still work for simple speech-related classification tasks, such as speech command recognition, but its performance  would be weak in more advanced speech tasks, such as ASR. 

In the standard framework of SSL, labelled data is used in downstream tasks for fine-tuning. It has been shown that a small quantity of labelled data can already guide a pre-trained model to achieve very satisfying performance on downstream tasks. This inspired semi-supervised learning using very little human-labelled data from the target data domain, for closing the gap between source and target data. 
Specifically, the training objectives of SSL and supervised learning are combined and optimised simultaneously. 
For many audio applications, SSL approaches have shown promising performance and reached (or even  surpassed)  state-of-the-art results achieved through supervised learning. Still, when labels are available or partly available, like in CLAR \cite{al2021clar} and UniSpeech \cite{Wang2021UniSpeech}, combining SSL and SL together into a multi-task learning setting enables to learn better speech representations for some audio tasks.

\section{Conclusion}
\label{sec:conclude}
\noindent
This survey has provided an overview of the existing approaches and methods for 
%\am{\st{audio self-supervised learning and multi-modal self-supervised learning that use audio} 
uni-modal and multi-modal self-supervised learning approaches using audio. The success of these methods has been analysed in several classic audio tasks, including speech recognition, speaker identification, speech emotion recognition, and acoustic scene classification. Audio SSL methods, such as Wav2Vec 2.0, have shown to even surpass the performance of supervised learning methods on the same task. Moreover, the generalisation ability of representations learnt using audio SSL can  %relax the pressure 
decrease the urgency of searching for hand-crafted, engineered features. 
%From b
%\shuo{Both perspectives, the superior performance on downstream tasks and the generalisability of the representation learning method,} 
%\emi{? which are the 2 perspectives?} \sure{First, the performance of audio ssl outperforms supervised methods. Second, the generalisation ability, an ssl model can be used for multiple downstream tasks. Therefore focusing on developing SSL model is "achieve many things at one stroke".}, %it leads to out belief 
%EMI: You coudl simplify the beggining fo hte sentence by simply saying "The presented overview suggests" 
%suggesting that self-supervised learning is the present and future for audio processing. 
%\shuo{And focusing on developing SSL model is achieving many things at one %stroke.}
%\am{
The superior performances obtained using SSL-based approaches support the generalisation capabilities of this representation learning method, and encourage the use of this technique to shape the future and advance the state-of-the-art in the field of audio processing. 
%}

\section*{Acknowledgements}
\noindent
This project has been supported in part by the European Union's Horizon 2020 research and innovation programme under grant agreement No.\,826506 (sustAGE), and by the BIT Teli Young Fellow Program from the Beijing Institute of Technology, China.

% {\appendix[Proof of the Zonklar Equations]
% Use $\backslash${\tt{appendix}} if you have a single appendix:
% Do not use $\backslash${\tt{section}} anymore after $\backslash${\tt{appendix}}, only $\backslash${\tt{section*}}.
% If you have multiple appendixes use $\backslash${\tt{appendices}} then use $\backslash${\tt{section}} to start each appendix.
% You must declare a $\backslash${\tt{section}} before using any $\backslash${\tt{subsection}} or using $\backslash${\tt{label}} ($\backslash${\tt{appendices}} by itself
%  starts a section numbered zero.)}


% \section{References Section}
% You can use a bibliography generated by BibTeX as a .bbl file.
%  BibTeX documentation can be easily obtained at:
%  http://mirror.ctan.org/biblio/bibtex/contrib/doc/
%  The IEEEtran BibTeX style support page is:
%  http://www.michaelshell.org/tex/ieeetran/bibtex/
 
% You can manually copy in the resultant .bbl file and set second argument of $\backslash${\tt{begin}} to the number of references
%  (used to reserve space for the reference number labels box).
\bibliographystyle{IEEEtran}
\bibliography{citations}

\vfill

\end{document}